\documentclass[phd,10pt,noupper,appendixpart,oneside,useAMS]{ubcthesis}
\usepackage[hypertex]{hyperref}
\usepackage{amssymb}
\usepackage{graphicx}

\institution{The University Of British Columbia}
\institutionaddress{Vancouver, Canada}
\program{Physics}
\numberofsignatures{4}          

\previousdegree{B.Sc., The Chinese University of Hong Kong, 2001}
\previousdegree{M.Phil., The Chinese University of Hong Kong, 2003}

\title{Cosmological Recombination}
\author{Wan Yan Wong}
\copyrightyear{ 2008}
\ubclocation{Vancouver}
\submitdate{August, 2008}

\advisor{Douglas Scott}
\advisortitle{Professor of Astronomy}
\renewcommand\thepart         {\Roman{part}}
\renewcommand\thechapter      {\arabic{chapter}}
\renewcommand\thesection      {\thechapter.\arabic{section}}
\renewcommand\thesubsection   {\thesection.\arabic{subsection}}
\renewcommand\thesubsubsection{\thesubsection.\arabic{subsubsection}}
\renewcommand\theparagraph    {\thesubsubsection.\arabic{paragraph}}

\begin{document}
\frontmatter

\maketitle
\authorizationform

\begin{abstract}
In this thesis we focus on studying the physics of 
 cosmological recombination and how the details of 
 recombination affect the Cosmic Microwave 
 Background~(CMB) anisotropies.
We present a detailed calculation of the spectral line 
 distortions on the CMB spectrum arising from
 the Ly$\,\alpha$ and two-photon  
 transitions in the recombination of hydrogen~(H), 
 as well as the corresponding lines from helium~(He).  
The peak of these distortions mainly comes from 
 the Ly\,$\alpha$ transition and occurs at about 
 $170\,\mu$m, which is the Wien part of the CMB. 
The detection of this distortion would provide the most
 direct supporting evidence that the Universe was 
 indeed once a plasma.

The major theoretical limitation for extracting cosmological 
 parameters from the CMB sky lies in the precision with which
 we can calculate the cosmological recombination process.  
Uncertainty in the details of hydrogen and helium recombination
 could effectively increase the errors or bias the values of the
 cosmological parameters derived from microwave 
 anisotropy experiments.
With this motivation, we perform a multi-level calculation
 of the recombination of H and He with the addition of the 
 spin-forbidden transition for neutral helium (He\,{\sc i}),
 plus the higher order two-photon transitions 
 for H and among singlet states of He\,{\sc i}.  
Here, we relax the thermal equilibrium assumption among 
 the higher excited states to investigate the effect 
 of these extra forbidden transitions on the ionization
 fraction $x_{\rm e}$ and the CMB angular power spectrum $C_\ell$.  
We find that the inclusion of the spin-forbidden transition 
 results in more than a percent change in $x_{\rm e}$, while
 the higher order non-resonance two-photon transitions give 
 much smaller effects compared with previous studies.

Lastly we modify the cosmological recombination code
 {\sc recfast} by introducing one more parameter to
 reproduce recent numerical results for the speed-up 
 of helium recombination.  
Together with the existing hydrogen `fudge factor', 
 we vary these two parameters to account for the 
 remaining dominant uncertainties in cosmological 
 recombination.
By using a Markov Chain Monte Carlo method with 
 {\sl Planck} forecast data, we find that we need
 to determine the parameters to better than 
 10\% for He\,{\sc i} and 1\% for H, in order to 
 obtain negligible effects on the cosmological parameters.
\end{abstract}

\tableofcontents
\listoftables
\listoffigures
\chapter{List of Symbols}
\begin{center}
\begin{tabular}{c l}
$\alpha_{\rm x} $ & Case B recombination coefficient of species x \\
$\beta_{\rm x} $ & Case B photoionization coefficient of species x \\
$\Delta^2_{\rm R}$ & Variance of the comoving curvature perturbations \\
$\lambda$ & Wavelength of a photon\\
$\Lambda$ & Cosmological constant \\
$\Lambda_{\rm H}$ & Spontaneous 2s--1s two-photon rate of H\,{\sc i} \\
$\Lambda_{\rm He}$ & Spontaneous 2$^1$S$_0$--1$^1$S$_0$ two-photon rate of He\,{\sc i} \\
$\Lambda^{\rm x}_{j-i}$ & Spontaneous two-photon rate of species x from $j$th state to $i$th state  \\
$\mu$ &  Chemical potential in the radiation spectrum \\
$\nu$ &  Frequency of a photon\\
$\rho_{\rm cr}$ & Critical density (zero curvature) \\
$\Omega_{\Lambda}$ & Ratio of dark energy density to the critical density $\rho_{\rm cr}$ \\
$\Omega_{\rm b}$ & Ratio of baryon density to the critical density $\rho_{\rm cr}$ \\
$\Omega_{\rm c}$ & Ratio of cold dark matter density to the critical density $\rho_{\rm cr}$ \\
$\Omega_{\rm m}$ & Ratio of total matter density to the critical density $\rho_{\rm cr}$ \\
$\Omega_{\rm tot}$ & Ratio of total density of the Universe to the critical density $\rho_{\rm cr}$ \\
$\sigma_{\rm T}$  & Thomson scattering cross-section\\
$\sigma(\nu)$ & Ionization cross-section at frequency $nu$\\
$\tau$ & Optical depth \\
$a_{\rm R}$ & Radiation constant, $a_{\rm R}$\,$\equiv$\,$8 \pi^5 k_{\rm B}^4/(15 c^3 h_{\rm P}^3)$ \\
$a_{\ell, m}$ & Amplitude of spherical harmonic component \\
$A_{\rm s}$ & Scalar amplitude of the primordial perturbation \\
$A_{j-i}$ &  Einstein $A$ coefficient of transition from $j$th to $i$th state\\
$B_{j-i}$ &  Einstein $B$ coefficient of transition from $j$th to $i$th state\\
$b_{\rm He}$ & Fudge factor for He\,{\sc i} recombination \\
$c$   & Speed of light \\
$C_{\ell}$ & CMB anisotropies at angular moment $\ell$\\
$E_i$ & Ionization energy of the $i$th state in an atom \\
$E^{\rm int}$ & Total internal energy of a system with matter and radiation\\
$f_{\rm He}$ &  Number fraction of helium nuclei, $f_{\rm He}$\,$\equiv$\,$n_{\rm He}/n_{\rm H}$ \\
$F_{\rm H}$ & Fudge factor for speeding up the H\,{\sc i} recombination at low redshift \\
$G$   & Newton's gravitational constant \\
$g_i$ & Degeneracy of the $i$th state in an atom \\
$g(z)$ & Visibility function for the CMB photons\\
$H$   & Hubble parameter, expansion rate of the Universe, $H$\,$\equiv$\,$\dot{R}/R$  \\

\end{tabular}
\end{center}
\newpage
\begin{center}
\begin{tabular}{c l}
$H_0$ & Current value of Hubble constant  \\
$h$ & Dimensionless value of $H_0$, $h$\,$\equiv$\,$H_0$/100\,km\,s$^{-1}$\,Mpc$^{-1}$  \\ 
$h_{\rm P}$  & Planck's constant  \\
$I_{\nu}$ & Specific intensity per unit frequency \\
$I_{\lambda}$ & Specific intensity per unit wavelength\\
$\bar{J}$ & Specific intensity per unit frequency from a blackbody  \\
$k$ & Wavenumber or inverse scale of primodial fluctuation \\
$k_{\rm B}$ & Boltzmann's constant \\
$\ell$ & Multipole of the CMB temperature fluctuation \\
$l$ & Angular momentum of a level in an atom \\
Mpc & Mega-parsec ($10^6$\,pc), 1\,pc\,=\,3.26156 light years\,=\,3.0857$\times$\,$10^{16}$\,m \\
$m_{\rm e}$ &  Electron mass \\
$m_{\rm p}$ &  Proton mass \\
$m_{\rm H}$ &  Mass of hydrogen atom \\
$m_{\rm He}$ & Mass of helium $^4$He atom \\
$n$ & Principle quantum number of a level in an atom \\
$n_{\rm x}$ &  Number density of nucleus of species x\\
$n_{\rm e}$ &  Number density of electrons \\
$n^{\rm x}_i$ & Number density of electrons in the $i$th level of atom x \\
$n_{\rm s}$ &  Index of power spectrum of primodial fluctuations \\ 
$p_{\rm s}$ &  Sobolev escape probability of photons\\
$R(t)$ & Scale factor for universal expansion \\
$\Delta R^{\rm x}_{j-i}$ & Net transition rate from $j$th state to $i$th state  of species x\\
$T_{\rm M}$ & Matter temperature \\
$T_{\rm R}$ & Radiation temperature \\
$T_0$ & Current radiation temperature, $T_{\rm R}(z=0)$ \\
$U$ & Radiation energy density \\
$x_{\rm e}$ & Ionization fraction or free electron fraction, $x_{\rm e}$\,$\equiv$\,$n_{\rm e}/n_{\rm H}$  \\
$y$ &  Compton-scattering distortion parameter \\
$Y_{\rm p}$ & Primordial mass fraction of $^4$He \\
$Y_{\ell,m}$ & Spherical harmonics \\
$z$ & Redshift \\
\end{tabular}
\end{center}

\acknowledgements
First I would like to thank my supervisor, Douglas Scott 
for his ideas, encouragement and patience.
He introduced me to the field of cosmology and 
guided me through my research projects. 
I have learned a lot through stimulating discussions 
with him and he always shares his ideas openly 
in different aspects of physics and astronomy. 
 
I would also like to thank the other collaborators in this work.
Sara Seager generously shared her original numerical 
recombination code and shared with me her understanding
of recombination. She also provided me hospitality during
my stay at the Carnegie Institute of Washington.  
And Adam Moss helped me to make the C{\sc osmo}MC 
code work properly.

I would like to thank the Astronomy group at the University 
of British Columbia.  The professors provided an interactive,
warm and helpful environment for me to study here.  And 
the graduate students, especially my officemates, gave me a
sense of what is Canadian culture.
I would also like to thank the staff in St.\,John's College, 
especially the kitchen chefs.  They provided me with a comfortable 
stay and wonderful meals during my two years of living there.

Here I would also like to thank my friends for all their support.  
In particular, Kandy Wong and Cecilia Mak always help me 
out and bring lots of fun to my life in Vancouver. 

I owe my father and mother many thanks.  They brought me 
into this amusing world and allowed me to do whatever I like to do.
I thank my brother Ting Chun Wong for taking care of the family
when I am away from home.

And to my $\heartsuit$husband Henry Ling.  He always
supports and helps me through the difficult times.

\newpage

\chapter{Co-Authorship Statement}
This thesis is in the manuscripted format and Chapters~3 to 6 
are essentially reprints of individual published works~(see the footnotes of
the first page in each chapter for references).  
My supervisor, Professor Douglas Scott provided many useful 
discussions during all of these works and also gave me numerous
suggestions in editing the papers, but in each case 
the calculations and writing are on my own. \\
{\bf Chapter 3} \\
Professors Sara Seager and Douglas Scott are the co-authors of the work in
Chapter~3, and initiated this project.  
The numerical recombination code was originated and 
developed by Professor Sara Seager before this work started.  
I performed all the calculations by modifying the relevant parts in 
the numerical code, analyzed the results and wrote the manuscript.
\\
{\bf Chapter 4} \\
Professor Douglas Scott is the co-author of the work in Chapter~4, and 
motivated me to start this project.  I collected and updated the atomic data
in the numerical code originally developed by Professor Sara Seager.  
Modifications were made in the numerical code specifically for
the study in this Chapter.  I performed all 
the theoretical and numerical calculations, analyzed the results
and wrote the manuscript.
\\
{\bf Chapter 5} \\
Professor Douglas Scott is the co-author of the work in Chapter~5, and 
motivated me to clarify this previous claimed effect on the recombination
calculation.  I developed the consistent approach 
under the equilibrium assumption and also estimated the maximum 
effect in the real situation.  I also wrote the manuscript of this work.
\\
{\bf Chapter 6} \\
Professor Douglas Scott and Dr.~Adam Moss are the co-authors of the 
work in Chapter~6.  Professor Douglas Scott motivated me to start
this project.  I developed the method and modified the 
existing {\sc recfast} recombination code by including the 
recent updates and uncertainties.  Dr.~Adam Moss provided the
{\sl Planck} forecast data and helped me in running the 
C{\sc osmo}MC code.  I performed all the numerical calculations,
analyzed the results and wrote the paper.

\mainmatter

\chapter{Introduction}

The detection of the 2.725\,K Cosmic Microwave Background~(CMB)
 is one of the strongest pieces of supporting evidence for the Big Bang 
model, which is the widely accepted theory for the history of the Universe. 
Together with other observations, we know that the Universe
is expanding implying that it was much denser
and hotter in the past and used to be a plasma of ions
and electrons.  The CMB, which is the remnant of the early radiation,
was last scattered when the atoms became neutral. 
This period is called cosmological recombination, and it happened
when the Universe was a few hunderd thousand years old.
In the decades following its discovery, 
the CMB was found to be remarkably homogeneous and isotropic, but 
its tiny temperature fluctuations give us the most distant 
image we have of the Universe.  This carries important information 
about the geometry, the expansion rate and contents of the Universe,
as well as clues about the origin  of all the structure it contains
(see, for example, \cite{1Peacock:1999,1Peebles:1993}).
Exploiting this information requires an extremely precise
undertanding of the process of cosmological recombination, 
which is the main topic of this thesis.  
In order to explain why this is the case we should
first review the physics of the standard cosmological
model.

\section{A brief history of the Universe}
In the late 1920s, Hubble\,\cite{1Hubble:1929} discovered that
 the Universe is expanding. 
He found that atomic lines in the spectrum of nearly all
distant galaxies are redshifted~(or shifted to longer wavelengths) 
compared with the laboratory values.  This means that
the galaxies are moving away from us due to the expansion
of the Universe.  The redshift $z$ is defined as  
\begin{equation}
1+z \equiv \frac{\lambda_{\rm obs}}{\lambda_{\rm emit}} 
= \frac{R(t_{\rm obs})}{R(t_{\rm emit})} \, ,
\end{equation}
where $\lambda_{\rm obs}$ and $\lambda_{\rm emit}$ are the observed
 and emitted wavelengths, respectively.  
Here $R(t)$ is a time-dependent scale 
 factor, which gives infinitestimal distances in space 
 when multiplied by the comoving distance $dr$.
This idea of a uniform scale factor for the expansion
is consistent with Hubble finding that the velocity of 
galaxies $v$ increases linearly with distance $r$, 
which is the famous Hubble's law:
\begin{equation}
v = H r\, .
\end{equation}
Here $H$ is the Hubble constant and represents the rate
of expansion so that
\begin{equation}
H = \frac{\dot{R}}{R}\, ,
\end{equation}  
and today we have $H_0$\,$\equiv$\,$H(t_0)$.
Although the actual value of the constant determined by Hubble is far from 
our current estimates, the Hubble diagram nevertheless
proves that the Universe is expanding,
and the same principle is used for
today's measurements: measure the redshifts and
estimate the distances of distant objects to determine $H$.
Redshift, can be easily estimated from the
shifting of the spectral lines, but  
it is hard to determine the distances without 
any information of on the intrinsic brightness or the intrinsic size
of an object, so that precision measurements of Hubble's
constant have been elusive.

The current value of the Hubble constant $H_0$~(the 
subscript `0' represents the present value, that is at $z$\,=\,0)
was determined by the Hubble Key Project\,\cite{1Freedman:2001}  
using `standard candles', which basically have the 
same intrinsic brightness or have a correlation between
some observables and the intrinsic brightness.  For example,    
Cepheid variables and Type~Ia supernovae are commonly used 
standard candles.  The measured value of $H_0$ is equal to 
$72 \pm 8$\,km\,s$^{-1}$Mpc$^{-1}$\,\cite{1Freedman:2001}.
We usually define a dimensionless constant for $H_0$, which is
\begin{equation}
h \equiv \frac{H_0}{100 \,{\rm km}\,{\rm s}^{-1}{\rm Mpc}^{-1}}\, .
\end{equation}
and therefore, $h$\,=\,0.72\,$\pm$\,0.08.
Assuming $\dot{R}(t)$ is constant, the age of the Universe is 
then equal to $1/H_0$, which is about 13.7\,Gyr.

The Universe appears to be homogeneous and isotropic on large scales
(distances greater than about 300\,Mpc) from observations of the distribution of
galaxies\,\cite{1Peebles:1980}. 
 This is the Cosmological Principle; based on that we can 
build a simple model of the expanding Universe within General Relativity.
Here we temporarily ignore the density fluctuation on small scales, 
which are of small amplitude in the early Universe but 
important later for the formation of galaxies and clusters (the structure formation).
On large scales, the Universe can be described by the
Friedmann-Robertson-Walker (FRW) metric and the geometry of the
 Universe depends on the total density~(see,
 for example, \cite{1Peacock:1999,1Peebles:1993}).  
Given the expansion rate~$H$, there is a critical 
density $\rho_{\rm cr}$ that determines whether the Universe 
has flat geometry.
 This critcal density is
\begin{equation}
\rho_{\rm cr} = \frac{3 H^2}{8 \pi G} \, ,
\end{equation}
and we usually define a density parameter 
\begin{equation}
\Omega_{i} = \rho_i / \rho_{\rm cr}\, ,
\end{equation}
where $i$ represents different components (e.g. matter, radiation 
and dark energy) in the Universe.
The Universe is spatially closed 
 if the total density of the Universe is larger than $\rho_{\rm cr}$,
 and  spatially open 
if its density is lower than $\rho_{\rm cr}$.
  
\begin{figure} [!ht]
\begin{center}
\includegraphics[width=0.8\textwidth]{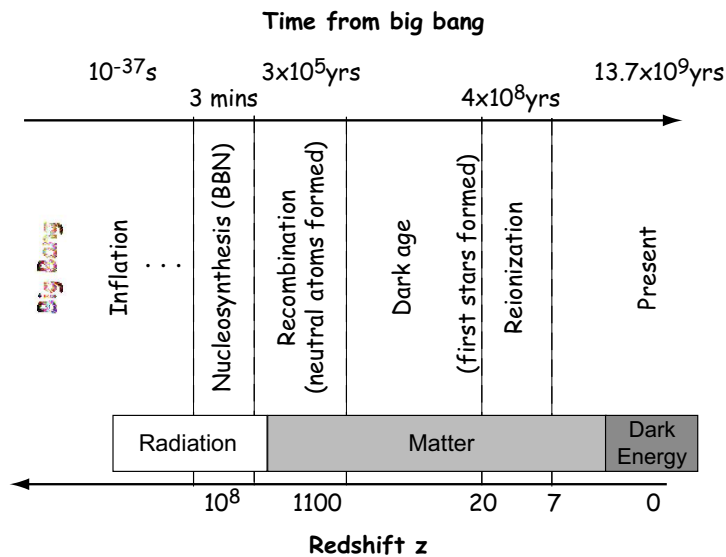}
\caption[A schematic picture of a brief history of the Universe.]
{A schematic picture of a brief history of the Universe. 
 Boxes indicate the periods
when radiation, matter or dark energy are dominant.}
\label{universe}
\end{center}
\end{figure}

From the combined results of recent observations of 
the CMB, acoustic signatures in galaxy clustering
 and Type~Ia supernovae, the Universe is found to 
be very close to flat with the total density 
$\Omega_{\rm tot}$\,$=$\,1.0052$\pm$\,0.0064\,\cite{1Hinshaw:2008}.
At the present time the Universe consists of
about 4\% baryons ($\Omega_{\rm b}$), 20\% cold dark matter 
($\Omega_{\rm c}$), 76\% dark energy ($\Omega_{\Lambda}$) and 
a tiny portion of photons ($\Omega_{\rm R}$\,$=$\,4.17$\times$\,$10^{-5} h^{-2}$).  
Here `baryon' means ordinary matter, for example atoms, 
nuclei and electrons.  
`Dark matter' is some gravitationally interacting
(weakly interacting with baryons) and non-luminous substance. 
The dark matter is considered to have velocity dispersion
which is negligible for structure formation, meaning that
it decoupled when it was non-relativistic (cold);
its fluctuations are the seeds of structure formation.
The concept of dark matter was first proposed by 
Zwicky\,\cite{1Zwicky:1933} in 1933 through observations at the 
rotational curves of stars in galaxies.  This dark
matter was introduced in order to explain the increasing
rotational velocity of material with increasing distance from 
the centres of galaxies.  As we will see later, since 
Big Bang Nucleosynthesis (as well as the CMB) gives a 
very low limit on the baryon density,
some non-baryonic matter must exist in the Universe. 
In general, the density of matter $\rho_{\rm m}$ is proportional 
to $(1+z)^3$, while that of the radiation $\rho_{\rm R}$ is 
proportional to $(1+z)^4$.  
The dark energy provides the negative pressure 
responsible for the recent accelerated expansion of 
the Universe and its density is constant over redshift 
(this is Einstein's cosmological constant, $\Lambda$) 
or very nearly so.  

Due to different scalings of the density of each species with 
redshift, the components dominate the Universe at different times. 
Figure~\ref{universe} shows a brief history of the Universe
and indicates the important epochs using both time and 
redshift as coordinates.
In cosmology, redshift $z$ is usually used instead of time,
 since it is (in principle at least) directly observable 
and so independent of the cosmological model.  
In the Big Bang picture, the very early times are still
quite uncertain, but the physics of the thermal history of
the Big Bang Nucleosynthesis~(BBN) and recombination 
are well understood and firmly established.

The earliest times were radiation dominated. The Universe was 
very hot (the background radiation temperature $T$\,=\,$T_0 (1+z)$, 
where $T_0$\,=\,2.725\,K) and dense.
Due to the strong and highly energetic photon background, there 
were no bound nuclei until BBN occured at about 
3 minutes after the Big Bang ($z \simeq 10^8-10^9$).
During BBN, the temperature decreased to about 100 keV/$k_{\rm B}$,
which is lower than the typical binding energy of the nuclei. 
Therefore, nuclei of  deuterium (D), helium ($^3$He, $^4$He) and 
lithium ($^7$Li) were able to form without being destroyed by the photons.
Given the baryon density $\Omega_{\rm b}$, the theoretical calculation
of standard BBN can predict the abundance of different species of 
nuclei with very small uncertainties due to nuclear and weak-interaction
rates~(see Figure~1 in \cite{1Burles:1999} or Figure~5 in \cite{1Steigman:2007xt}).
In particular, the abundance of D is very sensitive to $\Omega_{\rm b}$.
By measuring the primodial abundance of D through the absorption lines in 
the hydrogen clouds at redshift $z \simeq 3-4$, we can put tight constraints
on $\Omega_{\rm b}$ using the theoretical BBN prediction~(see 
\cite{1Burles:1999} and references therein).
BBN gives a limit that the baryons can contribute at most 
5\% of the critical density, and therefore the rest of the matter
must be non-baryonic.

At about $3 \times 10^5$\,years ($z \simeq 1100$) after the 
Big Bang, the radiation temperature dropped to 
around 1\,eV/$k_{\rm B}$, which is lower
 than the ionization energy of typical atoms.  
 This period is called cosmological recombination. 
 During this time, the ions and electrons were able to 
 bind together without being ionized by the background photons.
After the Universe became neutral, the photons were no longer 
scattered by the electrons and could basically travel freely to 
the present, being redshifted in the expanding Universe. 
These are the CMB photons that we detect today. 
The CMB has been found to be remarkably smooth, the amplitude of the 
temperature deviations $\Delta T/T$ is only about $10^{-5}$, 
which is a strong contrast to the non-linear structure formed
by the galaxies and clusters we observe today.
Therefore this fluctuation amplitude of temperatures 
in the CMB gives us an idea about the strength of the matter
 density fluctuations at the time of recombination,
which evolved into the large scale structures we observe now.  

After recombination, the Universe remained dark and 
neutral $(20 \leq z \leq 900)$ until the first stars formed.
There has not been any detection of informtion from this 
`dark age' and we are still not sure how and when 
exactly the first stars formed.  
Up until now, the most distant quasar that has been observed 
is at about $z=6.5$\,\cite{1Jiang:2008,1Willott:2007}.
From the hydrogen absorption line spectra from such high-$z$
quasars\,\cite{1Becker:2001} we know that the Universe was
fully ionized by ultraviolet radiation from hot 
stars at $z \lesssim 6$.  
Moreover the CMB provides a constraint on 
the optical depth $\tau_{\rm reion}$ during this
reionization epoch through the Thomson scattering effect
on the photons.  The integrated optical depth is 
\begin{equation}
\tau_{\rm reion} = 
\int_0^{z_{\rm reion}} c\, \sigma_{\rm T} n_{\rm e}(z) \frac{dt}{dz} dz \, ,
\label{1eqtau}
\end{equation}
where $\sigma_{\rm T}$ is the Thomson scattering cross-section, 
$n_{\rm e}$ is the number density of free electrons and 
$z_{\rm reion}$ is the redshift at which the Universe became ionized.
From the latest CMB measurement and assuming that
the Universe became fully ionized instantaneously,
the current estimate is
 $z_{\rm reion}$\,$\simeq$\,11\,\cite{1Hinshaw:2008}. 
Stars and galaxies are created basically due to the 
gravitation collapse of dense regions, but the process 
is non-linear and also involves the pressure of the gas.   
Therefore, although the current matter inhomogeneites 
in the Universe and the temperture fluctuations of 
the CMB originated from the same source, they appear 
very different today.

In inflationary models, the primodial perturbations are generated by quantum
fluctuations~(see \cite{1Peacock:1999,1Peebles:1993} for a general review).  
For the simplest model, by assuming the matter is adiabatic and 
its fluctuations are Gaussian, the initial conditions for 
density perturbations can be described by only two parameters: the
scalar amplitude $A_{\rm s}$ and the spectrum index $n_{\rm s}$ 
(the slope of the power spectrum; the subscript `s' distinguishes
these scalar perturbations from possible tensor, or gravity wave, 
contributions).  
The variance of the comoving
curvature perturbations is usually defined as\,\cite{1Scott:2006}
\begin{equation}
\Delta_{R}^2 = A_{\rm s} \left( \frac{k}{k_0} \right)^{n_{\rm s}-1},
\end{equation}
where $A_{\rm s}= \Delta_{R}^2(k_0)$, $k$ is the wavenumber and 
$k_0=0.05$\,Mpc$^{-1}$.

Since the CMB photons come from the time before stars formed,
the anisotro-pies in the CMB provide us with information about
density perturbations at the recombination time and 
in combination with measurements made today, they are a powerful
 tool for constraining the parameters of the cosmological model. 
From the above discussion, and assuming a flat Universe, 
the standard cosmological model
(the $\Lambda$ Cold Dark Matter model, $\Lambda$CDM)
 consists of six parameters:
$\Omega_{\rm b}$, $\Omega_{\rm m}$, $h$, 
$\tau_{\rm reion}$, $A_{\rm s}$ and $n_{\rm s}$.
There could of course be more parameters in the cosmological model
(see \cite{1Lahav:2006} for a review), for example, including
the tensor mode of the primodial perturbations or allowing 
the Universe to deviate from flatness ($\Omega_{\rm tot} \neq 1$).

Since the CMB photons were mostly last scattered during the 
epoch of cosmological recombination, we need to understand 
in detail how the photons decoupled from the matter 
during that period in order to obtain the correct CMB 
anisotropy power spectrum for constraining the 
cosmological parameters using the observations.
In this thesis, we focus on the physics of
 recombination and how the details of the
 recombination process affects the CMB.
We now therefore present an introduction to the 
physics of cosmological recombination (the last 
scattering surface of the CMB photons), and also 
the basic principles of the formation of the CMB anisotropies.

\section{Cosmological recombination}
\begin{figure} 
\begin{center}
\includegraphics[width=0.9\textwidth]{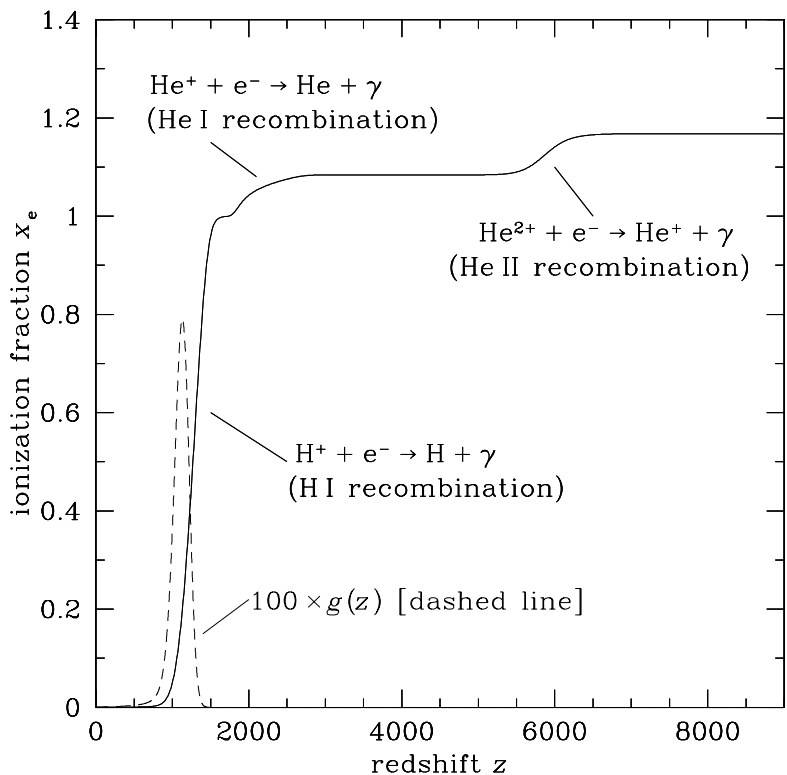}
\caption[The ionization history for cosmological recombination
generated by the current version of {\sc recfast}.]
{The ionization history for cosmological recombination
generated by the current version of {\sc recfast}.  
The dashed line shows the visibility function $g(z)$
as a function of redshift (multipied by 100 for better
illustration).
The cosmological $\Lambda$CDM model used here has: $\Omega_{\rm b}$\,=\,0.04;
$\Omega_{\rm m}$\,=\,0.24; $\Omega_{\Lambda}$\,=\,0.76;
$h$\,=\,0.70; $Y_{\rm p}$\,=\,0.25; and $T_0$\,=\,2.725\,K.}
\label{1xe_recom}
\end{center}
\end{figure}

Recombination in an expanding Universe is not an instantaneous process.
It is basically controlled by the recombination
 time and by the Hubble expansion time.
If the recombination time is much shorter than the expansion time, 
 then the electrons and ions follow an equilibrium distribution.
For the ionization of a plasma, the equilibrium situation is 
described by the Saha equation. 
Taking hydrogen as an example~(see Equation~(13) in 
\cite{1Seager:1999km} and references therein),
\begin{equation}
\frac{n_i}{n_{\rm e} n_{\rm p}} 
= \left( \frac{h_{\rm P}^2}{2 \pi m_{\rm e} k_{\rm B} T_{\rm R}} \right)^{3/2} 
\frac{g_i}{4} \, e^{E_i/k_{\rm B}T_{\rm R}} \, .
\end{equation}
Here $n_i$ is the number density of electrons in the $i$th
 energy level of the H atom, $n_{\rm p}$ is the number 
 density of free protons, $m_{\rm e}$ is the mass of the electron, 
$k_{\rm B}$ is the Boltzmann constant, $h_{\rm P}$ is Planck's constant, 
$g_i$ is the degeneracy of the energy level $i$ 
and $E_i$ is the ionization energy of level $i$.
Due to the higher ionization energy, helium recombined at 
higher redshifts, first by forming He$^+$ (He\,{\sc ii})
and then neutral He (He\,{\sc i}).  
Hydrogen started to recombine shortly after.  
Figure~\ref{1xe_recom} shows the full ionization 
history of recombination by plotting the ionization
 fraction ($x_{\rm e}$\,$\equiv$\,$n_{\rm e}/n_{\rm H}$, 
where $n_{\rm H}$ is the number density of H nuclei) versus $z$.
Based on standard BBN, about 8\% (by number) of the atomic 
nuclei are helium.  
And since the ionization fraction $x_{\rm e}$ is 
normalized to the total number density of hydrogen, 
$x_{\rm e}$ is equal to about 1.16 when the Universe is fully
ionized.

Peebles~(1968)\,\cite{1Peebles:1968} and 
Zeldovich~(1968)\,\cite{1Zeldovich:1968}
first calculated the H\,{\sc i} recombination evolution 
in detail and found that the recombination process is 
much slower than  Saha equilibrium 
(for example, see Figure~6.8 in\,\cite{1Peebles:1993}).
The Saha equation is good for describing the initial 
departure from full ionization, but the equilibrium situation breaks
down shortly after recombination starts.
When the temperature of the Universe reached 
about 0.3\,eV/$k_{\rm B}$ at $z$\,$\simeq$\,1700, 
there were not enough photons in the Wien tail to keep ionizing the H atoms.
Due to the high photon to baryon ratio 
$n_{\gamma}/n_{\rm b}$\,$\simeq$\,$10^{9}$, 
direct recombinations to the ground state were highly prohibited.  
The `spectral distortion' photons emitted from direct recombination 
are highly energetic and easily re-ionize the nearby neutral atoms. 
This is very similar to the `Case B' recombination familiar
in other areas of astrophysics (see e.g. \cite{1Osterbrock:2006}), 
in which the electrons mostly cascade down to the ground state through 
the first excited state $n$\,=\,2.  However, in cosmological H\,{\sc i}
recombination, the resonant 2p--1s Ly\,$\alpha$ transition is also 
strongly suppressed, because the line is optically thick.  These line photons
can only escape reabsorption through redshifting out of the line 
and the probability for this is very low.  
The other way for the electrons to move from the first 
excited state to the ground state is through the 2s--1s 
two-photon forbidden transition.
Almost half of the electrons cascade down from
the $n$\,=\,2 state through this process~(see Chapter~2 \& 3 for details).  
Overall, the net recombination rate to ground state from $n$\,=\,2
state is lower than the recombination rate into the $n$\,=\,2 state, 
and this causes a `bottleneck', which is responsible for making 
the net recombination rate  much smaller than the one 
given by Saha equilibrium.

In the next chapter, we will discuss details of the radiative 
processes during recombination and also recent development
in performing the numerical calculations. 
However, all the updates are based on the basic picture 
of the standard recombination given here.
We have already discussed how H\,{\sc i} recombination is not
an equilibrium process.  The situation is similar for helium 
recombination.  He\,{\sc i} recombination is also slower
than Saha equilibrium due to the `bottleneck' at the
first excited state, but He\,{\sc ii} deviates from 
the Saha value at only the 0.2\% level due to the relatively 
fast two-photon rate to the ground state\,\cite{1Seager:1999km,1Switzer:2007sq}.

The ionization fraction $x_{\rm e}$ affects the CMB anisotropies 
$C_\ell$ (see Equation (\ref{1dcl}) for the definition of $C_\ell$) 
through the shape of the last scattering surface which is given by the 
visibility function $g(z)$,
\begin{equation}
g(z) = e^{-\tau} \frac{d \tau}{d z}\, ,
\end{equation}
where $\tau$ is the Thomson optical depth during recombination
(excluding the effects of reionization if we are only
considering primary anisotropies).  Here $\tau$ is defined the same as
in Equation~(\ref{1eqtau}), but with different integration limits 
(say, from $z$\,=\,$\infty$ to 100).
One can consider $g(z)$ as the probability that a photon last 
 scattered at redshift $z$.
In Figure~\ref{1xe_recom}, the function $g(z)$ is plotted
 on top of the ionization history of cosmological recombination. 
Since $\tau$ changes rapidly with $z$, $g(z)$ is sharply peaked, 
 and its width gives us the thickness of the last scattering surface
 (which means that the CMB photons we see last scattered in the specific range of 
  redshift 600\,$\lesssim$\,$z$\,$\lesssim$\,1500).
It is usual to define the location of the peak of $g(z)$ as the
 redshift of the recombination epoch, when the radiation 
effectively decoupled from the matter $z_{\rm dec}$.
This is approximately equal to 1100 in the current 
 cosmological $\Lambda$CDM model.  
From the profile of $g(z)$, we can see that H\,{\sc i} 
recombination  affects the $C_{\ell}$ much more than He.  
The later stages of He\,{\sc i} recombination can also 
 change the high-$z$ tail of $g(z)$~(see Chapter~6 for more details),
 but He\,{\sc ii} recombination occurs too early to 
bring any significant effects on $C_\ell$.  

\section{Cosmic microwave background}
\begin{figure} 
\begin{center}
\includegraphics[width=0.85\textwidth]{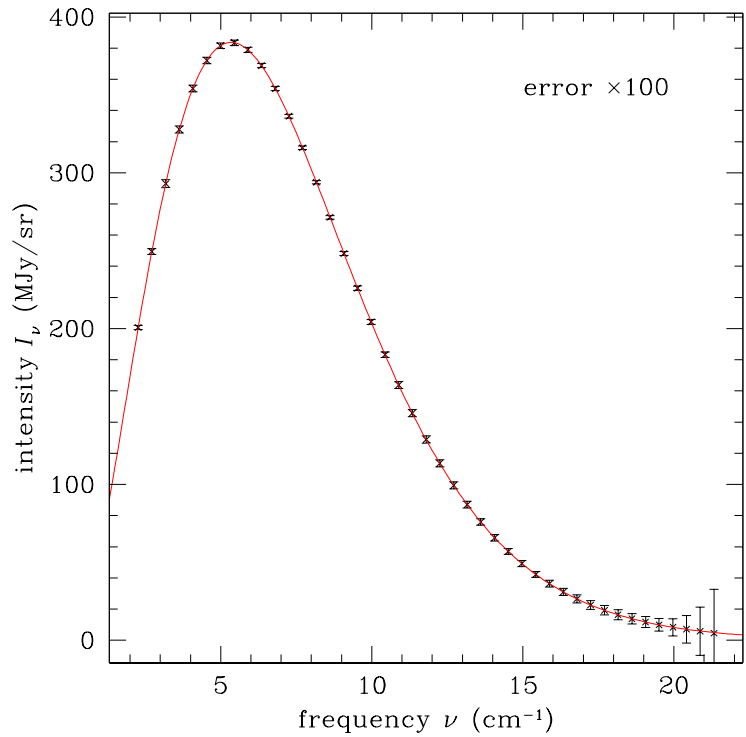}
\caption[Intensity of Cosmic Microwave Background radiation 
as a function of frequency with FIRAS data.]
{Intensity of cosmic microwave background radiation 
as a function of frequency.
The crosses are the data points from FIRAS\,\cite{1Fixsen:1996,1Fixsen:2002}
and the solid line is the expected intensity from a pure blackbody spectrum
with $T_{\rm R}=2.725$\,K.
Note that the plotted one-sigma error bars have been magnified by 100.
Other experiments extend the frequency range, but typically
with much larger errors, and add nothing substantially 
new to the constraints on the spectral shape.}
\label{1firas}
\end{center}
\end{figure}

From many measurements, particularly those of 
the Far-InfraRed Absolute Spectrophotometer~(FIRAS)
on board with the Cosmic Background Explorer
({\sl COBE})\,\cite{1Fixsen:1996,1Fixsen:2002,1Mather:1994}, 
the CMB was found to be very close to a pure blackbody spectrum, which is
described by the Planck function $\bar{J}$:
\begin{equation}
\bar{J} = \frac{2 h_{\rm P} \nu^3/c^2}{e^{h_{\rm P} \nu/k_{\rm B} T_{\rm R}} -1}\, .
\end{equation}
Figure~\ref{1firas} shows the data points from 
 FIRAS\,\cite{1Fixsen:1996,1Fixsen:2002},
 with error bars multiplied by 100 and compared with the theoretical
 blackbody spectrum with $T_{\rm R}$\,=\,2.725\,K.
We can see that the data points match the blackbody shape incredibly
 well within the frequency $\nu$ range from 2 to 20\,cm$^{-1}$
 (i.e. 60 to 600\,GHz).
The deviation is less than 5\,$\times$\,$10^{-5}$ at the peak 
of the CMB spectrum\,\cite{1Fixsen:1996}.
The background photons originate from an 
epoch much earlier than that of recombination,
coming from the electron-positron annihilations 
before BBN and from when the energy of the photons was so high
that bremsstrahlung and double Compton scattering
could create and destroy photons so that they were rapidly
thermalized into a blackbody spectrum\,\cite{1Sunyaev:1970}.
Hence spectral distortion constrain any energy injection
later than that epoch.
The FIRAS data put strong limits on 
the chemical potential $|\mu|$\,$<$\,9\,$\times$\,$10^{-5}$ and the 
Compton-scattering distortion parameter
 $|y|$\,$<$\,1.5\,$\times$\,$10^{-5}$\,\cite{1Fixsen:1996,1Sunyaev:1980}.
These strong constraints eliminated many earlier
competing cosmological models and provide 
strong evidence that the radiation temperature $T_{\rm R}$
scales accurately as (1\,+\,$z$) (see, for example, 
\cite{1Peacock:1999,1Wright:1994} for more details).
The small value of $y$ shows that the hydrogen remained
neutral for quite a long time, otherwise distortions of 
the blackbody spectrum due to Compton scattering by the 
hot electrons would be observed~(see \,\cite{1Sunyaev:1980} 
and references therein).

The other main feature of the CMB is the dipole variation of
the temperature across the sky, with an amplitude equal to
3.358\,mK~(see \,\cite{1Scott:2006} for a review).  
This anisotropy is determined by the
Doppler shift from the solar system's motion relative to 
the `rest frame' of the radiation, which is supported by 
measurements of the radial velocities of relatively local galaxies.
When we talk about the temperature anisotropies of the CMB, 
this contribution from our relative motion is usually removed.

The first detection of the CMB temperature anisotropies was made
by the {\sl COBE} Differential Microwave Radiometer~(DMR; \,\cite{1Smoot:1992}).
The variations in temperature, $\Delta T /T$, were 
found to be of the order of $10^{-5}$.
We usually decompose maps of the CMB temperature fluctuations 
using the spherical harmonic expansion:
\begin{equation}
\frac{\Delta T }{T} \equiv \frac{T(\theta, \phi) - \bar{T}}{\bar{T}} 
= \sum_{\ell, m} a_{\ell,m} Y_{\ell m}(\theta, \phi)\, .
\end{equation}
If the fluctuations are Gaussian and the sky is statistically 
isotropic (independent of $m$), then the temperature field is
 fully charaterized by the amplitudes $C_\ell$, 
\begin{equation}
\langle a^*_{\ell,m} a_{\ell',m'} \rangle = \delta_{\ell \ell'} \delta_{m m'} C_{\ell} \, .
\label{1dcl}
\end{equation}
We usually plot $\ell (\ell +1) C_\ell /2 \pi$, since this is 
the contribution to the variance of the power spectrum per
logarithmic interval in $\ell$~(see, for 
example, \cite{1Hu:2002,1Scott:2006,1White:1994}).
The radiation temperature itself corresponds to the monopole
 $\ell =0$, while the dipole variation corresponds to $\ell =1$.  

Temperature fluctuations in the CMB are essentially
 a projection of the matter density perturbations 
 at the recombination time.
There are many reviews covering details of the 
formation of the CMB anisotropies~(see~\cite{1Hu:2002,1Scott:2006} 
 and references therein)
 and we just briefly recount the basic mechanism here.  
Photons from high density regions were redshifted when
they climbed out of the potential wells (the Sachs-Wolfe effect).
And the adiabaticity between matter and photons also gives a 
higher temperature in higher density regions.
 The other primary source is the oscillating density 
 and velocity of the photon fluid itself.
Before the epoch of recombination, the baryons and the 
radiation are tightly coupled as a single photon-baryon fluid, 
 through Thomson and Compton scatterings.  
The structure seen in the anisotropy power spectrum
 is mainly due to the acoustic oscillations 
 in this photon-baryon fluid, driven by the evolving 
 perturbations in the gravitational potential.  
One can think of these oscillations as
standing waves in a harmonic series, with the 
fundamental mode being the scale which has reached
maximal compression at the time of last scattering.
After recombination, when the Universe became neutral, 
the photons decoupled from the atoms and could 
propagate freely to us~(although there are some 
secondary anisotropies formed when the photons 
travel along the line of sight).  
Therefore, the correct interpretation of the
relationship between the underlying matter fluctuation
spectrum and the photon distribution depends strongly
on the angular diameter distance between us and the last
scattering surface.  
This distance depends on the expansion and curvature 
of the Universe or equivalently, the energy content of 
the Universe.
Therefore, the CMB temperature anisotropies can provide 
precise constraints on the cosmological expansion model,
as well as the scale dependence of the primodial fluctuations.

In addition, the Thomson scattering between electrons and 
photons also leaves a characteristic signature in the 
polarization of the CMB photons.
The quadrupole temperature anisotropy in the photon field
generates a net linear polarization pattern through Thomson 
scattering.  
It has became conventional to decompose the polarization 
pattern into two modes: a part that comes from a divergence 
(`{\sl E}-mode'); and another part from a curl 
(`{\sl B}-mode').  Scalar perturbations (i.e. spatial
variations in density) coming from the inflation 
epoch only give an E-mode signal, while tensor perturbations
(i.e. gravity waves) produce both {\sl E} and {\sl B}-modes.  
Much current activity in CMB experiments is focussed
on trying to measure these {\sl B}-modes, in order
to probe the physics of inflation.
In fact, there are 6 possible cross power spectra from
the full temperature and polarization anisotropy data set.
Cross-correlation between the {\sl B}-mode and either the
{\sl T} or {\sl E}-mode is zero due to having opposite parity.
This leaves us with 4 possible observables: $C_\ell^{TT}$, $C_\ell^{EE}$, $C_\ell^{TE}$ and $C_\ell^{BB}$.

\begin{figure} 
\begin{center}
\includegraphics[width=0.8\textwidth]{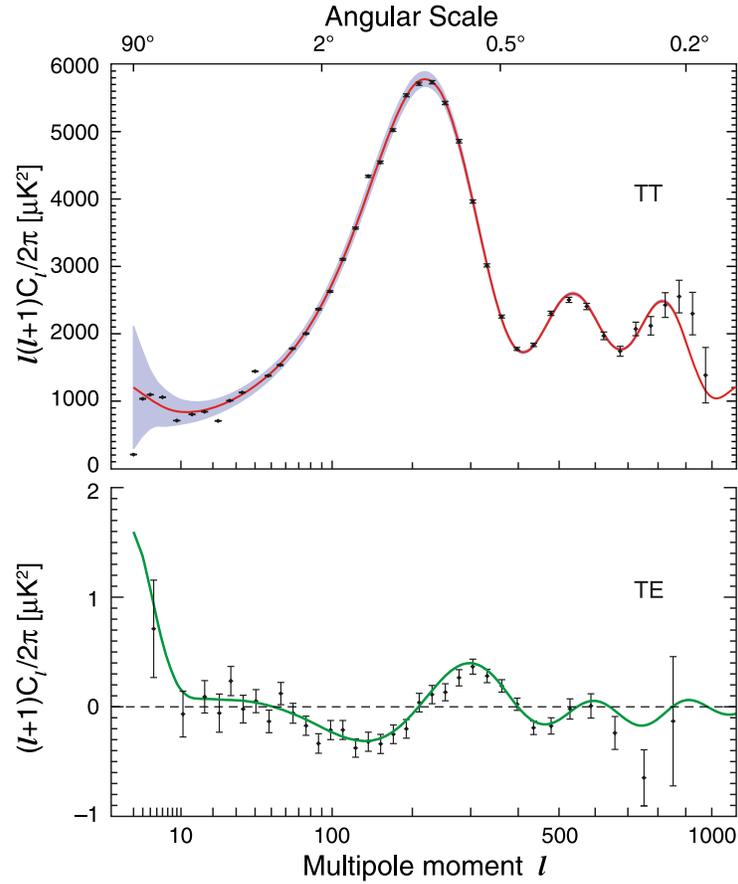}
\caption[The temperature auto-correlation ({\sl TT}) and temperature-polarization cross-correlation ({\sl TE}) 
power spectra with 2\,$\leq$\,$\ell$\,$\leq$ 1000 
from the 5 year WMAP data.]
{The temperature auto-correlation ({\sl TT}) and temperature-polarization 
cross-correlation ({\sl TE}) power spectra with $2 \leq \ell \leq 1000$.
The points are from the 5 year WMAP data and 
the error bars are the noise errors only.  
The solid line is the best-fit 6 parameter $\Lambda$CDM model, fit
to the {\sl WMAP} data only\,\cite{1Hinshaw:2008}.
The grey shaded area shows the 1 $\sigma$ error band 
due to cosmic variance
(i.e. the fact that our realization of the CMB sky can
vary from the underlying expectation value). 
This figure is taken from Hinshaw et al.~(2008)\,\cite{1Hinshaw:2008}.}
\label{1WMAP5}
\end{center}
\end{figure}

Figure~\ref{1WMAP5} shows the anisotropies $C_\ell^{TT}$
and $C_\ell^{TE}$ with $\ell$\,$\geq$\,2 from recent result based 
on the Wilkinson Microwave Anisotropy Probe~({\sl WMAP}; \cite{1Hinshaw:2008})
 5-year data.  
The points show the {\sl WMAP} data, while the solid
 line is the best-fit $\Lambda$CDM model. 
We can see that the first two acoustic peaks of the 
 temperature spectrum are well measured and there is 
 clearly a rise for the third peak.  
Together with other ground based experiments
 (see~\cite{1Scott:2006} and references therein), 
 perhaps the first five acoustic peaks have now been 
 localized.
 
The amplitude of the polarization signal is about 2 
 orders of magnitude smaller than the temperature one 
 and so it is much harder to detect.  
The DASI\,\cite{1Kovac:2002} experiment first
 demonstrated the existence of CMB polarization 
 in 2002 and the {\sl WMAP} experiment has
 measured the {\sl TE} power spectrum to high 
 precision\,\cite{1Hinshaw:2008}.  
Figure~\ref{1WMAP5} shows the recent measurements
of $C_\ell^{TE}$ from the {\sl WMAP} 5 year results.

\section{Why are we interested in recombination?}
\subsection{Distortion photons from recombination}
From the previous section, we know that the photons in 
the radiation background were thermalized to a nearly perfect
blackbody spectrum by bremsstrah-lung and double Compton
 scattering processes before recombination.  
As well as the photons from this blackbody background,
 there were some extra distortion photons produced during 
 the epoch of cosmological recombination.  
When an electron combined with an ionized atom and 
cascaded down to the ground state, there was at least one 
distortion photon emitted for each recombination.
These recombination photons give a distinct series of spectral line 
distortions on the nearly perfect blackbody CMB spectrum.
The main contribution to the distortion comes from the 
H\,{\sc i} Ly\,$\alpha$ transition at about $z \simeq 1500$,
and this line will be observed in the Wien tail~($\sim 100$\,$\mu$m)
 of the CMB spectrum today (see Figure~\ref{3sumline} in Chapter~3).
Since these distortion photons are produced directly from each
recombination of the atoms, the overall shape and amplitude of the 
line are very sensitive to the details of the recombination process. 
Therefore the detection of this spectral distortion would provide
direct contraints on the physics of recombination and also
provide incontrovertible evidence that the Universe was once a hot,
dense plasma which recombined.

FIRAS showed that the CMB spectrum around the peak
is well-modelled by a 2.725\,K  Planck spectrum.  
It was found that there is also a Cosmic Infrared Background~(CIB; 
see \cite{1Dole:2006,1Hauser:2001,1Puget:1996}), which peaks at 
about 150\,$\mu$m, right above the recombination 
distortion on the CMB spectrum~(see Figure~\ref{3sumline}).
This background is mainly due to luminous infrared galaxies 
at fairly recent epochs and it makes the detection of the 
 recombination distortion even more challenging.
The first calculations of the line distortion on the CMB tail 
were presented by Peebles~(1968)\,\cite{1Peebles:1968} and 
by Zeldovich et al.~(1968)\,\cite{1Zeldovich:1968}.  
However, they provided no details about the line shape, and 
since then there have been no explicit calculations showing 
different contributions to the line shape.  
Today we have a better understanding of the cosmological 
model as well as improved detection techniques, and so 
it is time to calculate these spectral distortion 
lines to much higher accuracy, in order to investigate 
whether they could be detected and whether such a 
detection would be cosmologically interesting.
A detailed study of this line distortion on the CMB
spectrum coming from the recombination time will be 
presented in Chapter~3.

\subsection{Precision cosmology}
The CMB anisotropies have been well studied theoretically,
 and the calculations are robust, because they can be based on 
 linear perturbation theory~(see \cite{1Hu:2002} 
 and references therein), given that the primordial 
fluctuations are of small amplitude.  
CMBFAST\,\cite{1Seljak:1996} is one of the most widely
used numerical Boltzmann codes for calculating the $C_{\ell}$.
It has been tested over a large set of cosmological models and 
 is consistent with other codes with an accuracy at
 better than the 1\% level\,\cite{1Seljak:2003th}.  
We have already entered the era of precision 
 cosmology\,\cite{1Hu:2000,1Planck:2006,1Spergel:2003,1Turner:2001}.  
With the release of the {\sl WMAP} 5 year data, 
 we can constrain the cosmological parameters extermely well
 from the shape of the anisotropy power spectrum\,\cite{1Hinshaw:2008}.
The next generation of CMB satellites, {\sl Planck}\,\cite{1Planck:2006}, 
 which will be launched in early 2009, has been designed to 
 sensitively measure the $C_{\ell}$ of the {\sl TT}- and {\sl TE}-modes 
 up to $\ell$\,=\,2500 and the {\sl EE}-mode for $\ell \leq 2000$.
In order to extract the correct cosmological parameters from the
 experimental data, theoretical calculations with consequently 
 higher accuracy are required.  
It now seems clear that we need to obtain the theoretical $C_\ell$s 
 to better than the 1\% level. 
And the main theoreical uncertainty comes from details of
the ionization history during recombination\,\cite{1Seljak:2003th}.
 
{\sc recfast}\,\cite{1Seager:1999bc} is the most common 
numerical code for calculating the evolution of the ionization 
fraction $x_{\rm e}$ during recombination; it is embedded into
all of the widely-used Boltzmann codes.  
It is written to be a short and quick program for reproducing 
 the results from a multi-level atom calculation\,\cite{1Seager:1999km},
 which follows the evolution of the number density of electrons 
 at each of more than 100 atomic levels for each species of atom.
The accuracy of the $x_{\rm e}$ obtained from {\sc recfast} is 
at the percent level, which is sufficient for {\sl WMAP},
 but may not be good enough for {\sl Planck}.
This fact has recently motivated many researchers to investigate 
 several detailed physical processes during recombination which 
 may cause roughly percent level changes on $x_{\rm e}$.
Although the basic physical picture for standard cosmological
 recombination is quite well established, the non-equilibrium
 details of recombination are unexpectedly complicated to solve.
That is because it must be done consistently with the interaction 
 between the matter and radiation field, in order to reach the 
 required sub-\,1\% accuracy in $x_{\rm e}$~(see Chapter~2 for a review).
In one specific example~(Chapter~4), we investigated
the effect of inclusion of the higher order non-resonant 
two-photon transitions and the semi-forbidden transitions in 
a multi-level atom calculation, which was first
suggested by Dubrovich \& Grachev~(2005)\,\cite{1Dubrovich:2005fc}
using a three-level atom model.  

There have recently been comprehensive studies of calculations of 
the He\,{\sc i} reombination, with all relevent radiative 
processes to the 0.1\% accuracy level\,\cite{1Hirata:2007sp,1Switzer:2007sn,1Switzer:2007sq}.
However there still has not been a single numerical 
calculation which includes all the improvements in H\,{\sc i}
recombination (which of course has greater effect on the $C_\ell$
than for He).
From another point of view, given the precision of the experimental 
$C_\ell$ measurement, we may want to ask how accurate the theoretical 
model needs to be in order not to bias the determination 
of the cosmological parameters.
In another of our projects~(Chapter~6), we investigated how the remaining 
uncertainties in recombination affects the constraints 
on the cosmological parameters using {\sc Planck} forecast
data. We do this through use of the C{\sc osmo}MC code\,\cite{1Lewis:2002ah}, 
which is a numerical code for exploring the multi-dimensional cosmological 
parameter space with the Markov Chain Monte Carlo method.

\section{Outline of the thesis}
This thesis focuses on the study of cosmological recombination and its effects
on the CMB.  Here we have briefly reviewed the standard model for the
evolution of Universe, including the basic picture of cosmological 
recombination and the formation of the CMB.  
Chapter~2 provides an overview of progress in the theoretical 
calculation of recombination, and the recent updates for obtaining
the ionization fraction $x_{\rm e}$ to better than 0.1\% accuracy.
In Chapter~3 we present a calculation of the spectral distortions in the CMB 
due to H\,{\sc i} Ly\,$\alpha$ and the lowest 2s--1s line transitions,
as well as the corresponding lines of He\,{\sc i} and He\,{\sc ii}, 
during the epoch of recombination.
Next, in Chapter~4, we investigate the effects of including 
non-resonant two-photon transitions and the semi-forbidden transitions
 in the process of H\,{\sc i} and He\,{\sc i} recombination.
Chapter~5 is a brief study to clarify that the previously claimed 
effect of the reheating of matter due to the distortion photons 
emitted during recombination is neligible.
In Chatper~6 we investigate how uncertainties in the recombination 
calculation affects the determination of the cosmological parameters
 in future CMB experiments.
Finally we present our conclusion and ideas for future directions in Chapter~7.

\chapter[Progress in recombination calculations]
{Progress in recombination calculations\footnote[1]{A version of this 
chapter will be submitted for publication: 
Wong W.~Y. `Progress in recombination calculations'.}}
In this chapter, we will give a review of progress 
 in controlling the accuracy of the recombination 
 calculation, starting from a traditional three-level 
 atom model and going up to the recent multi-level 
 atom models including interactions between matter and radiation.
We will also describe the remaining uncertainties which 
 will need to be tackled in the numerical codes in order
 to obtain the ionization fraction to better than 1\%.

\section{Standard picture of recombination}
Cosmological recombination calculations were first performed 
 forty years ago
 by Peebles~(1968)\,\cite{Peebles:1968} and Zeldovich, Kurt 
 and Sunyaev~(1968)\,\cite{Zeldovich:1968} using a 
 3-level atom model in hydrogen~(${\rm H}^+$\,+\,e$^-$\,$\rightarrow$\,H\,+\,$\gamma$).
In this simplified model, one only follows the detailed 
 rates of change of electrons in the continuum, the first 
 excited state and also the ground state of the atom.
The higher excited states are assumed to be in thermal 
 equilibrium with the first excited state.  
The cosmological recombination of H is slower than 
 that via the Saha equation; it is `Case B' recombination,
 since direct recombination to the ground state is highly
 prohibited and the Ly\,$\alpha$ line is optically thick. 
Due to the short mean free time of the ionizing photons 
 compared to the expansion time of the Universe
 \mbox{(by a factor of $\simeq 10^{-9}$)}, the ionizing 
 photons emitted from direct recombination to the ground
 state easily photoionize the surrounding neutral atoms.  
Therefore, the electrons recombine mainly through the 
 first excited state ($n$\,=\,2) and cascade down to the 
 ground state by the Ly\,$\alpha$ or the 2s--1s two-photon 
 transition.
The two-photon transition plays an important role in 
 recombination and the net rate is comparable to the 
 net Ly\,$\alpha$ rate (see Fig.\,\ref{3ratesH}), because 
 only a tiny amount of the  Ly\,$\alpha$ photons redshift 
 out of the line and escape to infinity without getting 
 absorbed or scattered.  
To account for the redshifting of the Ly\,$\alpha$ resonance 
 photons, Peebles~(1968)\,\cite{Peebles:1968} approximated the 
 intensity distribution as a step and scaled the Ly\,$\alpha$
 rate by multiplying by the ratio of the rate of redshifting 
 of photons through the line to the expansion rate of the Universe.
The radiation field and the matter are strongly coupled 
 through Compton scattering, and therefore the matter 
 temperature $T_{\rm M}$ can be well approximated as 
 the radiation temperature $T_{\rm R}$.  
These two temperatures start to depart only in the
 very late stages of recombination 
 (at $z\simeq 200$)\,\cite{Hu:1995fqa,Peebles:1968,Switzer:2007sn}, 
 when most of the electrons have already recombined. 
After that the matter temperature decreases adiabatically,
 $T_{\rm M} \propto (1+z)^2$, while $T_{\rm R}$ decays as $(1+z)$.

The above description gives us the standard picture for 
 the H\,{\sc i} recombination.  
It has been argued that we should also include stimulated
 recombination in the three level atom model\,\cite{Jones:1985},
 but the effect is quite negligible.
An analogous physical situation was proposed for He\,{\sc i} 
 recombination~(${\rm He}^+$\,+\,e$^-$\,$\rightarrow$\,He\,+\,$\gamma$)
 by Matsuda et al.~(1969, 1971)\,\cite{Matsuda:1969,Matsuda:1971}, 
 and a slower recombination than Saha equilibrium was then found.
However, it was later argued that He\,{\sc i} recombination 
 should be well approximated by the Saha equation by taking into 
 account the tiny amount of neutral hydrogen formed at the same 
 time.  
Since these H\,{\sc i} atoms can capture the He\,{\sc i}~$2^1$P--$1^1$S 
 resonant line photons as well as the photons from direct 
 recombination to the $1^1$S ground state~\cite{Hu:1995fqa}, 
 this speeds up He\,{\sc i} recombination.  
This issue was not entirely cleared up until some recent 
 calculations included the continuum opacity of H\,{\sc i} in 
 the He\,{\sc i} recombination evolution\,\cite{Kholupenko:2007qs,Switzer:2007sn}, 
 as will be discussed in Section~\ref{sec:c2rad}. 

He\,{\sc ii} recombination~(${\rm He}^{2+}$\,+\,e$^-$\,$\rightarrow$\,He$^+$\,+\,$\gamma$)
 was found to remain very close to Saha equilibrium\,\cite{Seager:1999km,Switzer:2007sq}
 due to the fast radiative rates. 
This, together with the fact that He\,{\sc ii} recombination 
 occurs too early to have any effects on the CMB anisotropies, 
 means that we do not discuss He\,{\sc ii} recombination in detail 
 in this chapter.

\section{Multi-level atom model}
Thirty years later, after the first calculations there was an 
 increased demand for an accurate ionization history for modeling 
 the CMB power spectrum for new experiments, for example, {\sl WMAP}.  
Seager et al.~(1999, 2000)\,\cite{Seager:1999bc,Seager:1999km} 
 set a benchmark precision for the numerical recombination 
 calculation by following the evolution of the occupation numbers
 of 300 atomic energy levels in H\,{\sc i} and 200 levels in He\,{\sc i} 
 without any thermal equilibrium assumption between each state.  
This multi-level H\,{\sc i} atom consisted of maximum 300 separated 
 quantum number energy levels ($n$-states), while the He\,{\sc i} 
 atom included the first four angular momentum states ($l$-states)
 up to $n$\,=\,22 and just the separated $n$-states above that.  
The rate equation for each level was written down using the 
 photoionization and photorecombination (bound-free) rates,
 the photoexcitation (bound-bound) rates and the collision rates.
The bound-bound rates included all the resonant transitions 
 but only one forbidden transition, the lowest spontaneous 
 two-photon transition (2s--1s for H\,{\sc i} and $2^1$S--$1^1$S for He\,{\sc i}).  
The bound-bound rate of the Lyman-series transitions were scaled 
 with the Sobolev escape probability $p_{\rm s}$ \cite{Rybicki:1984} 
 to account for the redshifting and trapping of the distortion photons
 in the radiation field; this reduces to Peebles' step method 
 when $p_{\rm s} \propto 1/\tau$~(see Equation~(3.14) for an
 explicit expression for $p_{\rm s}$) and $\tau \gg 1$ ($\tau$~is 
 the optical depth of the line).  
All the rates were obtained with the radiation background approximated 
 as a perfect blackbody spectrum. 

In this multi-level atom model, Seager et al.~(1999, 2000)\,\cite{Seager:1999bc,Seager:1999km} 
 found a speed-up in H\,{\sc i} recombination at low redshift 
 compared with the standard Case~B recombination~\cite{Peebles:1968,Zeldovich:1968} 
 at low redshift, and a delayed He\,{\sc i} recombination in 
 contrast to that from the Saha equation. 
The H\,{\sc i} recombination was faster than previously estimated
 because of the non-zero bound-bound rates among higher 
 excited~($n$\,$\geq$\,2) states.
In the three-level atom model, the higher excited states 
 are assumed to be in thermal equilibrium and the bound-bound 
 rates between these states are negligible.
However, these bound-bound rates are actually dominated 
 by spontaneous de-excitations due to the strong but cool 
 radiation field, which means that the electrons prefer to 
 cascade down to the lower energy states rather than staying
 at the higher excited states.  
This results in faster H\,{\sc i} recombination.  

On the other hand in this study, the He\,{\sc i} was found to 
 follow a standard hydrogen-like Case~B recombination,  
 agreeing with the earlier study of Matsuda et al.~(1969,
  1971)\,\cite{Matsuda:1969,Matsuda:1971},
 which adopted a three-level atom model for He\,{\sc i} 
 by considering the singlets only.  
There are two sets of states in the He\,{\sc i} atom: the singlets 
 and the triplets.
In this multi-level atom model calculation, the triplet states 
 were found to be highly unpopulated because the collisional transitions 
 between singlets and triplets are weak, and therefore the 
 electrons mainly cascade down to the ground state via the 
 singlet states.  
Concerning any mechanisms which might bring the He\,{\sc i} recombination 
 into Saha equilibrium, Seager et al.~(2000)\,\cite{Seager:1999km} found 
 that the photoionization rate of H\,{\sc i} was much lower than the 
 He\,{\sc i} $1^1$S--$2^1$P photoexcitation rate, and concluded that 
 the H\,{\sc i} atoms have negligible effect on stealing the He\,{\sc i} 
 resonance line photons in order to speed up the He\,{\sc i} recombination.

Evolution of the matter temperature with all the relevant cooling 
 processes (specifically, Compton, adiabatic cooling, free-free, 
 photorecombination and line cooling) and the formation of 
 hydrogen molecules~(for example, H$_2$) was also considered, 
 but these effects, along with the collisional transitions, were 
 found to be negligible for the ionization fraction $x_{\rm e}$~(see 
 Table~\ref{tab2.1} for the magnitude of the change in $x_{\rm e}$).  
Seager et al.~(2000)\,\cite{Seager:1999km} also discussed the effects 
 of the secondary distortions due to photons emitted from the H\,{\sc i} 
 Ly\,$\alpha$ and 2s--1s transitions, and also from the corresponding 
 transitions in He\,{\sc i} and He\,{\sc ii}.  
These distortion photons can be redshifted into a frequency 
 range where they could photoionize the electrons in the first 
 excited ($n=2$) state and the ground state of H\,{\sc i} during 
 the time of H\,{\sc i} recombination.  
Again, the effect, was found to be very small.

In order to reproduce the accurate numerical results 
 without going through the full multi-level calculation, 
 the authors used a so-called `effective three-level model'~\cite{Seager:1999bc}
 by multiplying by a `fudge factor' $F_{\rm H}$ the recombination 
 and ionization rates in the standard Case~B recombination 
 calculation to reproduce the speed-up of H\,{\sc i} recombination.
For He\,{\sc i}, the result can be well approximated by considering
 the standard Case~B recombination situation in a three-level 
 atom model (singlets only), consisting of the continuum, 
 the first excited singlet state and the ground state 
 (see Section~3.2 for details).  
{\sc recfast}~\cite{Seager:1999bc} is the publicly available 
 computer code which calculates the ionization fraction 
 $x_{\rm e}$ (the detailed profile of the last scattering surface)
 using the effective three-level model discussed above.  
It is currently adopted in most of the commonly used 
 Boltzmann codes, for example, {\sc cmbfast}\,\cite{Seljak:1996}, 
 {\sc camb}\,\cite{Lewis:2000} and {\sc cmbeasy}\,\cite{Doran:2005}, 
 to numericallly evolve the accurate CMB anisotropy
 spectrum for different sets of cosmological parameters.

\section{Recent improvements and suggested modifications}
Recently, driven mainly by the up-coming high-$\ell$ CMB 
 experiments\,\cite{Kosowsky:2003sw,Planck:2006} and also 
 the possibility of detecting spectral
 distortions\,\cite{RubinoMartin:2006ug,RubinoMartin:2007cq,Sunyaev:2008ts,Wong:2005yr}, 
 there have been many suggested updates and improvements to
 the the multi-level atom model suggested by Seager et al.~(2000)\,\cite{Seager:1999km}.
In this section, we discuss these new physical  
 processes included in recombination, concentrated on those
 which may lead to more than 1\% level change in the 
 ionization fraction $x_{\rm e}$.

\subsection{Energy levels}
In the Seager et al.~(2000)\,\cite{Seager:1999km} calculation, 
 they only considered separated $n$-states for H\,{\sc i},
 while the $l$-states  are assumed to be in thermal equilibrium
 within each $n$-shell.  
{Rubi{\~n}o-Mart{\'{\i}}n} et al.~(2006)\,\cite{RubinoMartin:2006ug}
 first tried to relax this assumption by resolving all the 
 $l$-states up to $n$\,=\,30, and later the same authors\,\cite{Chluba:2006bc}
 even pushed the maximum level to $n$\,=\,100.
They found that the total population of any shell is smaller
 than the value obtained from the Saha equation during H\,{\sc i} 
 recombination.
The deviation of the populations from Saha was claimed to increase 
 from $\sim$\,0.1\% at $z$\,=\,1300 to $\sim$\,10\% at $z$\,=\,800 
 and this in general led to a slower recombination
 at lower redshift compared with previous studies~(see Table\,\ref{tab2.1}). 
There seems to be no need to further consider the separate 
 states with different spin orientations~(for example, hyperfine splitting),
 since the rates of the resonant transitions connecting individual hyperfine 
 splitting states (or only one $l$-state to the other states) are the same, 
 even if these splitting states are not in equilibrium.

Note that there is a serious problem in the {Rubi{\~n}o-Mart{\'{\i}}n} et al.~(2006)
 model: the ionization fraction $x_{\rm e}$ does not converge when 
 increasing numbers of $n$-shells are included. 
Due to computational limitations, the most intensive calculation 
 involved 5050 separate $l$-states with maximum $n$\,=\,100.  
Although the $l$-changing and $n$-changing collisional 
 transitions were additionally considered in their model, 
 these authors found that these transitions were not strong
 enough to bring the higher excited states back into thermal 
 equilibrium and the divergence problem remained.  
Although there is undoubtedly still some physics missing in this model 
 (which will be discussed later in this chapter), it is still 
 worth asking how many levels we need to consider to solve for
 the recombination of  H\,{\sc i} atom.
Using the thermal equilibrium assumption in each $n$-shell, 
 Seager et al.~(2000)\,\cite{Seager:1999km} found that the 
 ionization fraction $x_{\rm e}$ converges well when considering 
 a maximum of 300 energy levels, and claimed that this should 
 be the maximum number of levels which needs to be considered
 by arguing that for such a large $n$ state the thermal 
 broadening width of the level is larger than the gap between 
 that level and continuum.

For He\,{\sc i} recombination, no such convergence problem exists.
Switzer \& Hirata~(2008)\,\cite{Switzer:2007sn,Switzer:2007sq} 
 performed a similar multi-level atom model calculation including
 the interaction of matter with the radiation field by resolving 
 all the $l$-states with $n \leq 10$.  
The number of resolved $l$-states are limited by the availability 
 of the atomic data of He\,{\sc i} and so this calculation is the 
 best that can be carried out for now (the limited availablility 
 of the atomic data will be discussed in Section\,\ref{sec:c2atom}).
Switzer \& Hirata~(2008)\,\cite{Switzer:2007sq} found that the 
 change of $x_{\rm e}$ was smaller than $0.004$\% when reducing 
 the maximum principal number $n$ of the levels from 100 to 45.
In their model, the effect was not significant because the 
 feedback of the spectral distortion from these highly excited 
 states suppressed the net recombination to the ground state
 via these states (See Section\,\ref{sec:c2rad} for more details).

\subsection{Bound-bound transitions}
In the formerly standard recombination model, the lowest 
 two-photon transition is the only forbidden transition considered.   
Dubrovich \& Grachev~(2005)\,\cite{Dubrovich:2005fc} first suggested
 that it might be important to include more intercombination (i.e. transitions
 connecting triplets and singlets), the He\,{\sc i} $2^3$P$_1$--$1^1$S$_0$
 spin-forbidden transition specifically and the non-resonant two-photon 
 transitions from the higher excited states 
 (\mbox{$n$s, $n$d $\rightarrow$ 1s} for H\,{\sc i} and  
  \mbox{$n^1$S, $n^1$D$ \rightarrow 1^1$S$_0$} for He\,{\sc i}).
They demonstrated that these additional transitions significantly 
 speed up the recombination in an effective three-level atom model 
 calculation.
We first focus on the effect of including the intercombination
 $2^3$P$_1$--$1^1$S$_0$ transition.
With the addition of this transition in a standard multi-level atom 
 model, Wong \& Scott~(2007)\,\cite{Wong:2006iv}~(see also Chapter~4)
 found that more than 40\% of the photons from $n=2$ state cascaded 
 down to the ground state through the triplet $2^3$P$_1$ 
 state~(see Table\,\ref{tab:HeI}).  
This is almost the same as the amount of electrons going from 
 $2^1$P$_1$ through the resonant transition, and the net rates
 of the $2^3$P$_1$--$1^1$S$_0$ and $2^1$P$_1$--$1^1$S$_0$ 
 transitions are comparable (see Figure\,\ref{4grateHeI} 
 and Section\,\ref{sec:c4result} for details) under the 
 Sobolev photon escape approximation.  
The He\,{\sc i} recombination speeds up due to this extra channel 
 through the triplets to the ground state and the change on 
 $x_{\rm e}$ is about 1.1\% at $z \simeq 1750$.
Switzer \& Hirata~(2008)\,\cite{Switzer:2007sq} also found that 
 the $2^3$P$_1$--$1^1$S$_0$ transition is important for He\,{\sc i} 
 recombination in their improved multi-level model calculation 
 including the evolution of the radiation field; the effect 
 of the radiative feedback between the $2^1$P$_1$--$1^1$S$_0$ and 
 $2^3$P$_1$--$1^1$S$_0$ transitions was found to bring a 1.5\% 
 change in $x_{\rm e}$ (the details of the feedback effect 
 will be discussed in Section\,\ref{sec:c2rad}).

For the higher order two-photon transitions, 
 Dubrovich \& Grachev~(2005)\,\cite{Dubrovich:2005fc}
 attempted to include the corresponding rates in an analogous 
 way to the lowest 2s--1s two-photon transition.  
However, they found that these two-photon transitions from high 
 $n$ states are more complicated than the 2s--1s one.
That is because the matrix elements for these transition rates 
 have poles when the intermediate states are not virtual 
 (i.e.~\mbox{$n$s, $n$d $\rightarrow$ $m$p $\rightarrow$ 1$s$} with
 1\,$<$\,$m$\,$<$\,$n$) and this is in-distinguishable from 
 the resonant one-photon transitions themselves.  
If we include all the poles in calculating the two-photon decay 
 rate, then we obtain a very fast rate since the process is dominated
 by the resonant Lyman-series transitions.  
Additionally we double count the number of electrons recombining
 through those one-photon resonance transitions.  
In order to avoid these problems due to the resonance poles, 
 Dubrovich \& Grachev~(2005)\,\cite{Dubrovich:2005fc} 
 approximated the non-resonant two-photon rate by considering 
 only one pole (the $n$p state) as the intermediate state 
 in the matrix element.  
Their estimated rate was very fast~(scaling as $n$ for large $n$)
 and this dramatically sped up the recombination process, 
 with $\Delta x_{\rm e}$ equal to a few percent.

Wong \& Scott~(2007)\,\cite{Wong:2006iv} proposed an improved,
 net non-resonant two-photon rate for H\,{\sc i} from 
 $n$\,=\,3\,\cite{Cresser:1986,Florescu:1988}, and this was 
 significantly lower.
This calculation included all the non-resonant poles 
 (i.e.~all the $n \geq 3$ intermediate states).  
By comparing with the Dubrovich \& Grachev~(2005)\,\cite{Dubrovich:2005fc} 
 estimate at $n=3$, the rate obtained is an order of magnitude smaller,
 due to the destructive interference of some matrix elements, 
 which was ignored in Dubrovich \& Grachev~(2005)\,\cite{Dubrovich:2005fc}
 (since they only considered one pole).  
Using this rate with the same $n$ scaling given by 
 Dubrovich \& Grachev~(2005) for the higher $n$ two-photon 
 rates, Wong \& Scott~(2007)\,\cite{Wong:2006iv} found that 
 the maximum change in $x_{\rm e}$ was only 0.4\%.
Chluba \& Sunyaev~(2007)\,\cite{Chluba:2007qk} performed a 
 more detailed calculation of the high $n$ two-photon rates 
 for H\,{\sc i} by studying the frequency distribution profile
 of the photons from these transitions. 
They estimated the effective two-photon rates by subtracting 
 Lorentz profiles of the possible resonant transitions directly 
 from the full two-photon profile in order to avoid the 
 double-counting from the one-photon resonant transitions.  
The rates they found were lower than the ones given by 
 Dubrovich \& Grachev~(2005)\,\cite{Dubrovich:2005fc},
 also due to destructive interference of the matrix elements.  
With their effective rates, Chluba \& Sunyaev~(2007)\,\cite{Chluba:2007qk}
 obtained essentially the same value for the maximum change in 
 $x_{\rm e}$ at similar redshift range as found by
 Wong \& Scott~(2007)\,\cite{Wong:2006iv}.

Hirata \& Switzer~(2008)\,\cite{Hirata:2007sp} and
 Hirata~(2008)\,\cite{Hirata:2008ny} further studied the 
 role of these high $n$ two-photon transitions in He\,{\sc i}
 and H\,{\sc i} recombination, respectively, by including 
 the related two-photon scattering (Raman scattering) 
 and the possibility of re-absorption of the photons 
 from the resonant intermediate states.
In their model, they separated the spectrum of the 
 photons into non-resonant~(photons emitted through 
 a virtual intermediate state) and resonant regions.  
They added an additional rate due to these non-resonant 
 photons in analogy to the lowest 2s--1s two-photon rate.
The higher order non-resonant two-photon rates were also 
 found to be much lower than those estimated by 
 Dubrovich \& Grachev~(2005)\,\cite{Dubrovich:2005fc}, 
 again because of the destructive interference of the 
 matrix elements, and the rates scale as $n^{-3}$.
The resonant transitions were considered as being photons
 from the corresponding one-photon resonant transitions,  
 but with a modified line profile; these photons were highly probable 
 to be scattered or absorbed by other atoms.  
For He\,{\sc i}, Hirata \& Switzer~(2007)\,\cite{Hirata:2007sp} 
 found that inclusion of these higher order two-photon transitions 
 brought no more than a 0.04\% change in $x_{\rm e}$.  
But for H\,{\sc i}, with the additional consideration of 
 the feedback between the Ly$\alpha$ line and the two-photon 
 transitions\,\cite{Hirata:2008ny}~(see Section\,\ref{sec:c2rad} 
 for details), the change in $x_{\rm e}$ was found to 
 be more than a percent around the peak of the 
 visibility function.

Some other forbidden transitions were also included  
 in He\,{\sc i} recombination, specifically, the magnetic 
 dipole $2^3$S--$1^1$S$_0$ transition~\cite{Lin:1977,Switzer:2007sq,Wong:2006iv}, 
 the electric dipole transitions with 
 $n$\,$\leq$\,10 and $l$\,$\leq$\,7\,\cite{Wong:2006iv},
 the intercombination $n^3$P$_1$--$1^1$S$_0$ transitions, 
 the electric quadrupole \mbox{$n^1$D--$1^1$S$_0$ ($n \geq 4$)} 
 transitions, the magnetic quadrupole $2^3$P$_2$--$1^1$S$_0$ 
 transition and the electric octupole \mbox{$n^3$F--$1^1$S$_0$ ($n \geq 4$)} 
 transitions\,\cite{Switzer:2007sq}.
However, the effect of the inclusion of all the above transitions
 is very small ($\Delta x_{\rm e}$\,$\leq$\,0.001\%) and can
 therefore be neglected. 
One may ask whether we should include the one-photon
 2s--1s magnetic dipole transition for H\,{\sc i}.  
The rate of this transition is equal to 
 2.49\,$\times$\,$10^{-6}$\,s$^{-1}$\cite{Barut:1988,Parpia:1982},
 which is about 6 orders of magnitude smaller than the 2s--1s 
 two-photon transition.  
Therefore we expect that the effect of the inclusion of this magnetic
 dipole transition should be negligible.

It is worth remembering that there are two electrons 
 bound in each He\,{\sc i} atom.  
In the standard He\,{\sc i} recombination calculation, 
 we usually consider the inner electron to be in the 
 ground state.  
One may wonder whether these other electrons might sometimes
 leave the ground state by stealing photons and getting
 excited to higher levels.  
However, He\,{\sc ii} recombination occurs much 
 earlier than He\,{\sc i} recombination, and therefore 
 almost all of the inner electrons were already in the 
 ground state based on the Boltzmann distribution at the 
 time when He\,{\sc i} recombination began.  
In order to excite the inner electrons from the ground 
 state to the first excited state, the energy of the
 incident photons would need to be about 40\,eV, which
 is almost double the ionization energy of He\,{\sc i}.
This means that the abundance of such 40\,eV photons 
 is $10^{-14}$ of the He\,{\sc i} ionization photons
 at $z$\,=\,2500, based on the blackbody spectrum, 
 implying that the inner electrons have almost
 no chance to get excited from the ground state during
 the He\,{\sc i} recombination. 
Hence we can completely neglect all such transitions.

\subsection{Bound-free transitions}
\label{sec:c2bf}
One of the approximations adopted in the standard 
 recombination calculation is that there are no direct 
 recombinations to the ground state.  
This is because the photons emitted in this transition 
 immediately reionize another neutral atom (the same 
 situation applies to both H\,{\sc i} and He\,{\sc i}).  
Chluba and Sunyaev~(2007)\,\cite{Chluba:2007yp} revisited
 this approximation by calculating the net rate of direct 
 recombinations to the ground state for H\,{\sc i}  
 through detailed consideration of photon escape.  
Although the escape probability of a photon emitted from 
 the continuum to the ground state is about 10--100 times 
 larger than that of the Ly\,$\alpha$ photons, the inclusion 
 of these direct recombinations only brings about a
 $0.0006$\% change in $x_{\rm e}$. 
Hu et al.~(1995)\,\cite{Hu:1995fqa} had earlier argued 
 that the direct recombination of He\,{\sc i} should be 
 possible, due to the absorption of the continuum photons 
 by the tiny amount of H\,{\sc i} atoms in the later stages
 of He\,{\sc i} recombination.  
However, Switzer \& Hirata~(2008)\,\cite{Switzer:2007sn} 
 showed that this effect on the speed-up of He\,{\sc i} 
 recombination is negligible \mbox{($\Delta x_{\rm e} \simeq 0.02$\%)}
 by calculating the effective cross-section of the 
 bound-free transition to the ground state due to the 
 presence of H\,{\sc i}.
From the above disscusion, we can therefore safely neglect 
 direct recombinations to the ground state for both 
 H\,{\sc i} and He\,{\sc i}.

\subsection{Radiative transfer}
\label{sec:c2rad}
In the standard multi-level calculation of recombination, 
 the radiation background field is approximated as a perfect 
 blackbody spectrum.  
For the interaction between atoms and the radiation field, 
 the Sobolev approximation is adopted to account for the escape 
 probability of the photons redshifting out of the line.  
But in order to calculate $x_{\rm e}$ to better than the 1\% level,
 the above approximations are not sufficient, and fundamentally
 we need to solve for the evolution of the number densities 
 of the atomic levels and the radiation field, with the 
 distortion photons from recombination process solved 
 consistently in an expanding environment.  
Several recent studies\,\cite{Chluba:2005uz,Chluba:2007yp,Grachev:2008xj,
 Hirata:2008ny,Hirata:2007sp,Kholupenko:2006jm,Kholupenko:2007qs,Switzer:2007sn,Switzer:2007sq}
have suggested that additional radiative transfer processes 
 (for example, the feedback between lines) might cause 
 significant effects on recombination. 
In particular, Switzer \& Hirata~(2008)\,\cite{Hirata:2007sp,Switzer:2007sn,Switzer:2007sq}
 have performed the most complete and systematic 
 multi-level He\,{\sc i} atom model calculation, with 
 the consideration of both coherent and incoherent scattering
 process between atoms and photons.  
They specifically included the feedback between lines,
 absorption due to the continuum opacity of H\,{\sc i}, 
 stimulated and induced two-photon transitions, the 
 collisional transitions and Thomson scattering (all 
 examples of incoherent scattering), together with  
 partial redistribution of the line profile due to coherent 
 scattering.  
We will discuss each of these processes in turn.

\subsubsection{Feedback from spectral distortions}
In the standard multi-level atom calculation, no feedback 
 between resonant lines is considered.  
However, in practice distortion photons escaping from 
 the higher order resonance transitions will redshift to 
 a lower line frequency and excite electrons in the
 corresponding state.  
For example, photons emitted from Ly\,$\gamma$ transitions 
 can excite electrons in the ground state after redshifting 
 to Ly\,$\beta$ or Ly\,$\alpha$ line frequencies.
In general, this feedback process will suppress the net 
 recombination rate to the ground state thereby slowing 
 down recombination.
Switzer \& Hirata~(2008)\,\cite{Switzer:2007sn} used an 
 iterative method to include the feedback between transitions 
 connecting the excited states and the ground state during
 He\,{\sc i} recombination.  
They only considered the radiation being transported from 
 the next higher transition [$(i+1)$th state to $1^1$S$_0$]
 to the $i$th transition~(to the ground state in the same species).
They found that the most significant change to the 
 ionization fraction~($\Delta x_{\rm e} = 1.5$\%) is 
 due to the feedback between the $2^3$P--$1^1$S$_0$ and 
 $2^1$P$_1$--$1^1$S$_0$ transitions.  
Chluba \& Sunyaev\,\cite{Chluba:2007yp} also studied the 
 same feedback effects among the Lyman-series transitions during 
 H\,{\sc i} recombination.  
They found that feedback from the Ly\,$\beta$ transition 
 on the Ly\,$\alpha$ line accounts for most of the contribution,
 and the maximum change in $x_{\rm e}$ is about $0.35$\%, 
 this appearing to be a convergent result when including
 Lyman-series transitions up to $n$\,=\,30.
For H\,{\sc i} recombination, we also need to consider
 the distortion photons from He\,{\sc i} recombination 
 feeding back to the H\,{\sc i} line transitions~(especially
  the Lyman series), which brings about a 0.1\% 
  change in $x_{\rm e}$~\cite{Sunyaev:2008ts}.  
In the Seager et al.~(1999)\cite{Seager:1999km} recombination 
 model, only the photons from He\,{\sc i} $2^1$P$_1$--$1^1$S$_0$
 and $2^1$S$_0$--$1^1$S$_0$ transitions were considered as 
 secondary distortions on the H\,{\sc i} recombination.
This should clearly be extended by calculating a detailed 
 He\,{\sc i} line spectrum, including all the released photons.

\subsubsection{Stimulated and induced two-photon transitions}
The standard recombination model only includes the 
 {\it spontaneous} 2s--1s two-photon emission rate 
 and the corresponding two-photon absorption rate 
 coming from detailed balance.
Taking H\,{\sc i} as an example, the 
spontaneous two-photon decay is
\begin{equation}
{\rm H(2s)} \rightarrow {\rm H(1s)} + \gamma_{\rm spon} + \gamma_{\rm spon}\, ,
\end{equation}
and the two-photon excitation is
\begin{equation}
{\rm H(1s)} + \gamma_{\rm bb} + \gamma_{\rm bb} \rightarrow {\rm H(2s)}\, .
\end{equation}
Here $\gamma_{\rm spon}$ represents a spontaneously emitted
photon and $\gamma_{\rm bb}$  represents a photon taken 
from a blackbody radiation spectrum.
Chluba \& Sunyaev~(2005)\,\cite{Chluba:2005uz} suggested
 that one should include the stimulated H\,{\sc i} 2s--1s
 two-photon emission due mainly to the low frequency 
 background photons.  The two stimulated decays are 
\begin{equation}
{\rm H(2s)} \rightarrow {\rm H(1s)} + \gamma_{\rm spon} + \gamma_{\rm stim}\, 
\end{equation}
and 
\begin{equation}
{\rm H(2s)} \rightarrow {\rm H(1s)} + \gamma_{\rm stim} + \gamma_{\rm stim}\, ,
\end{equation}
where $\gamma_{\rm stim}$ refers to a photon from stimulated emission.
The recombination is found to speed up, and these authors 
 claimed that the effect can be more than 1\% in $x_{\rm e}$.  
Later, Kholupenko \& Ivanchik~(2006)\,\cite{Kholupenko:2006jm}
 pointed out that the induced H\,{\sc i} 2s--1s two-photon
 absorption of a thermal background photon and a redshifted
 distortion photon from the H\,{\sc i} Ly\,$\alpha$ transition 
 should also be considered, i.e.
\begin{equation}
{\rm H(1s)} + \gamma_{\rm bb} + \gamma_{\rm dist} \rightarrow {\rm H(2s)}\, ,
\end{equation}
where $\gamma_{\rm dist}$ represents a spectral distortion photon.
By including this absorption process, recombination is
 actually delayed overall, and the maximum change in
  $x_{\rm e}$ is about 0.6\%\cite{Hirata:2008ny,Kholupenko:2006jm}. 
Hirata~(2008)\,\cite{Hirata:2008ny} extended the above
 ideas further to include the higher order two-photon
 transitions (H\,{\sc i} $n$d, $n$s--1s) using the 
 steady-state approximation.
Instead of adopting an effective rate, he performed a 
 radiative transfer calculation to account for the
 emitted line photons, whether they are being re-absorbed or 
 scattered later.   
The result showed that the recombination speeds up after 
 inclusion of the stimulated and induced higher order 
 two-photon transitions, the maximum change being 1.7\% 
 in $x_{\rm e}$ at $z \simeq 1250$, which is bigger than 
 the result of using only the effective rates in the 
 previous studies\,\cite{Chluba:2007qk,Wong:2006iv}.
Hirata~(2008)\,\cite{Hirata:2008ny} also investigated
 the effect of two other relevant two-photon process: 
 Raman scattering
\begin{equation}
{\rm H}(nl) + \gamma \rightarrow {\rm H(1s)} + \gamma'\, ,
\end{equation}
where $\gamma'$ is a photon with higher energy compared
with $\gamma$;
and direct two-photon recombination to the ground state
\begin{equation}
{\rm H}^+ + {\rm e}^- \rightarrow {\rm H(1s)} + \gamma + \gamma' \, .
\end{equation}
The direct two-photon recombination process was found
 to be negligible, but on the other hand, the Raman 
 scattering brought about dramatic effects on 
 H\,{\sc i} recombination.  
Raman scattering is dominant in the 2s--1s transition,
 since the 2s state is the most populated among  all the
 $n$s and $n$d states with $n$\,$\geq$\,2.  
Through Raman scattering, the CMB photons can excite atoms
 in the 2s state and the atoms will decay down to the 
 ground state by emitting photons with frequencies between
 the Ly\,$\beta$ and Ly\,$\alpha$ lines.  
Therefore, Raman scattering provides another channel for the
 electrons to get down to the ground state and this 
 initially speeds up recombination.  
However, the photons emitted from the Raman scattering
 process having energy larger than Ly\,$\alpha$ 
 will redshift and feed back on the Ly\,$\alpha$ and 
 2s--1s transitions.
This additional feedback delays recombination and 
 $x_{\rm e}$ {\it increases} by about 1\% at $z \simeq 900$.  

For He\,{\sc i} recombination, a similar study was 
 performed by Hirata \& Switzer ~(2008)\,\cite{Hirata:2007sp}
 and a much smaller effect was found on $x_{\rm e}$~($<$\,0.01\%).
The reason is that the abundance of H is much greater than
 for He (about a factor of 12 in number) which leads to a 
 lower optical thickness in the case of the He\,{\sc i} 
 $2^1$P$_1$--$1^1$S$_0$ line than the H\,{\sc i} 
 Ly\,$\alpha$ line for the resonant photons from two-photon
 transitions.  
The other reason comes from the different shapes of the
 frequency spectra of the lowest two-photon transition
 at low frequencies.  
The frequency spectrum for He\,{\sc i} $2^1$P$_1$--$1^1$S$_0$
 is proportional to $\nu^3$, while that at H\,{\sc i}  2s--1s 
 is proportional to $\nu$.  
This is because the H\,{\sc i} 2p and 2s states are 
 essentially degenerate (actually the 2p state is slightly lower
 than 2s due to the Lamb shift~\cite{Lamb:1947}, but the shift is 
 only 4.372\,$\times$\,$10^{-6}$\,eV)
 and so there is a pole in the matrix element 
 at zero frequency when 2p is the intermediate state~\cite{Hirata:2007sp}.
The stimulated and induced two-photon transitions dominate
 at low frequencies and therefore there is a larger probability
 in the H\,{\sc i} two-photon spectrum at both ends
 (where one of the photons has a small frequency) . 
As a result the effect is more significant in H\,{\sc i}
 recombination.

\subsubsection{Photon absorption due to continuum opacity of H\,{\sc i}}
The other important improvement in He\,{\sc i} recombination
 is inclusion of the continuum opacity of H\,{\sc i}\,\cite{Switzer:2007sn}.
In the later stages of He\,{\sc i} recombination a 
 tiny but significant amount of neutral hydrogen 
 H\,{\sc i} is formed \mbox{($n_{\rm H I}/n_{\rm H} < 10^{-4}$ at $z$\,$\simeq$\,2000)}, 
 and these H\,{\sc i} atoms can absorb (through photoionization)
 the distortion photons emitted during He\,{\sc i} recombination.
In Section \ref{sec:c2bf}, we have already discussed how
 the effect on the direct recombination of He\,{\sc i} due to this
 continuum opacity of H\,{\sc i} is negligible.  
However, the presence of the H\,{\sc i} continuum opacity
 significantly affects the transitions connecting the 
 excited states and the ground state,
 particularly the $2^1$P$_1$--$1^1$S$_0$ transition.  
The $2^1$P$_1$--$1^1$S$_0$ transition, which is the 
 lowest He\,{\sc i} resonance transition, is also one of 
 the main paths for the electrons to cascade down to the 
 ground state.
In the standard multi-level atom model, about 60\% of the 
 electrons in the $n=2$ state reach the ground state 
 through this transition (see Table\,\ref{tab:HeI}).
The energy of the photons emitted from the $2^1$P$_1$--$1^1$S$_0$ 
 transition is equal to 21.2\,eV, which is much larger 
 than the ionization energy of H\,{\sc i}.  
Therefore, the H\,{\sc i} atoms can absorb: 
(1) the He\,{\sc i} $2^1$P$_1$--$1^1$S$_0$ line photons
 directly; or (2) the redshifted line photons from the 
 next higher transitions before they redshift down to 
 the $2^1$P$_1$--$1^1$S$_0$ line and excite another atom.
This process removes these distortion photons and 
prevents them from re-exciting other He\,{\sc i} atoms:
\begin{eqnarray}
{\rm He}(2^1 {\rm P}) &\rightarrow& {\rm He}(1^1 {\rm S}) + \gamma \, \nonumber \\
{\rm H}(1 {\rm s}) + \gamma &\rightarrow& {\rm H}^+ + {\rm e}^- \, .
\end{eqnarray}  
For process (1), the usual Sobolev escape probability 
 can be modified due to the direct line photon absorption 
 by the H\,{\sc i} atoms instead of  
 He\,{\sc i}~\cite{Kholupenko:2006jm,Switzer:2007sn}.  
The  modified escape probability, which is in 
 general larger than the Sobolev value, has been applied to 
 the He\,{\sc i} resonant  $n^1$P--$1^1$S$_0$, intercombination 
 $n^1$D--$1^1$S$_0$ and quadrupole $n^1$F--$1^1$S$_0$ 
 lines~\cite{Switzer:2007sn}.  
Recombination is significantly sped up mainly due to the 
 extra continuum opacity of H\,{\sc i} within the 
 $2^1$P$_1$--$1^1$S$_0$ line. 
This effect gives more than a 2\% change in $x_{\rm e}$, while 
 the opacity in other lines only contributes about $0.05$\%.
For process (2), the absorption of the redshifted line photons
 suppresses feedback between the lines.  
For example, there are some distortion photons from 
 He\,{\sc i} $2^3$P--$1^1$S$_0$ which are absorbed by 
 H\,{\sc i} before they can redshift down to the He\,{\sc i} 
 $2^1$P$_1$--$1^1$S$_0$ line frequency to excite electrons in 
 the ground state of He\,{\sc i} atoms.  
Therefore, the number of redshifted distortion photons
 available for the feedback between He\,{\sc i} lines 
 is smaller, and hence the He\,{\sc i} recombination 
 speeds up a little.  
Overall, the continuum opacity of H\,{\sc i} modified to 
 include these feedback process brings about a 
 $0.5$\% change in $x_{\rm e}$.

\subsubsection{Coherent scattering}
In the Sobolev escape probability method, a Voigt 
 profile is assumed for both the frequency spectra 
 of the emitted and absorbed photons in the line transitions.
However, this will only be true when the system is very 
 close to thermal equilibrium.  
For an optically thick line (for example, 
 He\,{\sc i} $2^1$P$_1$--$1^1$S$_0$) without the 
 continuum opacity of other species of atoms, 
 the radiation field in the region of the line 
 frequency is in thermal equilibrium with the population
 ratio of the corresponding two levels relevant for 
 this transition.
However, in the presence of the continuum opacity of 
 H\,{\sc i}, the H\,{\sc i} and He\,{\sc i} atoms 
 complete for the the distortion photons from the 
 line transitions and so no such thermal equilibrium 
 exists. 
The emission and absorption line profiles may not be 
 the same as each other or equal to a Voigt function, 
 since there is no complete redistribution in the line profile.  
In such a non-equilibrium situation, we need to consider 
 all the possible paths for an electron at each state
 to go after it is excited by a resonant photon from a 
 lower state.  
Therefore, besides the incoherent scattering processes,
 we also need to consider coherent scattering 
 (relative to the atom's rest frame).  
An electron excited by a resonant photon to a higher 
 state can decay to the original lower state by 
 emitting a photon with the same energy {\it without}
 any intermediate interaction.  
The emitted and absorbed photons have no energy difference
 in the atom's rest frame, but there is a small 
 fractional change in the photon's frequency (on the order of $v/c$, 
 where $v$ is the atomic velocity) in the comoving frame. 
If the effects of coherent scattering are significant, 
 the line profile is only partially redistributed and the 
 frequency spectrum of the emission line photons
 depends on the radiation background.
Switzer \& Hirata~(2008)\,\cite{Switzer:2007sn} 
 performed a Monte Carlo simulation for the partial 
 redistribution of the profile in the He\,{\sc i} $n^1$P$_1$--$1^1$S$_0$ 
 resonance line due to coherent scattering. 
The effect they found was about $0.02$\% in $x_{\rm e}$,
 compared with the model having feedback and continuous 
 opacity of H\,{\sc i}, as discussed above.
 
\subsubsection{Thomson scattering and collisional transitions}
Thomson scattering and collisional transitions were also 
 considered by Switzer \& Hirata~(2008)\,\cite{Switzer:2007sq}
 in the He\,{\sc i} recombination calculation, but both 
 of these processes were found to be negligible.
During He\,{\sc i} recombination, Thomson scattering 
 may be significant, since a large fraction of
 electrons have not yet recombined.  
The photons can gain energy after multiple electron 
 scatterings  and the photons which had previously 
 redshifted out of the line can be scattered back 
 into the line.  
This reduces the escape probability of the line 
 and hence delays the recombination.  
However, in the presence of feedback between lines 
 and the continuum opacity of H\,{\sc i}, the 
 distortion photons are more likely to get re-absorbed
 instead, and therefore Thomson scattering
 is strongly suppressed.  
The net effect becomes only $0.03$\% in $x_{\rm e}$.
During H\,{\sc i} recombination, Thomson scattering should 
 also be considered, because the optical depth of the 
 Lyman-series lines is very high ($\sim$\,$10^9$ for 
 Ly$\alpha$, which is $10^3$ times that of the He\,{\sc i} 
 $2^1$P$_1$--$1^1$S$_0$ line).  
Due to this high optical depth, a similar calculation 
 is necessary for studying the partial redistribution of the 
 Lyman-series line profiles with all the possible 
 coherent and incoherent scattering processes.
However, no such systematic calculation (similar to
 the He\,{\sc i} one) has been performed yet.  
Since the rate of H\,{\sc i} recombination is mainly 
 controlled by the trapping of the Ly\,$\alpha$ photons,
 there are several studies concerning only the line 
 profile of the Ly\,$\alpha$ transition.
Rybicki \& Dell'Antonio\,\cite{Rybicki:1994} studied 
 the time-dependent spectral profile of the Ly\,$\alpha$
 transition in an expanding environment using the 
 Fokker-Planck equation and, found that the quasi-static 
 assumption is an adequate approximation for  
 this transition.
Several other works \cite{Grachev:1989,Grachev:2008xj,Krolik:1989,Krolik:1990}
 have also included the effect of the frequency 
 shift of Ly\,$\alpha$ due to the recoil of the H atoms,
 with suggestions that the effect on $x_{\rm e}$ may be 
 at the level of 1\%.

Collisional transitions, caused by the collisions between 
 atoms and ions, are usually neglected in recombination,  
 because of the high photon to baryon ratio ($\simeq$\,$10^9$).
Such collisional processes tend to bring the species into
 equilibrium and maintain statistical balance between 
 the energy levels.  
The bound-free transitions, bound-bound transitions, 
 and charge exchange (He$^+$\,+\,H\,$\rightarrow$\,He\,+H$^+$\,+\,$\gamma$) 
 between H\,{\sc i} and He\,{\sc i}
 due to the collisions are found to be too slow to 
 have any effect on He\,{\sc i} recombination\,\cite{Switzer:2007sq}.
In the later stage of H\,{\sc i} recombination, 
 the separated $l$-states fall out of equilibrium and 
 the collisional transitions become very important for 
 redistributing the electrons within each $n$ shell, 
 at least for the higher excited states 
 ($n$\,$\geq$\,50)\,\cite{Chluba:2006bc}.  
For the lower excited states, radiative transitions
 are dominant and the effect of collisional processes 
 between the H\,{\sc i} 2s and 2p states was found to 
 be negligible\,\cite{Burgin:2006}.

The electrons, ions and neutral hydrogen are 
 well approximated as a single tightly coupled component
 in the standard recombination picture\,\cite{Hannestad:2000fy}.  
They are considered as a single `baryon' fluid and 
 described by a single temperature, the matter temperature $T_{\rm M}$.  
The matter temperature is very close to the radiation temperature
 $T_{\rm R}$ during recombination, due to the strong effects of
 Compton scattering.  
During He\,{\sc i} recombination ($z$\,$\geq$\,1600), for 
 example, the fractional temperature difference 
 $(T_{\rm R}-T_{\rm M})/T_{\rm R}$ is smaller 
 than $10^{-6}$\,\cite{Seager:1999km,Switzer:2007sn}.  
The effects of the Compton scattering become weaker
 during H\,{\sc i} recombination, since most of the 
 electrons are captured to form neutral neutral atoms.  
Adiabatic cooling starts to become important for 
 matter when Compton scattering effects become
 slow compared with expansion time.  
The matter temperature then starts to depart 
 from the radiation temperature, because the 
 matter cools faster.  
But actually, even during H\,{\sc i} recombination
 (700\,$<$\,$z$\,$<$\,1500), the fractional difference between 
 these two temperatures is no more than $1$\% 
 (see Fig.\,2 in \cite{Switzer:2007sn}).  
This summarizes the general picture for the 
 evolution of matter temperature.
Several authors\,\cite{Hannestad:2000fy,Seager:1999km,Switzer:2007sn}
 have performed detailed calculations of the evolution
 of $T_{\rm M}$ by including all the relevant heating
 and cooling processes between the matter and radiation
 fields, in addition to Compton and adiabatic cooling. 
The results found are basically the same as in previous 
 studies\,\cite{Hu:1995fqa,Peebles:1968}, with the 
 additional processes bringing negligible change on 
 the matter temperature.
One study suggested that one should include the 
 heating of matter due to the distortion photons
 emitted during H\,{\sc i} recombination, and that
  this effect delayed recombination\,\cite{Leung:2003je}.
However, the coupling between matter and these distortion 
 photons is very weak. 
Almost all of these photons go into
 the radiation field and form the spectral distortion lines
 on the CMB blackbody spectrum\,\cite{Wong:2006wh}
 (see Chapter\,\ref{ch5:TM} for details).
This additional suggested effect is therefore negligible.

\subsection{Atomic data} 
\label{sec:c2atom}
In the multi-level atom calculation of cosmological 
 recombination, the non-equilibrium situation existing
 between states is important, since radiative processes
 are much stronger than collisional ones\,\cite{Hu:1995fqa,Seager:1999km}.
Based on recent studies, it is necessary to include
 energy levels with principal quantum number  
 $n \leq 50$ for He\,{\sc i} and $n \leq 300$ for H\,{\sc i}
 in the multi-level atom model.
Therefore, for the numerical recombination calculation, 
 detailed and accurate atomic data are required for the 
 energies of the states, and for the bound-free and 
 bound-bound transition rates, not only for the lower states,
 but also for the higher excited states. 

For the H\,{\sc i} atom, there is an exact solution 
 for the non-relativistic Schr\"odinger equation, 
 and the energies of each $(n,l)$ state are 
 given by $E_n = -R_{\rm H}/n^2$, where $R_{\rm H}$ 
 is the hydrogen Rydberg constant and 
 $h_{\rm P} c R_{\rm H} = 13.5984$\,eV\,\cite{PDBook}.
With the exact wavefunctions, the rates of the 
 bound-bound resonant (electric dipole) transitions
 between resolved $l$-states can also be
 determined to very high accuracy\,\cite{Switzer:2007sn}.
This is also true for the lowest two-photon 2s--1s 
 transition, and there are many papers in the 
 literature determining the theoretical value of this 
 spontaneous rate, $\Lambda^{\rm H}_{2s-1s}$\,\cite{Breit:1940,Goldman:1989,Goppert:1931,Labzowsky:2005,Nussbaumer:1984,Santos:1998,Spitzer:1951}.  
The latest value of $\Lambda^{\rm H}_{2s-1s}$
 is 8.2206\,s$^{-1}$, given by Labzowsky et al.~(2005)\,\cite{Labzowsky:2005},
 and this agrees with other calculations to 
 about the 0.1\% level of accuracy.
This small uncertainty has negligible effect on 
 recombination.
For two-photon transitions from the higher excited 
 states ($n$\,$>$\,2) to the ground state,
 we need to have the detailed spectra of the 
 emitted photons in order to avoid double counting
 the photons in the resonant transitions.
By direct summation of the matrix elements or by
 by using the Green functions method, the spectra 
 can be calculated to 0.1\% accurarcy
 (see \cite{Chluba:2007qk,Hirata:2007sp} and 
 references therein).
 
For $n\leq 10$, TOPbase\,\cite{Cunto:1993} provides spectra
 for the photoionization cross sections $\sigma(\nu)$ 
 for each $(n,l)$ level.
And we can use the Gaunt factor approximation\,\cite{Menzel:1935,Switzer:2007sn}
 to calculate the photoionization cross section for
 the states with $n >10$.  
The Gaunt factor is the ratio of the photoionization
 cross-section from a quantum-mecahnical calculation 
 to the value obtained from the semi-classical 
 electromagnetism formalism~(see, for example, 
 Chapter\,6 in \cite{Dopita:2003}). 
Rubi{\~n}o-Mart{\'{\i}}n et al.~(2006)\,\cite{RubinoMartin:2006ug}
 compared three numerical methods\,\cite{Boardman:1964,Burgess:1958,Karzas:1961} 
 for obtaining these cross-sections and found that 
 the results agree to the percent level.

The atomic physics of He\,{\sc i} is more complicated
 than H\,{\sc i} because it is a two-electron system.  
Morton et al.~(2006)\,\cite{Morton:2006} have provided
 the largest and most recent set of ionization energies
 of resolved $l$ states for $n$\,$\leq$\,10 and $l$\,$\leq$\,7,
 with accurarcy better than $10^{-5}$, combined with both
 experimental and theoretical results. 
For the other states, it is usual to adopt re-scaled 
 hydrogenic values; it should be a good approximation 
 to consider an electron orbiting a pointlike He$^+$ ion
  for $l$\,$\geq$\,2.
For the bound-bound transition rates, 
 Drake \& Morton~(2007)\,\cite{Drake:2007} have also 
 presented the most up-to-date data-set of the emission 
 oscillator strengths $f$ for the electric dipole 
 transitions, and also the intercombination (spin-forbidden)
 transitions between the singlets and triplets with 
 $n$\,$\leq$\,10 and $l$\,$\leq$\,7.
Bauman et al.~(2005)\cite{Bauman:2005ng,Porter:2007hs} 
 developed a computer code for generating $f$s and the 
 Einstein coefficients $A$ for the bound-bound transitions
 for even higher excited states ($n$\,$\leq$\,13 and 
 $l$\,$\leq$\,11) by combining different data sources 
 and approximations.  
The accuracy of these two approaches is about 5\% to 10\%, respectively,
 which is estimated by comparing the results with experimental 
 data using the adopted approximations\,\cite{Bauman:2005ng,Drake:2007}. 
For the higher order resonant transitions, $n$\,$>$\,12,
 the rescaled hydrogenic values are used for the 
 bound-bound resonant rates and the uncertainty should be
 at least at the 10\% level.
Since the resonant lines are optically thick, 
 the intercombination transitions 
 ($n^3$P$_1$--$1^1$S$_0$ with $n\geq 2$) play an 
 important role in recombination, especially the 
 $2^3$P$_1$--$1^1$S$_0$ transition. 
The theoretical value of the $2^3$P$_1$--$1^1$S$_0$
 spontaneous transition rate $A_{2^3 {\rm P}_1-1^1{\rm S}_1}$
 ranges from 171\,s$^{-1}$ to 233\,s$^{-1}$ in different 
 calculations\,\cite{Drake:1969a,Lach:2001,Laughlin:1978,Lin:1977}.
The latest $A_{2^3 {\rm P}_1-1^1{\rm S}_1}$ value is 
 177\,s$^{-1}$ given by {\L}ach \& Pachucki~(2001)\,\cite{Lach:2001}.  
Although variations of among estimates of this rate 
 are about 30\%, the effect on $x_{\rm e}$ is only at 
 the 0.1\% level\,\cite{Switzer:2007sq}.  
The most important forbidden transition for 
 He\,{\sc i} is the lowest two-photon 
 $2^1$S$_0$--$1^1$S$_0$ transition.  
The latest value of the spontaneous rate 
 $\Lambda^{\rm He}_{2^1 {\rm S}_0-1^1 {\rm S}_0}$
 is 51.02\,s$^{-1}$\,\cite{Derevianko:1997}, which
 agrees with other theoretical values\,\cite{Drake:1986,Drake:1969b} 
 at the 1\% level.  
For the higher order two-photon transitions 
 ($n^1$S$_0$,\,$n^1$D$_2$--$1^1$S$_0$), 
 Hirata \& Switzer~(2008)\,\cite{Hirata:2007sp} have 
 tried to estimate the corresponding rates and also the 
 freqency spectrum of the emitted photons by direct
 summation of the matrix elements.  
The accuracy of their method is at about the 10\% level. 
But this uncertainty brings almost no
 change on recombination, since the effect of the 
 inclusion of the higher order two-photon transitions
 was found to be insignificant for He\,{\sc i} 
 recombination\,\cite{Hirata:2007sp}.

For the bound-free cross-sections, 
 Hummer \& Storey~(1998)\,\cite{Hummer:1998} 
 provided the largest set of data for the 
 spectrum of the cross-section  $\sigma(\nu)$ with 
 $n$\,$\leq$\,25 and $l$\,$\leq$\,4, while 
 Topbase\,\cite{Cunto:1993} only contains 
 $\sigma(\nu)$ with $n$\,$\leq$\,10 and $l$\,$\leq$\,2. 
Bauman et al.~(2005)\,\cite{Bauman:2005ng,Porter:2007hs} 
 have combined these two results with other 
 approximations in a computer code  which can generate
 $\sigma(\nu)$ up to $n$\,=\,27 and $l$\,=\,26.  
These three sets of data (although not entirely
 independent) agree at the few percent level.  
For higher excited 
states ($n \geq 10$), the re-scaled hydrogenic 
 cross-section\,\cite{Storey:1991} can be used.
This is a reasonable approximation, giving accuracy 
 at about the 10\% level\,\cite{RubinoMartin:2007cq}.
Overall there is about a 10\% error in the atomic 
 data of He\,{\sc i}, but the effect on $x_{\rm e}$ 
 should be no more than the 0.1\% level, 
 helped considerably by the low abundance of 
 He\,{\sc i} (about 8\% of the total
 number of H and He atoms).
 
\subsection{Fundamental constants, cosmological parameters and other uncertainties}
The accuracy of the avaliable fundamental physical constants is
 of course important for the numerical recombination calculation.  
The biggest uncertainty comes from the gravitational constant
 $G$\,\cite{Hu:1995fqa}, due to the inconsistency among 
 different experimental measurements~(see Chapter~10 in \,\cite{Mohr:2008} for details).  
The latest recommended value by the Committee on Data for 
 Science and Technology~(CODATA) is 
 \mbox{$G= 6.67428(67) \times 10^{-11}$\,m$^3$kg$^{-1}$s$^{-2}$},
 with a fractional uncertainty equal to $10^{-4}$\,\cite{Mohr:2008}. 
The gravitational constant mainly affects the overall time 
 scale of the expanding Universe.
However, this uncertainty brings almost no effect 
 on $x_{\rm e}$ \mbox{($\Delta x_{\rm e} \ll 10^{-3}$)}\,\cite{Wong:2006iv}.
All other relevant physical constants are measured to
 much higher accuracy and their effects on recombination
 can be ignored. 

The CMB monopole temperature $T_{\rm CMB}$ is one of 
 the few cosmological parameters that can be measured
 {\it directly} by experiments, and  is usually 
 considered as one of the fundamental `input' parameters
 in the standard six parameter $\Lambda$CDM cosmological
 model for calculating the CMB temperature and 
 polarization anisotropies.
Given $T_{\rm CMB}$, we can determine the radiation density
 or the photon background field of the Universe, and this 
 strongly affects the speed of recombination.
The latest value of $T_{\rm CMB}$ is $2.725 \pm 0.001$\,K,
 which is the final assessment, including calibration
 and other systematic effects, coming from measurements
 made with FIRAS instrument (on the {\sl COBE} satellite)\,\cite{Fixsen:2002}.
Although the relative uncertainty in $\Delta T / T$ is only
 0.04\% , it leads to a 0.5\% change in 
 $x_{\rm e}$\,\cite{Chluba:2007zz} at $z$\,$\simeq$\,900.
But the corresponding effect on the $C_{\ell}$ is only 
 at the 0.1\% level for $\ell$\,$\simeq$\,2500\,\cite{Chluba:2007zz,Hamann:2007sk}.

The other uncertainty among the input cosmological
 parameters is the primordial helium abundance $Y_{\rm p}$ 
 (defined to be the mass fraction of helium)\,\cite{Chluba:2007zz}.  
In the standard Big Bang Nucleosynthesis (BBN) calculation,
 the derived value of $Y_{\rm p}$ only depends on 
 the baryon to photon ratio 
 $\eta_{10}$\,$\equiv$\,$10^{10}$\,($n_{\rm B}/n_{\gamma}$), 
 and can be numerically calculated to about the 0.2\% level 
 of accuracy \,\cite{Steigman:2007xt}.
Note that the number of neutrino species $N_{\nu}$ is
 assumed to be 3 in standard BBN 
 (although it is not quite correct if there is 
 mixing between different kinds of neutrinos).  
The number of neutrino species affects the He abundance 
 because of the change in the expansion rate of the Universe  
 ($\Delta Y_{\rm p}$\,$\approx$\,0.013\,$\Delta N_{\nu}$)\,\cite{Steigman:2007xt}.
Based on standard BBN and the {\sl WMAP} five-year results, 
 $Y_{\rm p}$ is determined to be equal to 
 $0.2486 \pm 0.0005$\,\cite{Dunkley:2008}, which is a little larger than the 
 value estimated  from the direct observational results
 $Y_{\rm p} = 0.240 \pm 0.006$\,\cite{Steigman:2007xt}.
After the BBN epoch, helium can be produced in all
 H-burning stars, while some other heavier elements,
 such as oxygen O, are produced only in short-lived 
 massive stars.  
In low-metallicity regions, the measured He abundance should 
 be close to $Y_{\rm p}$  if the oxygen to hydrogen ratio 
 O/H is very low.  
Therefore, the observed value of $Y_{\rm p}$ is usually
 determined by studying line emission from the recombination
 of ionized H and He in low-metallicity extragalactic 
 H\,{\sc ii} regions.
However, the observed value of $Y_{\rm p}$ is still quite 
 uncertain, due to the sysmatic errors and the lack
 of evidence for the correlation between helium and
 oxygen abundances 
(see Section~3.3 in \cite{Steigman:2007xt} for details).
Due to discrepancies between the theoretical and 
 observational results, the uncertainty of $Y_{\rm p}$ 
 should be considered to be about 5\% and this
 brings a change in $x_{\rm e}$ at about the 1\% level at 
 redshifts around the peak of the visibility function.

In most recombination codes, only the masses
 of the constituents of atomic hydrogen and helium 
 are taken into account for converting the baryon 
 density $\Omega_{\rm B}$ to the number of 
 hydrogen atoms $n_{\rm H}$, i.e.
\begin{equation}
n_{\rm H} = \frac{3 H_0^2 \Omega_{\rm B}}{8 \pi G} \frac{1-Y_{\rm p}}{m_{\rm H}}.
\end{equation}
It has been argued that we should also consider the 
 binding energy in each atom\,\cite{Steigman:2006nf} 
 as well as the abundance of lithium in the above formula.  
However, the binding enerygy is about $10^{-3}$ of the mass 
 of a proton and the mass fraction of lithium is only
 $10^{-9}$.  
Therefore, the effects on recombination should be very small.

When we calculate the ionization history of cosmological 
 recombination, we mainly talk about the hydrogen and helium
 because these two elements comprise more than 
 99\% of the total number of atoms in the Universe, 
 particularly in the primodial abundance.
However, from the standard BBN, there are also tiny 
 amount of deuterium (D) and lithium (Li) produced.
Since Li$^{2+}$ and Li$^{+}$ have higher ionization 
 energies (122.4 and 75.6\,eV respectively), they actually
 recombined {\it before} helium\,\cite{Lepp:2002}.  
On the other hand, neutral Li recombined at a much 
 later time ($z \lesssim 300$) than hydrogen 
 recombination\,\cite{Switzer:2005}.
However, the lithium recombination brings negligible
 effect on $x_{\rm e}$, because of its low abundance. 
Deuterium recombined at the same time as the rest of the
 hydrogen, due to having almost the same atomic structure,
 but with a heavier nucleus.
Similar to Li, the abundace of D is also low ($\simeq 10^{-5}$) 
 and therefore, the recombination of D brings a negligible
 effect on $x_{\rm e}$.

There is a tiny fraction of free electrons left 
 ($n_{\rm e}/n_{\rm H} \simeq 10^{-5}$) after
 hydrogen recombination, and this allows for the 
 formation of molecules in the later stages of 
 evolution~(see \cite{Galli:1998,Lepp:2002,Seager:1999km}
 and reference therein).
Due to the high photon to baryon ratio, then at early
 times there are huge numbers of photons about the
 dissociation energy of H$_2$ and hence collisional
 processes (e.g. three-body reactions) are inefficient
 in molecule formation.  
The molecules are only produced through radiative 
 association~\cite{Lepp:2002}.
For example, H$_2$ (H$^-$\,+\,H$\rightarrow$\,H$_2$\,+\,e$^-$)
 is produced via the formation
 of H$^-$~(H\,+\,e$^-$\,$\rightarrow$\,H$^-$\,+\,$\gamma$; 
 see \cite{Hirata:2006,Lepp:2002} for the latest
 calculations).
Since the process of radiative association requires
 the existence of H\,{\sc i}, significant production of 
 molecules occurs only after recombination.
These primordial molecules are important coolants in
 the star formation process and hence are crucial
 for understanding the formation of the first stars
 and galaxies, but they again have negligible 
 effect on $x_{\rm e}$. 
Due to the very low fraction of free electrons 
 available for molecule formation, the abundance
 of these molecules is very low and they are also 
 produced too late ($z \lesssim 300$) to 
 significantly affect the CMB photons.

In all of this discussion we have focussed
 on the {\it standard} picture of recombination.
Of course it is possible that we are still missing
 important pieces of the big picture.
Some other non-standard physics could also easily 
 alter the ionization fraction $x_{\rm e}$ at more than 
 the percent level. 
Examples include a non-negligible interacting cross-section  
 of dark matter\,\cite{Hu:1995fqa,Padmanabhan:2005}, 
 strong primordial magnetic fields\,\cite{Gopal:2005,Hu:1995fqa}, 
 strong spatial inhomogeneities\,\cite{Hu:1995fqa,Novosyadlyj:2006fw}, 
 extra Ly\,$\alpha$ emission from primordial black 
 holes\,\cite{Peebles:2000pn} 
 and a time-varying fine structure constant\,\cite{Avelino:2000}.

\section{Discussions}
In this chapter, we have briefly reviewed the 
 recent updates and remaining uncertainties
 in the numerical recombination calculation.
In order to obtain the ionization fraction 
 to better than the percent level, then complicated
 details of the  non-equilibrium situation need to
 be included.  
Most of the significant improvements have been 
 mainly from additional radiative processes
 controlling the population of the $n=2$ states.
This is because there is no direct recombination to 
 the ground state and cascading down through 
 $n=2$ states is the main path for electrons to
  reach the ground state.
If the existing studies have already considered 
 all the relevant physical processes in He\,{\sc i}
 recombination, then we currently have the corresponding
 numerical calculation to an accuracy better 
 than 1\%.
However, for H\,{\sc i} recombination, 
 there is still no single computational 
 code which includes all of the suggested 
 improvements.
Hydrogen recombination is even more important for 
 calculating the CMB anisotropies $C_{\ell}$, 
 because it dominates the detailed profile of 
 the visibility function.  
A comprehensive numerical calculation of 
 H\,{\sc i} recombination, including at least all 
 the suggested processes here, is neccessary and 
 urgent in order to obtain high accuracy $C_{\ell}$
 for future experiments.

\begin{center}
\begin{table}
\centering
\begin{tabular}{l r c l}
Effect & $\Delta x_{\rm e}/x_{\rm e}$ & $z_{\rm max}$ &References \\
\hline 
\hline
\multicolumn{3}{l}{\bf Energy level} \\
Separate $l$-states in H\,{\sc i} atom & $-0.7$\% & 1090 & \cite{Chluba:2006bc,RubinoMartin:2006ug} \\
 & $+1$\% &  $\leq 900$  & \\
\multicolumn{3}{l}{\bf Bound-bound transitions} \\
Inclusion of He\,{\sc i} $2^3$P$_1$--$1^1$S$_0$ & $-1.1$\% & 1750 & \cite{Dubrovich:2005fc,Wong:2006iv} \\
  & $-0.3$\%*& $1900$ & \cite{Switzer:2007sq} \\
Inclusion of He\,{\sc i} $n^3$P$_1$--$1^1$S$_0$ ($n \geq 3$) & $-0.004$\%* & 2000 &
 \cite{Switzer:2007sq} \\
Inclusion of H\,{\sc i} $n$s, $n$d--1s ($n \geq 3$): & & &\\
\ -- effective rate only  & $-0.4$\% & 1200 & \cite{Chluba:2007qk,Wong:2006iv} \\
\ -- with feedback  & $-1.2$\% & 1250 &\cite{Hirata:2008ny} \\
\ -- with feedback and Raman scattering & +1.3\% & 900 &\cite{Hirata:2008ny} \\
Inclusion of He\,{\sc i} $n^1$S, $n^1$D--$1^1$S$_0$ ($n \geq 3$): & &  &\\
\ -- effective rate only  & $-0.5$\% & 1800 &\cite{Dubrovich:2005fc,Wong:2006iv} \\
\ -- with feedback and Raman scattering & $-0.05$\%  & 2000  &\cite{Hirata:2007sp} \\
\multicolumn{3}{l}{\bf Bound-free transitions} \\
Direct recombination for H\,{\sc i} & $-0.0006$\% & 1280 & \cite{Chluba:2007yp} \\
Direct recombination for He\,{\sc i} & $-0.02$\% & 1900 & \cite{Hu:1995fqa,Switzer:2007sn}\\
\multicolumn{3}{l}{\bf Radiative transfer} \\
Continuum opacity of H\,{\sc i} &  $-2.5$\%* & 1800 & \cite{Kholupenko:2007qs,Switzer:2007sn,Switzer:2007sq} \\
\ in He\,{\sc i} $2^1$P$_1$--$1^1$S$_0$  & & &\\
Feedback between He\,{\sc i} $2^3$P--$1^1$S$_0$ & $+1.5$\%* & 1800 & \cite{Switzer:2007sn,Switzer:2007sq} \\ 
\ and $2^1$P$_1$--$1^1$S$_0$ &   & to 2600 &\\
Stimulated and induced H\,{\sc i} 2s--1s & $+0.6$\% & 900 & \cite{Chluba:2005uz,Hirata:2008ny,Kholupenko:2006jm} \\
Diffusion of Ly$\alpha$ line profile & $\sim -1$\% & 900 & \cite{Grachev:1989,Grachev:2008xj}\\
\ (with recoil of H atoms) & & & \cite{Krolik:1989,Krolik:1990} \\
Continuum opacity of H\,{\sc i} modified & $-0.5$\%* & 1800 & \cite{Switzer:2007sn,Switzer:2007sq} \\
\ to feedback in He\,{\sc i} lines & & & \\
Continuum opacity of H\,{\sc i} in He\,{\sc i}& $-0.05$\%* & 1900 & \cite{Switzer:2007sn,Switzer:2007sq} \\
\ $n^1$P--$1^1$S$_0$, $n^3$P--$1^1$S$_0$ ($n \geq 3$), $n^1$D--$1^1$S$_0$ &  & & \\
Coherent scattering in $n^1$P--$1^1$S$_0$ & $-0.02$\%* & 2000 & \cite{Switzer:2007sn} \\
Evolution of $T_{\rm M}$ & $\pm 0.001$\% & -- & \cite{Hu:1995fqa,Seager:1999km,Switzer:2007sn} \\
Secondary distortions from He\,{\sc i}  & $+0.1$\% & -- &\cite{Seager:1999km,Sunyaev:2008ts} \\ 
\ \& H\,{\sc i} in H\,{\sc i} recombination & & & \\
\multicolumn{3}{l}{\bf Other} \\
He\,{\sc i} $2^3$P$_1$--$1^1$S$_0$ spontaneous rate & $\pm$\,$0.1$\%  & 1900 & \cite{Switzer:2007sq} \\ 
CMB monopole uncertainty $T_{\rm CMB}$ $\pm 1$\,mK & $\pm 0.5$\% & 900 & \cite{Chluba:2007zz} \\ 
Primordial He abundance $Y_{\rm p}$ $\pm 1$\%  & $\pm 1$\% & $<1200$ & \cite{Chluba:2007zz}  \\
Formation of hydrogen molecules & $-1$\% & $<150$  & \cite{Seager:1999km} \\
\end{tabular}
\caption[Summary of improvements and uncertainties in the 
 numerical recombination calculation.]
{Summary of the improvements and uncertainties in the numerical
 recombination calculation.
 Here $\Delta x_{\rm e}/x_{\rm e}$ is the maximum ratio difference
 of the ionization fraction $x_{\rm e}$ from the value given by 
 {\sc recfast} Version~1.3\,\cite{Seager:1999bc} and
 $z_{\rm max}$ is the approximate redshift at which this occurs.
 *Note: This is the relative change compared
 with the full radiative model in Switzer \& Hirata~(2008)\,\cite{Switzer:2007sq}.}
\label{tab2.1}
\end{table}
\end{center}

\chapter[Spectral distortions]
{Spectral Distortions\footnote[2]{A version of this 
chapter~(except Section~\ref{ch3:remark}) has been published: 
Wong W.~Y., Seager~S. and Scott D.~(2006) `Spectral distortions to the cosmic 
microwave background from the recombination of hydrogen and helium',
Monthly Notices of the Royal Astronomical Society, 367, 1666--1676.}}

\section{Introduction}
Physical processes in the plasma of the hot early Universe thermalize
the radiation content, and this redshifts to become the observed
Cosmic Microwave Background (CMB; see \cite{3scott04} and references 
therein).  Besides the photons from the radiation background,
there were some extra photons produced from the transitions when the
electrons cascaded down to the ground state after they recombined with
the ionized atoms.  The transition from a plasma to mainly neutral gas
occurred because as the Universe expanded the background temperature
dropped, allowing the ions to hold onto their electrons.  The photons
created in this process give a distortion to the nearly perfect
blackbody CMB spectrum.  Since recombination happens at redshift
$z\,{\sim}\,1000$, then Ly$\,\alpha$ is observed at ${\sim}\,100\mu$m
today.  There are ${\sim}\,1$ of these photons per baryon, which
should be compared with the ${\sim}\,10^9$ photons per baryon in the
entire CMB.  However, the recombination photons are superimposed on
the Wien part of the CMB spectrum, and so make a potentially
measurable distortion.

From the Far-Infrared Absolute Spectrophotometer (FIRAS) measurements,
Fixsen et al.~(1996)\,\cite{3fixsen96} and 
Mather et al.~(1999)\,\cite{3mather99} showed that the CMB is well
modelled by a $2.725\pm0.001\,$K blackbody, and that any
deviations from this spectrum around the peak are less than 50 parts per
million of the peak brightness.  Constraints on smooth functions, such as
$\mu$- or $y$-distortions are similarly very stringent.
However, there are much weaker constraints on narrower features in the CMB
spectrum.  Moreover, within the last decade it has been discovered
\cite{3puget96} that there is a Cosmic Infrared 
Background~(CIB; see~\cite{3hauserdwek01} and references therein), 
which peaks at $100$--$200\mu$m
and is mainly comprised of luminous infrared galaxies at moderate redshifts.
The existence of this background makes it more challenging to measure the
recombination distortions than would have been the case if one imagined them
only as being distortions to Wien tail of the CMB.  However, as we shall see,
the shape of the recombination line distortion is expected to be much narrower
than that of the CIB, and hence the signal may be detectable in a future
experiment designed to measure the CIB spectrum in detail.

The first published calculations of the line distortions occur in the
seminal papers on the cosmological recombination process by
Peebles~(1968)\,\cite{3peebles1968} and 
Zel'dovich et al.~(1968)\,\cite{3zks68}.  One of
the main motivations for studying the recombination process was to
answer the question: `Where are the Ly$\,\alpha$ line photons from the
recombination in the Universe?'~(as reported in~\cite{3rubino05}).
In fact these studies found that for hydrogen recombination (in a
cosmology which is somewhat different than the model favoured today)
there are more photons created through the two-photon 2s--1s
transition than from the Ly$\,\alpha$ transition.  Both
Peebles~(1968)\,\cite{3peebles1968} and 
Zel'dovich et al.~(1968)\,\cite{3zks68} plot the distortion of the CMB
tail caused by these line photons, but give no detail about the line
shapes.  Other authors have included some calculation or discussion of
the line distortions as part of other recombination related studies,
e.g.~Boschan \& Biltzinger~(1998)\,\cite{3boschan98}, 
and most recently Switzer \& Hirata~(2005)\,\cite{3switzer05}.  However,
the explicit line shapes have never before been presented, and the
helium lines have also been neglected so far.  The only numerical
study to show the hydrogen lines in any detail is a short conference
report by Dell'Antonio \& Rybicki~(1993)\,\cite{3dell93}, 
meant as a preliminary version of a more
full study which never appeared.  Although their calculation appears
to have been substantially correct, unfortunately in the one plot they
show of the distortions (their figure~2) it is difficult to tell
precisely which effects are real and which might be numerical.

Some of the recombination line distortions
from higher energy levels, $n>2$, have also been
calculated\,\cite{3bur94,3burgin03,3dell93,3dub75,
3dubs95,3dubs97,3fahr91,3kho05,3lyu83}.  
However, these high $n$ lines lie near the peak of 
the CMB and therefore are extremely weak
compared with the CMB (below the $10^{-6}$ level), while the
Ly\,$\alpha$ line is well above the CMB in the Wien region of the spectrum.

As trumpeted by many authors, we are now entering into the era of
precision cosmology.  Hence one might imagine that future delicate
experiments may be able to measure these line distortions.
Since the lines are formed by the photons emitted in each transitions of
the electrons, they are strongly dependent on the rate of recombination
of the atoms.  The distortion lines may thus be a more sensitive probe of
recombination era physics than the ionization fraction
$x_{\rm e}$, and the related visibility function which affects the CMB
anisotropies.  This is because a lot of energy must be injected in order
for any physical process to change
$x_{\rm e}$ substantially~(for example, \cite{3psh00}).  
In general that energy will go into spectral distortions, 
including boosting the recombination lines.

This also means that a detailed understanding of the
physics of recombination is crucial for calculating the
distortion.  The basic physical picture for cosmological recombination
has not changed since the early work of Peebles~(1968)\,\cite{3peebles1968} and 
Zel'dovich et al.~(1968)\,\cite{3zks68}.
However, there have been several refinements introduced since then, motivated
by the increased emphasis on obtaining an accurate recombination
history as part of the calculation of CMB anisotropies.
Seager et al.~(1999,2000)\,\cite{3sara99,3sara00} presented a detailed
calculation of the whole recombination process, with no assumption of
equilibrium among the energy levels.  This multi-level computation
involves 300 levels for both hydrogen and helium, and gives us the
currently most accurate picture of the recombination history.  In the
context of the Seager et al.~(2000)\,\cite{3sara00} recombination 
calculation, and with the
well-developed set of cosmological parameters provided by Wilkinson
Microwave Anisotropy Probe~({\sl WMAP}; \cite{3spergel03})
and other CMB experiments, it seems an appropriate time to calculate the distortion
lines to higher accuracy in order to investigate whether they could be
detected and whether their detection might be cosmologically useful.

In this Chapter we calculate the line distortions on the CMB
from the 2p--1s and 2s--1s transitions of H\,{\sc i} and the corresponding
lines of He~(i.e. the $2^1$p--$1^1$s and $2^1$s--$1^1$s transitions of
He\,{\small I}, and the 2p--1s and 2s--1s transitions of He\,{\small II})
during recombination, using the standard cosmological parameters and
recombination history.  In Section~\ref{3theory} we will describe the
model we used in the numerical calculation and give the equations used to
calculate the spectral lines.  In Section~\ref{3result} we will present our
results and discuss the detailed physics of the locations and shapes of
the spectral lines.  An approximate formula for the magnitude of the
distortion in different cosmologies will also be given.  Other
possible modifications of the spectral lines and their potential
detectability will be discussed in
Section~\ref{3discuss}.  And finally, we present our conclusions
in the last section.

\section{Basic theory}
\label{3theory}
\subsection{Model}
Instead of adopting a full multi-level code, we use a simple 3-level
model atom here.  For single-electron atoms (i.e. H\,{\sc i} and
He\,{\sc ii}), we consider only the ground state, the first excited
state and the continuum.  For the 2-electron atom (He\,{\sc i}), we
consider the corresponding levels among singlet states.  In general,
the upper level states are considered to be in thermal equilibrium
with the first excited state.  Case B recombination is adopted here,
which means that we ignore recombinations and photo-ionizations
directly to ground state.  
This is because the photons emitted from
direct recombinations to the ground state will almost
immediately reionize a nearby neutral H atom\,\cite{3peebles1968,
3sara00}.  We also include the two-photon rate from 2s to the ground
state for all three atoms, with rates: $\Lambda^{\rm H}_{\rm 2s-1s} =
8.229063\,$s$^{-1}$\,\cite{3gold89,3santos98}; $\Lambda^{\rm HeI}_{\rm 2^1s-1^1s}
= 51.02\,$s$^{-1}$\,\cite{3derev97}, although it makes no noticeable
difference to the calculation if one uses the older value of
$51.3\,$s$^{-1}$ from Drake, Victor \& Dalgarno~(1969)\,\cite{3drake69};
and $\Lambda^{\rm HeII}_{\rm 2s-1s} = 526.532\,$s$^{-1}$\,\cite{3gold89,3lip65}.
This 3-level atom model is similar to the one used in the program {\sc
recfast}, with the main difference being that here we
do not assume that the rate of change of the first
excited state $n_2$ is zero.

The rate equations for the 3 atoms are
similar, and so we will just state the hydrogen case as an example:
{\setlength\arraycolsep{1pt}
\begin{eqnarray}
(1+z) \frac{dn_1^{\rm{H}}(z)}{dz} &=& -\frac{1}{H(z)}
[\Delta R_{2\mathrm{p}-1\mathrm{s}}^{\rm{H}}
+ \Delta R_{2\mathrm{s}-1\mathrm{s}}^{\rm{H}} ] +3n_1^{\rm{H}} ; \\
(1+z) \frac{dn_2^{\rm{H}}(z)}{dz} &=& -\frac{1}{H(z)} [
n_{\rm{e}}n_{\rm{p}} \alpha_{\rm{H}}
- n_{2\rm{s}}^{\rm{H}} \beta_{\rm{H}} 
- \Delta R_{2\mathrm{p}-1\mathrm{s}}^{\rm{H}} \nonumber \\
&& \qquad \qquad \qquad \qquad \qquad \qquad - \Delta R_{2\mathrm{s} 
-1\mathrm{s}}^{\rm{H}} ] +3n_2^{\rm{H}} ; \\
(1+z) \frac{dn_{\rm{e}}(z)}{dz} &=& -\frac{1}{H(z)} \left[
n_{2\rm{s}}^{\rm{H}} \beta_{\rm{H}} - n_{\rm{e}}n_{\rm{p}} \alpha_{\rm{H}}
 \right] +3n_{\rm{e}} ; \\
(1+z) \frac{dn_{\rm{p}}(z)}{dz} &=& -\frac{1}{H(z)} \left[
n_{2\rm{s}}^{\rm{H}} \beta_{\rm{H}} - n_{\rm{e}}n_{\rm{p}} \alpha_{\rm{H}}
 \right] +3n_{\rm{p}}.
\end{eqnarray}}
\\*
Here the values of $n_i$ are the number density of the $i$th
state, where $n_{\rm e}$ and $n_{\rm p}$ are the number density of
electrons and protons respectively. $\Delta R^{\rm H}_{i-j}$ is the
net bound-bound rate between state $i$ and $j$ and the detailed form of
$\Delta R_{2\mathrm{p}-1\mathrm{s}}^{\rm{H}}$ and $ \Delta
R_{2\mathrm{s}-1\mathrm{s}}^{\rm{H}}$ will be discussed in the
next subsection.  $H(z)$ is the Hubble factor,
{\setlength\arraycolsep{2pt}
\begin{equation}
H(z)^2 =  H_0^2 \bigg[ \frac{\Omega_{\rm m}}{1+z_{\rm eq}}(1+z)^4
 + \Omega_{\rm m} (1+z)^3 + \Omega_K(1+z)^2 + \Omega_{\Lambda} \bigg].
\end{equation}}
\\*
Here $z_{\rm eq}$ is the redshift of matter-radiation equality\,\cite{3sara00},
\begin{equation}
1+z_{\rm eq} = \Omega_{\rm m} \frac{3H_0^2 c^2}{8 \pi G(1+f_{\nu}) U},
\end{equation}
where $U$ is radiation energy density $U= a_{\mathrm{R}}
T_{\mathrm{R}}^4$, $a_{\rm R}$ is the radiation constant,
 $f_{\nu}$~is the neutrino contribution to the 
energy density in relativistic species.
 Finally $\alpha_{\rm{H}}$ is the Case B recombination coefficient 
from Hummer~(1994)\,\cite{3hummer94},
\begin{equation}
\alpha_{\rm{H}} =  \ 10^{-19} \frac{at^b}{1+ct^d} \ \rm{m}^3 s^{-1},
\end{equation}
which is fitted by Pequignot et al.~(1991)\,\cite{3pequignot91},
 with $a=4.309$, $b=-0.6166$, $c=0.6703$, $d=0.5300$ and 
$t=T_{\rm{M}}/10^4$K, while $\beta_{\rm{H}}$ is the
photo-ionization coefficient:
\begin{equation}
\beta_{\rm{H}} = \alpha_{\rm{H}} \left( \frac{2 \pi m_{\rm e}
k_{\rm{B}} T_{\rm{M}}}
{h_{\rm p}^2} \right)^{\frac{3}{2}} {\rm exp} \left\{
-\frac{h_{\rm p} \nu_{2\rm{s,c}}}{k_{\rm{B}} T_{\rm{M}}} \right\},
\label{betaH}
\end{equation}
where $T_{\rm M}$ is the matter temperature and 
$\nu_{\rm 2s,c}$ is the frequency of the energy difference
between state 2s and the continuum.  For the rate of change of $T_{\rm M}$,
we only include the Compton and adiabatic cooling terms\,\cite{3sara00}, i.e.
\begin{equation}
(1+z) \frac{d T_{\rm{M}}} {dz} =
\frac{8 \sigma_{\rm{T}} U}{3 H(z) m_e c}
\frac{n_e}{n_e + n_{\rm{H}} + n_{\rm{He}} } (T_{\rm{M}}-T_{\rm{R}})
+ 2 T_{\rm{M}},
\label{eqTM}
\end{equation}
\\*
where $c$ is the speed of light and
$\sigma_{\mathrm{T}}$ is the Thompson scattering cross-section.

We use the Bader-Deuflhard semi-implicit numerical integration scheme
(see Section 16.6 in \cite{3nr}) to solve the above rate equations.
All the numerical results in this chapter are made using 
the $\Lambda$CDM model with parameters: 
$\Omega_{\rm b}=0.046$; $\Omega_{\rm m}=0.3$;
$\Omega_{\Lambda}=0.7$; $\Omega_{\rm K}=0$; $Y_p=0.24$; $T_0=2.725\,$K
and $h=0.7$~(see for examples,~\cite{3spergel03}).

\subsection{Spectral distortions}
We want to calculate the specific line intensity $I_{\nu_0}(z=0)$
(i.e. energy per unit time per unit area per unit frequency per unit
solid angle, measured in W$\,$m$^{-2}$Hz$^{-1}{\rm sr}^{-1}$) observed
at the present epoch, $z=0$.  The detailed calculation of
$I_{\nu_0}(z=0)$ for the Ly$\,\alpha$ transition and the two-photon
transition in hydrogen are presented as 
examples~(the notation follows Section 2.5 in \cite{3pad93}). 
A similar derivation holds for the
corresponding transitions in helium.  To perform this calculation we
first consider the emissivity $j_{\nu}(z)$ (energy per unit time per
unit volume per unit frequency, measured in W$\,$m$^{-3}$Hz$^{-1}$) of
photons due to the transition of electrons between the 2p and 1s
states at redshift $z$:
\begin{equation}
j_{\nu}(z)=h_{\rm P} \nu \Delta R_{2\rm{p}-1\rm{s}}^{\rm{H}}(z)
\phi[\nu(z)],
\end{equation}
where $\phi(\nu)$ is the frequency distribution of the emitted photons
from the emission process and $\Delta R_{2\rm{p}-1\rm{s}}^{\rm{H}}$ is
the net rate of photon production between the 2p and 1s levels, i.e.
\begin{equation}
\Delta R_{2\rm{p}-1\rm{s}}^{\rm{H}} = p_{12} \left(
n_{2 \rm{p}}^{\rm{H}} R_{21} - n_1^{\rm{H}} R_{12} \right).
\label{RLyH}
\end{equation}
Here $n_i^{\rm{H}}$ is the number density of hydrogen atoms having
electrons in state $i$, the upward and downward transition rates are
\begin{eqnarray}
R_{12} &=&  B_{12} \bar{J}, \\
\mathrm{and} \quad R_{21} &=& \left(A_{21} + B_{21} \bar{J}\right),
\end{eqnarray}
with $A_{21}$, $B_{12}$ and $ B_{21}$ being the Einstein coefficients
and $p_{12}$ the Sobolev escape probability~(see \cite{3sara00}),
which accounts for the redshifting of the Ly$\,\alpha$ photons due to
the expansion of the Universe.  As $n^{\rm H}_{1} \gg n^{\rm H}_{2
\rm{p}} $, $p_{12}$ can be expressed in the following form:
\begin{equation}
p_{12} = \frac{1- e^{-\tau_{\rm s}}}{\tau_{\rm s}}, \rm{with}
\label{es_prob}
\end{equation}
\begin{equation}
\tau_{\rm s}= \frac{A_{21} \lambda_{\rm 2p,1s}^3 \left(g_{2\rm{p}}
/g_1 \right) n_1}
{8 \pi H(z)}.
\end{equation}
We approximate the background radiation field $\bar{J}$ as a perfect
blackbody spectrum by ignoring the line profile of the
emission~(see \cite{3sara00}).  We also neglect secondary distortions
to the radiation field (but see the discussion in
Section~\ref{3sec:modifications}).
These secondary distortions come from photons
emitted earlier in time, during recombination of H or He, primarily
the line transitions described in this paper.
Assuming a blackbody we have
\begin{equation}
\bar{J}(T_{\rm{R}}) = \frac{2h_{\rm P} \nu_{\alpha}^3}{c^2}
\left[ {\rm exp}
\left( \frac{h_{\rm P} \nu_{\alpha}}
{k_{\mathrm{B}} T_{\rm{R}}} \right) -1 \right]^{-1},
\end{equation}
where $\nu_{\alpha}= c/121.5682\,$nm$=2.466 \times 10^{15}\,$Hz and
corresponds to the energy difference between states 2p and 1s, while the
frequency of the emitted photons is equal to $\nu_{\alpha}$.
Therefore, we can set $\phi[\nu (z)] = \delta[\nu(z) -\nu_{\alpha}]$,
i.e.  a delta function centred on $\nu_{\alpha}$, so that
\begin{equation}
j_{\nu}^{\rm{Ly} \alpha}(z)=h_{\rm P} \nu \Delta
R_{2\rm{p}-1\rm{s}}^{\rm{H}}(z)
\delta[\nu(z) -\nu_{\alpha}].
\end{equation}
The increment to the intensity coming from time interval $dt$ at redshift $z$ is
\begin{equation}
d I_{\nu}(z) = \frac{c}{4 \pi} j_{\nu} dt,
\end{equation}
which redshifts to give
\begin{equation}
d I_{\nu_0}(z=0) = \frac{c}{4 \pi} \frac{j_{\nu}}{(1+z)^3} dt.
\end{equation}
We assume that the emitted photons propagate freely until the present time.
Integration over frequency then gives
{\setlength\arraycolsep{2pt}
\begin{eqnarray}
I^{\mathrm{Ly} \alpha}_{\nu_0}(z=0) &=& \frac{c}{4 \pi} \int
\frac{j_{\nu}}{(1+z)^3} dt \label{inten} \\
&=& \frac{c h_{\rm P}}{4 \pi }
\frac{\Delta R_{2\rm{p}-1\rm{s}}^{\rm{H}} (z_{\alpha})}{H(z_{\alpha})
(1+z_{\alpha})^3}, 
\label{ILya}
\end{eqnarray}}
\\
with
\[
1+z_{\alpha} = \frac{\nu_{\alpha}}{\nu_0},
\]
using
\[
\nu(z) = \nu_0 (1+z) \quad \mathrm{and} \quad \frac{dt}{dz}
= -\frac{1}{H(z) (1+z)}.
\]
Equation~(\ref{ILya}) is the basic equation for determining the Ly$\,\alpha$
line distortion, using $\Delta R_{2\rm{p}-1\rm{s}}^{\rm{H}}(z)$ from
the 3-level atom calculation. 

\begin{figure} 
\begin{center}
\includegraphics[width=0.9\textwidth]{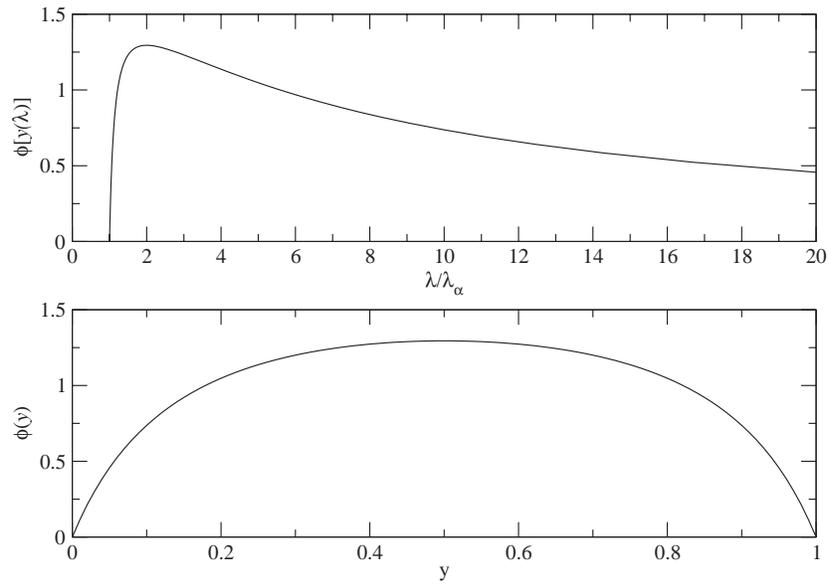}
\caption[The normalized emission spectrum for
the two-photon process (2s--1s) of hydrogen.]{The normalized emission spectrum for
the two-photon process (2s--1s) of hydrogen~\cite{3spitzer51,
3martinis00}.  The top panel shows $\phi(\lambda)$ vs $\lambda$, while
the bottom panel shows $\phi(y)$ vs $y$, where $\nu = y \nu_\alpha$.
Note that the spectrum is symmetric in $\nu$ about $\nu_{\alpha}/2$,
but the $\lambda$ spectrum is very asymmetric, being zero below
$\lambda_{\alpha}$, and having a tail extending to high $\lambda$.}
\label{3phi_H}
\end{center}
\end{figure}

For the two-photon emission between
the 2s and 1s levels, the emissivity at each redshift is
\begin{equation}
j_{\nu}(z)=h_{\rm P} \nu \Delta R_{2\rm{s}-1\rm{s}}^{\rm{H}}(z)
\phi[\nu(z)],
\end{equation}
and the calculation is slightly more complicated, since for
 $\phi(\nu)$ we need the frequency spectrum of the emission photons of
 the 2s--1s transition of H\,\cite{3martinis00,3spitzer51} as shown in
 Fig.~\ref{3phi_H}.  Here $\Delta R_{2\rm{s}-1\rm{s}}^{\rm{H}}$ is the
 net rate of photon production for the 2s--1s transition, i.e.
 \begin{equation}
 \Delta R_{2\rm{s}-1\rm{s}}^{\rm{H}} = \Lambda^{\rm H}_{\rm 2s-1s}
 \left(  n_{2 \rm{s}}^{\rm{H}}
- n_1^{\rm{H}} e^{-h_{\rm P}\nu_{\alpha}/k_{\rm{B}}T_{\rm{M}}}\right).
\label{R2phH}
\end{equation}
Therefore, using equation~(\ref{inten}), we have
\begin{equation}
I_{\nu_0}^{2 \gamma}(z=0) = \frac{c h_{\rm P} \nu_0}{4 \pi} \int_{0}^{\infty}
\frac{ \Delta R_{2\rm{s}-1\rm{s}}^{\rm{H}}  (z) \phi[\nu_0(1+z)]}{H(z)
 (1+z)^3} \ dz.
\label{dir_int}
\end{equation}
We use the simple trapezoidal rule~(see Section~4.1 in \cite{3nr})
to integrate equation~(\ref{dir_int}) numerically from $z=0$ to the time
when $\Delta R$ is sufficiently small that the integrand can be
neglected.

\section{Results}
\label{3result}
\begin{figure}
\begin{center}
\includegraphics[width=0.9\textwidth]{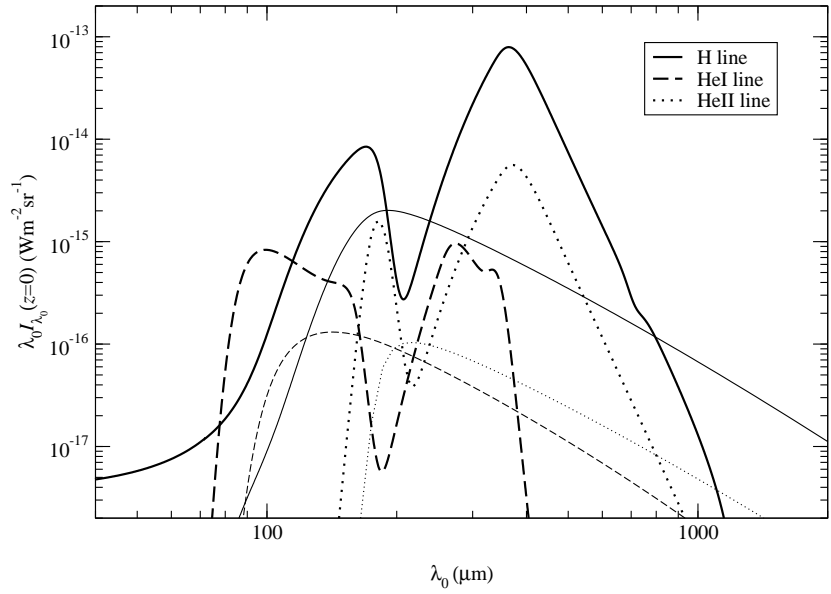}
\caption[The spectra of 8he individual line distortions from recombination.]
{The line intensity $\lambda_0 I_{\lambda_0}$ from the net
 Ly$\,\alpha$ emission of H (thick solid), the two-photon emission
 (2s--1s) of H\,{\sc i} with the spectrum $\phi(\nu)$ (thin solid), the
 2$^1$p--1$^1$s emission of He\,{\sc i} (thick dashed), the
 2$^1$s--1$^1$s two-photon emission of He{\sc i} (thin dashed), the
 2p--1s emission of He\,{\sc ii} (thick dotted) and the 2s--1s two-photon
 emission of He\,{\sc ii} (thin dotted).}
\label{3distortHHeII}
\end{center}
\end{figure}

\begin{figure}
\begin{center}
\includegraphics[width=0.9\textwidth]{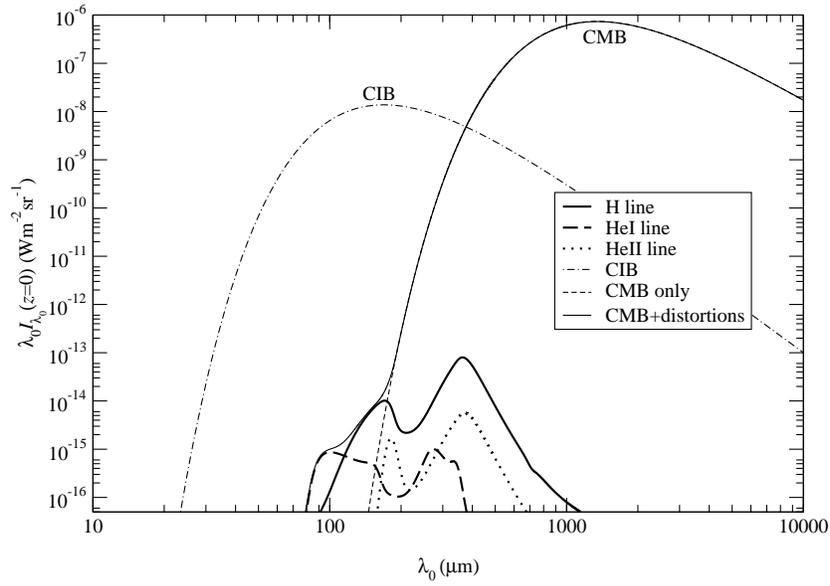}
\caption[The sum of all the emission lines of H and He 
plus the CMB as a function of frequency.]
{The line intensity $\lambda_0 I_{\lambda_0}$ from the sum
of the net Ly$\,\alpha$ emission and two-photon emission (1s--2s) of H
(thick solid), the sum of the $2^1$p--$1^1$s emission and
$2^1$s--$1^1$s two-photon emission of He{\small I} (thick dashed), and
the sum of the 2p--1s emission and 2s--1s two-photon emission of
He{\small II} (thick dotted), together with the background spectra:
CMB (long-dashed); and estimated CIB~(dot-dashed; \cite{3fixsen98}) 
The sum of all the above emission lines of H and He plus the CMB is
also shown (thin solid).}
\label{3sumline}
\end{center}
\end{figure}

Each of the line distortions is shown separately in
Fig.~\ref{3distortHHeII} and summed for each species in
Fig.~\ref{3sumline}.  The shape of the lines from H\,{\sc i}, He\,{\sc i} and
He\,{\sc ii} are fairly similar.  There are two distinct peaks to the
2p--1s emission lines.  We refer to the one located at longer
wavelength as the `pre-recombination peak', since the corresponding
atoms had hardly started to recombine during that time.  The physics
of the formation of this peak will be discussed in detail in
section~\ref{3pre_recom}.  The second (shorter wavelength)
peak is the main recombination
peak, which was formed when the atoms recombined.  While the longer
wavelength peak actually contains almost an order of magnitude more
flux, it makes a much lower relative distortion to the CMB.  The ratio
of the total distortion to the CMB intensity is shown in
Fig.~\ref{3ratioCMB}.  It is ${\sim}\,1$ for the main recombination peak,
but ${\sim}\,10^{-4}$ for the pre-recombination peak.

In Fig.~\ref{3sumline}, we plot the lines from H\,{\sc i} and He\,{\sc i}
together with the CMB and an estimate of the CIB.  We can see that the
lines which make the most significant distortion to the CMB are the
Ly$\,\alpha$ line and the 2$^1$p--1$^1$s line of He\,{\sc i}, and
that these lines form a non-trivial shape for the overall distortion.
The sum of all the spectral lines and the CMB is shown in
Fig.~\ref{3sumline}.  Note that these lines will also exist in the
presence of the CIB -- but the shape of this background is currently
quite poorly determined \cite{3fixsen98,3hauser98}.

We now discuss details of the physics behind the shapes of each of the
main recombination lines.

\subsection{Lines from the recombination of hydrogen}
During recombination, the Lyman lines are optically thick, which means
that nearly all photons emitted from the transition to $n=1$ are instantly
reabsorbed.  However, some of the emitted photons redshift out of the line due
to the expansion of the Universe and this makes the Ly$\,\alpha$
transition one of the possible ways for electrons to cascade down to
the ground state.  The other path for electrons going from $n=2$ to
$n=1$ is the two-photon transition between 2s and 1s.  Fig.~\ref{3ratesH}
shows the net photon emission rate of the Ly$\,\alpha$ and two-photon
transitions as a function of redshift for the standard $\Lambda$CDM
model.  The two-photon rate dominates at low redshift, where the bulk of
the recombinations occur.  This means that there are more photons
emitted through the two-photon emission process~(54\% of the total
number of photons created during recombination of H) than through the
Ly$\,\alpha$ redshifting process.  This conclusion agrees with
Zeldovich et al.~(1968)\,\cite{3zks68} -- although of course the balance depends on the
cosmological parameters~(see \cite{3sara00}) and for today's best fit
cosmology the two processes are almost equal.  Despite this fact, the
overall strength of the two-photon emission lines are weaker because the
photons are not produced with a single frequency, but with a wide
spectrum ranging from 0 to $\nu_{\alpha}$.  The
location of the two-photon peak (see Fig.~\ref{3distortHHeII}) is also
somewhat unexpected, since it is almost at the same wavelength as the
Ly$\,\alpha$ recombination peak, rather than at twice the wavelength.
The reason for this will be discussed in the following subsection.

We should also note that the tiny dip in our curves for
the long-wavelength tail of the
pre-recombination peak~(see Fig.~\ref{3distortHHeII}) is due to a
numerical error, when the number density of the ground state is very
small.  This can also be seen in the pre-recombination peak
for He{\sc ii}.

\subsubsection{The pre-recombination emission peak}
\label{3pre_recom}
The highest Ly$\,\alpha$ peak (shown in Fig.~\ref{3distortHHeII}) is
formed before the recombination of H has already started,
approximately at $z>2000$.  During that time the emission of
Ly$\,\alpha$ photons is controlled by the bound-bound Ly$\,\alpha$
rate from $n=2$ (i.e.\ the $n_2 R_{21}$ term in equation~(\ref{RLyH}))
and the photo-ionization rate ($n_2 \alpha_{\rm H}$).  From
Fig.~\ref{3prerec}, we can see that at early times the bound-bound
Ly$\,\alpha$ rate is larger than the photo-ionization rate.  This
indicates that when an electron recombines to the $n=2$ state, it is
more likely to go down to the ground state by emission of a
Ly$\,\alpha$ photon than to get ionized.  
The excess Ly alpha photons are not reabsorbed by ground state H, but
are redshifted out of the absorption frequency due to the expansion of
the Universe;  they escape freely and form the pre-recombination
 emission line.  Note that there is very little net
recombination of H, since the huge reservoir of $>13.6\,$eV CMB
photons keeps photo-ionizing the ground state H atoms~(see
Fig.~\ref{3noratio}).

\begin{figure} 
\begin{center}
\includegraphics[width=0.9\textwidth]{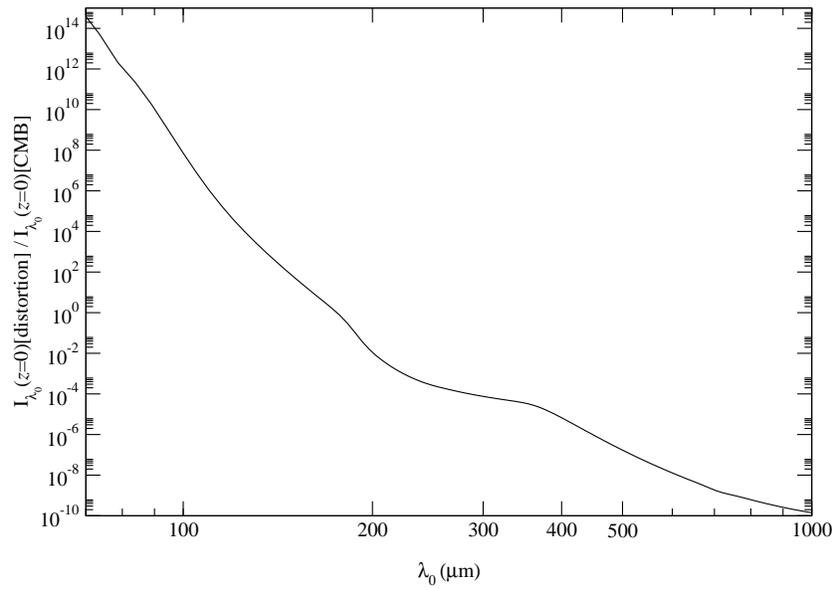}
\caption[The ratio of the total line distortion to the CMB intensity
as a function of redshift.]
{The ratio of the total line distortion to the CMB intensity
is plotted.  The ratio is larger than 1 (i.e. the intensity of the
distortion line is larger than that of the CMB) when $\lambda_0 \sim
170\,\mu$m which is just where the main Ly$\,\alpha$ line peaks.}
\label{3ratioCMB}
\end{center}
\end{figure}

\begin{figure} 
\begin{center}
\includegraphics[width=0.9\textwidth]{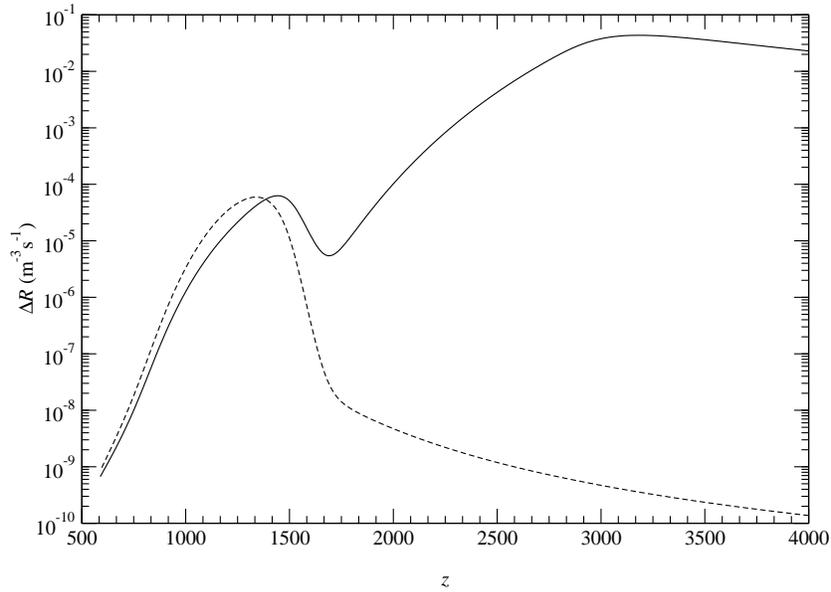}
\caption[Comparison of the net 2p--1s (solid) and 2s--1s (dashed) transition
rates of H.]
{Comparison of the net 2p--1s (solid) and 2s--1s (dashed) transition
rates of H.  The Ly$\,\alpha$ redshifting process dominates during the
start of recombination, while the 2-photon process is higher during most of the
time that recombination is occurring.  It turns out that in the standard
$\Lambda$CDM model about equal numbers of hydrogen atoms recombine through
each process, with slightly over half the hydrogen in the Universe recombining
through the 2-photon process.}
\label{3ratesH}
\end{center}
\end{figure}

\begin{figure}
\begin{center}
\includegraphics[width=0.9\textwidth]{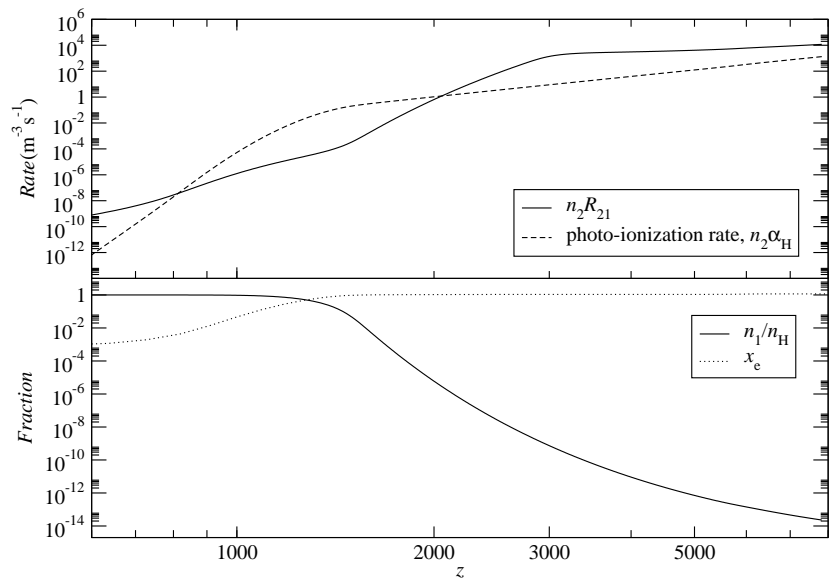}
\caption{The top panel shows the bound-bound Ly$\,\alpha$ rate $n_2
R_{21}$ and the photo-ionizing rate $n_2 \alpha_{\rm H}$ for $n$=2.
The lower panel shows the fraction of ground state H atoms $n_1/n_{\rm
H}$, and also the ionization fraction $x_{e}$.}
\label{3prerec}
\end{center}
\end{figure}

We now turn to a more detailed explanation of the pre-recombination
emission peak.  The bound-bound Ly$\,\alpha$ rate from $n=2$ is
initially approximately constant, as
it is dominated by the spontaneous de-excitation rate (the $A_{21}$
term in equation~(\ref{R2phH})).  At the same time the
photo-ionization rate is always decreasing as redshift decreases,
since the number of high energy photons keeps decreasing with the
expansion of the Universe.  Therefore, with a constant bound-bound
Ly$\,\alpha$ rate and the decreasing photo-ionization rate, the
emission of Ly$\,\alpha$ photons rises.  The peak of this
pre-recombination line of H occurs at around $z$\,=\,3000, by which time only a
very tiny amount of ground state H atoms have formed ($n_1/n_{\rm H} <
10^{-7}$, see Fig.~\ref{3prerec}).  These ground state H atoms build up
until they can reabsorb the Ly$\,\alpha$ photons and this lowers the
bound-bound Ly$\,\alpha$ rate.  The decrease of the bound-bound
Ly$\,\alpha$ rate is represented in the Sobolev escape probability
$p_{12}$ in equation~(\ref{es_prob}).  At high redshift, $p_{12}$ is 1
and there is no trapping of Ly$\,\alpha$ photons.  When H starts to
recombine, the optical depth $\tau_{\rm s}$ increases and the
Ly$\,\alpha$ photons can be reabsorbed by even very small amounts of
neutral H. For $\tau_{\rm s} \gg 1$, we can approximate $p_{12} \simeq
1/\tau_{\rm s}$ and $p_{12} \propto H(z)/n_1$.  Because of the
increase of the number density of the ground state and the decrease of
$H(z)$, the pre-recombination line decreases.  One can therefore think
of the `pre-recombination peak' as arising from direct Ly$\,\alpha$
transitions, before enough neutral H has built up to make the Universe
optically thick for Lyman photons.
This process occurs because the spontaneous
emission rate ($A_{21}$ term) is faster than the photo-ionization rate for
$n=2$; it increases as the Universe expands, due to the
weakening CMB blackbody radiation, and is quenched as the fraction of
atoms in the $n$\,=\,1 level grows.
The shorter wavelength peak, on the
other hand, comes from the process of redshifting out of the
Ly$\,\alpha$ line during the bulk of the recombination epoch.

By using the {\sc recfast} program~\cite{3sara99}, we can generate
the main Ly$\,\alpha$ recombination peak and also the two-photon emission
spectrum, by simply adding a few lines into the code.  However, the
pre-recombination peak cannot be generated from {\sc recfast},
since there the rate of change of the number density of the first excited
state $n_2$ is assumed to be negligible and is related to $n_1$ via
thermal equilibrium.  Moreover, in the effective 3-level formalism,
the Ly$\,\alpha$ line is assumed to be optically thick throughout the
whole recombination process of H (in order to reduce the calculation
into a single ODE), which is not valid at the beginning of the
recombination process.  Hence, one needs to follow the rate equations
of both states (i.e. $n$\,=\,1 and $n$\,=\,2) to generate the full
Ly$\,\alpha$ emission spectrum.  The pre-recombination peak of H
was mentioned and plotted in the earlier work of 
Dell'Antonio \& Rybicki~(1993)\,\cite{3dell93} as well,
although they did not describe it in any detail.

Another way to understand the line formation mechanism is to ask how
many photons are made in each process {\it per atom}.  We find that
for the main Ly$\,\alpha$ peak there are approximately 0.47 photons
per hydrogen atom (in the standard cosmology).  During the
recombination epoch, net photons for the $n=2$ to $n=1$ transitions
are only made when atoms terminate at the ground state.  Hence we
expect exactly one $n$\,=\,2 to $n$\,=\,1 photon for each atom, split between
the Ly$\,\alpha$ redshifting and 2-photon processes (and the latter
splits the energy into two photons, so there are 1.06 of these photons
per atom).  For the `pre-recombination peak', on the other hand, the
atoms give a Ly$\,\alpha$ photon when they reach $n=1$, but they then
absorb a CMB continuum photon to get back to higher $n$ or become
ionized.  The number of times an atom cycles through this process
depends on the ratio of the relevant rates.  If we take the rate per
unit volume from Fig.~\ref{3ratesH} and divide by the number density of
hydrogen atoms at $z\,{\simeq}\,3000$ then we get a rate which is
about an order of magnitude larger than the Hubble parameter at that
time.  Hence we expect about 10 `pre-recombination peak' photons per
hydrogen atom.  A numerical calculation gives the more precise value
of 8.11.

\subsubsection{The two-photon emission lines}
\label{2photonpara}
Surprisingly, the location of the peak of the line intensity of the
2s--1s transition is almost the same as that of the Ly$\,\alpha$
transition, as shown in Fig.~\ref{3distortHHeII}, while one might have
expected it to differ by a factor of 2.  In order to understand this
effect, we rewrite the equation~(\ref{dir_int}) in the following way:
\begin{equation}
I_{\nu_0}^{2 \gamma}(z=0) =
 \int_{0}^{\infty} \phi'(z') I^{\delta}_{\nu_0}[z=0;z'] \ d z' ,
\end{equation}
where $\phi'(z') = \nu_0 \phi(\nu')$, and
\begin{equation}
 I^{\delta}_{\nu_0}[z=0;z'] \equiv I^{\delta}_{\nu_0}[z=0;z'(\nu')] =
 \frac{c h_{\mathrm{p}}}{4 \pi } \frac{R_{2 \gamma}(z')}{H(z')(1+z')^3}, \quad
\label{singflux}
\end{equation}
with
\[
1+z' = \frac{\nu '}{\nu_0} .
\]
Equation~(\ref{singflux}) gives the redshifted flux (measured now at
$z$\,=\,0) of a single frequency $\nu'$ coming from redshift $z'$
and corresponding to the redshifted frequency $\nu_0$.

\begin{figure} 
\begin{center}
\includegraphics[width=0.9\textwidth]{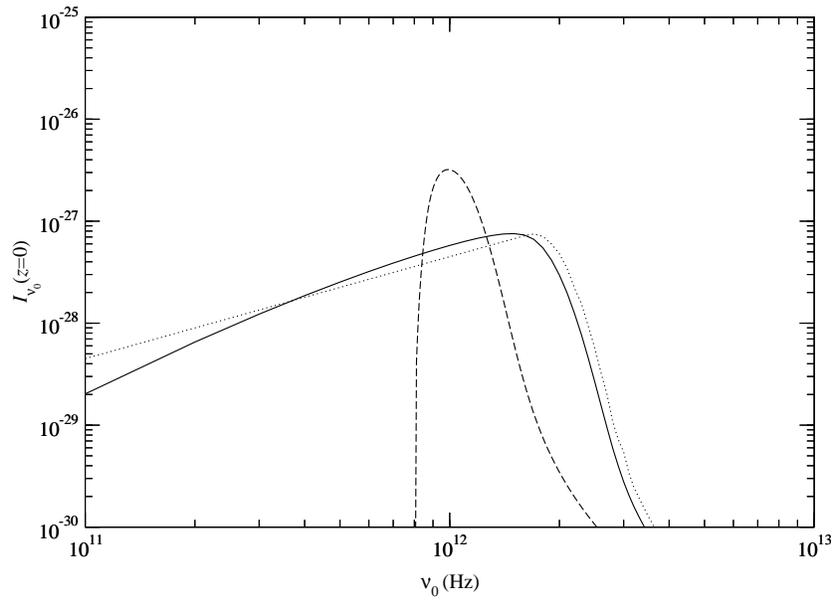}
\caption[The line intensity of the 2s--1s transition (two-photon
emission) $I_{\nu_0}(z=0)$ as a function of redshifted frequency
$\nu_0$ for three different assumptions.]
{The line intensity of the 2s--1s transition (two-photon
emission) $I_{\nu_0}(z=0)$ as a function of redshifted frequency
$\nu_0$ for three different assumptions: the correct frequency
spectrum of two-photon emission (solid); the delta function
approximation \mbox{$\delta(\nu - \nu_{\alpha} /2)$} (dashed); and the
flat spectrum approximation (dotted).}
 \label{3Icompare}
\end{center}
\end{figure}

We first calculate the line intensity of the two-photon emission with a
simple approximation: a delta function spectrum $\delta(\nu -
\nu_{\alpha}/2)$, where $\nu_{\alpha}/2$ is the frequency
corresponding to the peak of the two-photon emission spectrum
$\phi(\nu)$.  Fig.~\ref{3Icompare} shows the intensity spectrum of
two-photon emission using a delta frequency spectrum $\delta (\nu-
\nu_{\alpha}/2)$ compared with the two-photon emission using the correct
spectrum $\phi(\nu)$ .  We can see that there is a significant shift
in the line centre compared with the $\delta$-function case.  Where
does this shift come from?

\begin{figure} 
\begin{center}
\includegraphics[width=0.9\textwidth]{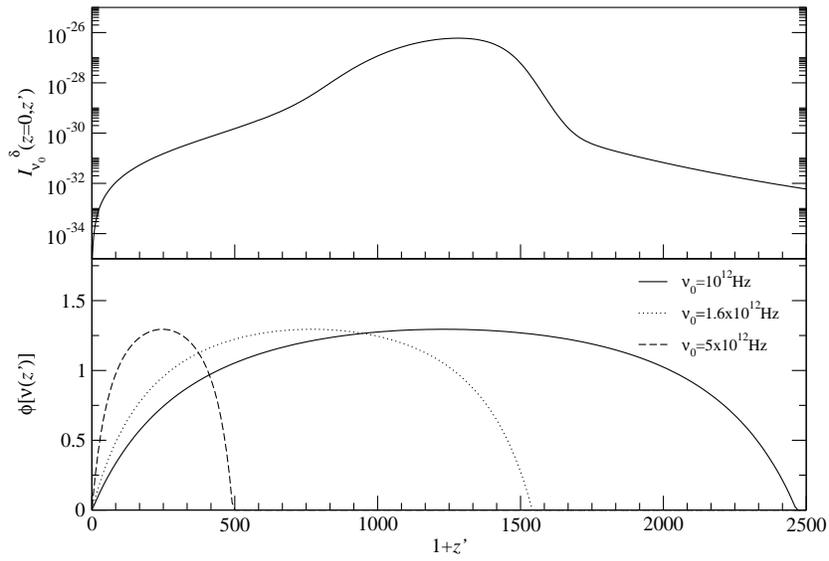}
\caption[The redshifted flux from single emission frequency
$I_{\nu_0}^\delta(z=0;z')$ plotted against the redshift of emission, $1+z'$.]
{The top panel shows the redshifted flux from single emission frequency
$I_{\nu_0}^\delta(z=0;z')$ plotted against the redshift of emission, $1+z'$.
The bottom panel shows the frequency spectrum of two-photon emission
$\phi[\nu(z')]$ plotted against $z'$ for three redshifted frequencies: $\nu_0=$
$10^{12}\,$Hz; $1.6 \times 10^{12}\,$Hz; and $5 \times 10^{12}\,$Hz.}
\label{3origI2ph}
\end{center}
\end{figure}

We know that the frequencies of emitted photons are within the range
of~0 to $\nu_{\alpha}$ at the time of emission.  For a fixed
redshifted frequency $\nu_0$ now, we can calculate the range of
emission redshifts contributing to $\nu_0$ (referred to as the
`contribution period' from now on), which is represented by
$\phi'(z')$ or $\phi(\nu')$ .  In Fig.~\ref{3origI2ph}, we show the
spectral distribution $\phi[\nu'(z')]$ as a function of redshift $z'$
for specific values of $\nu_0$.  For example, if we take $\nu_0\,=\, 5
\times 10^{12}\,\rm{Hz}$, then photons emitted between $1+z = 1$
(i.e. $\nu=\nu_0$) and ${\sim}\,500$ ($\nu=\nu_{\alpha}$) will give
contributions to $\nu_0$.  The smaller the redshifted frequency $\nu_0$,
the wider the contribution period.  We might expect that the line intensity
of this two-photon emission will be larger if the contribution period is
longer, as there are more redshifted photons propagating from earlier
times.  However, this is not the case, because the rate of two-photon
emission $R_{2 \gamma} $ also varies with time, and is sharply peaked
at $z \simeq$ 1300--1400.  Hence $I^{\delta}_{\nu_0}[z=0;z']$ is also
sharply peaked at $z \simeq$ 1300--1400. In Fig.~\ref{3origI2ph}, the
redshifted flux integrand $I_{\nu_0}^{\delta}(z=0, z)$ and the
emission spectrum $\phi[\nu(z)]$ are plotted on the same redshift
scale.  For $\nu_0= 5 \times 10^{12}\,\rm{Hz}$ (lowest panel), we can
see that the contribution period covers a redshift range when
$I_{\nu_0}^{\delta}(z=0, z)$ and $R_{2 \gamma}$ are small in value.
The contribution period widens with decreasing $\nu_0$ and covers more
of the redshift range when two-photon emission was high.  Therefore,
the flux $I_{\nu_0}(z=0)$ is expected to increase with decreasing
$\nu_0$ until the contribution period extends to the redshifts at
which the two-photon emission peaks.  As $\nu_0$ gets even smaller
(e.g.~$\nu_0=10^{12}$Hz), then the contribution period becomes larger
than the redshift range for two-photon emission and hence only lower
energy photons can be redshifted to that redshifted frequency.  As a
result, the flux $I_{\nu_0}(z=0)$ starts to decrease, and so we have a
peak. The flux peaks at $\nu_0 \simeq 10^{12}\,\rm{Hz}$ when we use
the $\delta$-function approximation.  However, from
Fig.~\ref{3origI2ph}, we can see that the contribution period for
$\nu_0 \simeq 10^{12}\,\rm{Hz}$ is much greater than that of the
two-photon emission period, and therefore this is not the location of
peak.  Based on the argument presented above, we expect the peak to be
at around $1.6 \times 10^{12}\,\rm{Hz}$, or $200\,\mu$m.

The basic mathematical point is that $\phi(y)$ is extremely poorly
represented by a $\delta$-function.  Since the spectrum $\phi(\nu)$ is
quite broad, it can be better approximated as a uniform distribution
than as a $\delta$-function.  Another crude approximation
would be to assume a flat spectrum for $\phi(\nu)$ in Fig.~\ref{3phi_H}.
Fig.~\ref{3Icompare} compares the intensity $I_{\nu_0}(z=0)$ found
using the correct form for $\phi(\nu)$ with the $\delta$-function and
flat spectrum approximations.  This shows that the flat spectrum gives
qualitatively the same results as the correct form of the spectrum,
and that the peak occurs fairly close to that of Ly$\,\alpha$,
but is much broader.  The same general arguments apply to the
two-photon lines of He\,{\sc i} and He\,{\sc ii} (as we discuss in
Section 3.3.2).

\subsubsection{Dependence of $\Omega_{\rm m}$ and $\Omega _{\rm b}$}
The largest distortion on the CMB is from the shorter wavelength
recombination peak of
the hydrogen Ly$\,\alpha$ line~(see Fig.~\ref{3ratioCMB}).
It may therefore be useful estimate the peak of this line's intensity
as a function of the cosmological parameters.  The relevant parameters
are the matter density ($\propto \Omega_{\rm m} h^2$) and the
baryon density ($\propto \Omega_{\rm b} h^2$).  This is because
$\Omega_{\rm m} h^2$ affects the expansion rate, while
$\Omega_{\rm b} h^2$ is related to the number density of
hydrogen.  No other combinations of cosmological parameters have a
significant impact on the physics of recombination.

We can crudely understand the scalings of these parameters through the
following argument.  Regardless of the escape probability $p_{12}$,
the remaining part of the rate $(n_{2 {\rm p}}^{\rm{H}} R_{21} -
n_1^{\rm{H}} R_{12})$ is roughly proportional to $n^{\rm H}_1$
$\propto \Omega_{\rm b}h^2(1-x_e)$.  The escape probability $p_{12}$
can be approximated as 1 at the beginning of recombination $(\tau_{\rm
s} \ll 1)$ and $1/\tau_{\rm s}$ during the bulk of the recombination
process (with $ \tau_{\rm s} \gg 1)$.  Note that $\tau_{\rm s} \propto
H(z)/n^{\rm H}_1 \propto (\Omega_{\rm m}h^2)^{1/2} [\Omega_{\rm
b}h^2(1-x_e) ]^{-1} $.  Therefore,
{\setlength\arraycolsep{2pt}
\begin{equation}
\Delta R_{\rm 2p-1s} \propto \left\{
\begin{array}{ll}
(\Omega_{\rm m}h^2)^{0} [\Omega_{\rm b}h^2(1-x_e) ] & {\rm for} \
\tau_{\rm s}\ll 1 \\ (\Omega_{\rm m}h^2)^{1/2} [\Omega_{\rm b} 
h^2(1-x_e)]^0 & {\rm for} \ \tau_{\rm s}\gg 1,
\end{array} \right.
\end{equation}}
\\
and thus {\setlength\arraycolsep{2pt}
\begin{equation}
I_{\lambda_0} \propto \frac{\Delta R}{H(z)} \propto \left\{
\begin{array}{ll}
(\Omega_{\rm m}h^2)^{-1/2} [\Omega_{\rm b}h^2(1-x_e) ] & {\rm for} \
\tau_{\rm s}\ll 1 \\ (\Omega_{\rm m}h^2)^{0} [\Omega_{\rm b}
h^2(1-x_e)]^0 & {\rm for} \ \tau_{\rm s}\gg 1.
\end{array} \right.
\end{equation}}
\\
From this rough scaling argument, we may expect that the
$\Omega_{\rm m}$ dependence of the peak of the Ly$\,\alpha$ line is an
approximate power law with index between $-1/2$ and 0, while for
$\Omega_{\rm b}$ the corresponding power-law index is expected to lie
between 0 and 1.  The dependence of $\Omega_{\rm m}$ is actually more
complicated when one allows for a wider range of
values~(see \cite{3dell93}).  The above estimation just gives a rough
physical idea of the power of the dependence.

A more complete numerical estimate of the peak of the recombination
Ly$\,\alpha$ distortion is:
{\setlength\arraycolsep{0pt}
\begin{equation}
\left( \lambda_0 I_{\lambda_0} \right)^{\rm{peak}}
 \simeq 8.5 \times 10^{-15}
\left( \frac{\Omega_{\rm b} h^2 }{0.0224}  \right)^{0.57}
\left( \frac{\Omega_{\rm m} h^2 }{0.147}  \right)^{0.15}
\mathrm{Wm}^{-2}\mathrm{sr}^{-1},
\label{peakLya}
\end{equation}}
\\
where we have normalized to the parameters of the currently
favoured cosmological model.
The peak occurs at
\begin{equation}
\lambda_0\simeq 170\,\mu{\rm m}
\end{equation}
for all reasonable variants of the standard cosmology.

\subsection{Lines from the recombination of helium (He\,{\sc i} and He\,{\sc ii})}
\begin{figure} 
\begin{center}
\includegraphics[width=0.9\textwidth]{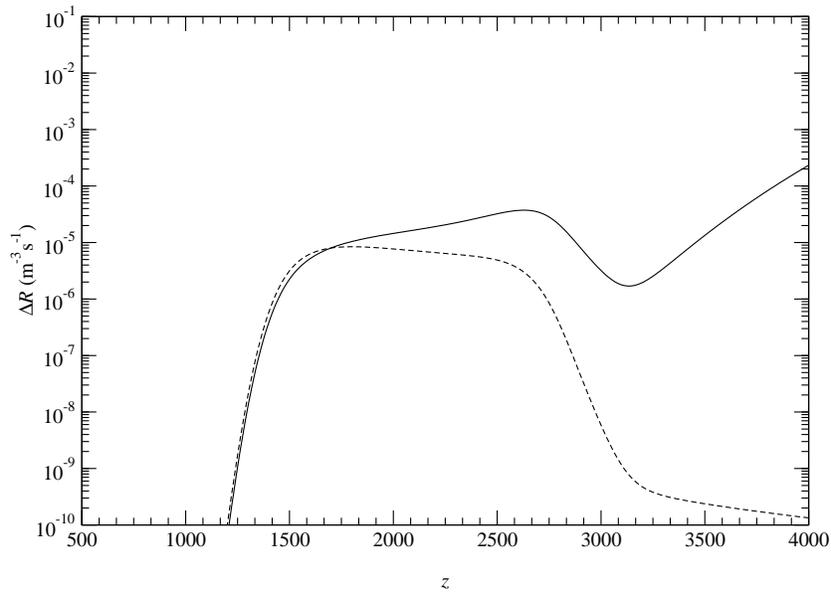}
\caption[Comparison of the net 2$^1$p--1$^1$s and
2$^1$s--1$^1$s two-photon transition rates of He\,{\sc i}.]
{Comparison of the net 2$^1$p--1$^1$s (solid) and
2$^1$s--1$^1$s two-photon (dashed) transition rates of He\,{\sc i}.
The two-photon rate is sub-dominant through most of the He\,{\sc i}
recombination epoch, and hence, unlike for hydrogen, most helium atoms did
{\it not} recombine through the two-photon process.}
\label{3ratesHeI}
\end{center}
\end{figure}

\begin{figure} 
\begin{center}
\includegraphics[width=0.9\textwidth]{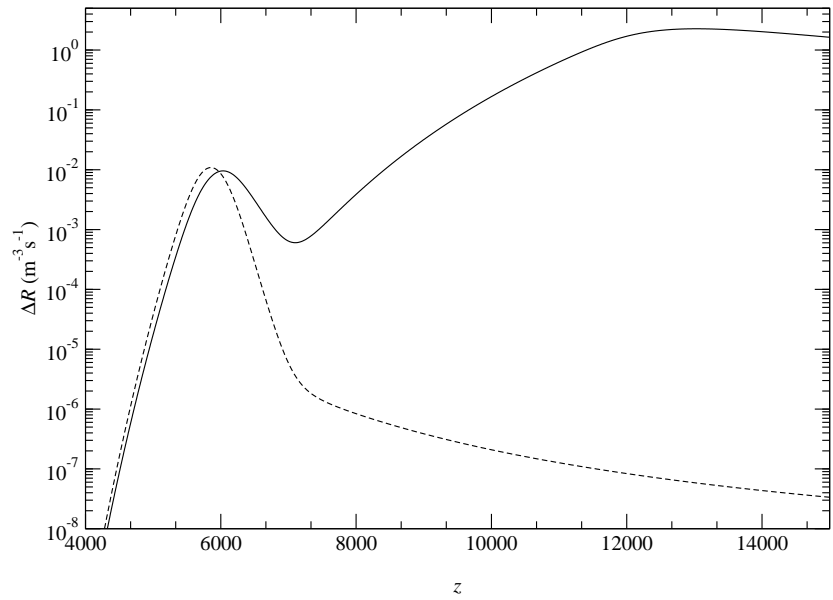}
\caption[Comparison of the net 2p--1s  and
2s--1s two-photon transition rates of He{\sc ii} as a function of
redshift.]
{Comparison of the net 2p--1s (solid) and
2s--1s two-photon (dashed) transition rates of He{\small II} as a function of
redshift.  The two-photon process is greater through most of the recombination
epoch, so that most of the cosmological He{\small III} $\to$ He{\small II}
process happens through the two-photon transition.}
\label{3HeIIrates}
\end{center}
\end{figure}

We compute the recombination of He\,{\sc ii} and He\,{\sc i} in the
same way as for hydrogen.  For the two-electron atom He\,{\sc i}, we ignore
all the forbidden transitions between singlet and triplet states due to
the low population of the triplet states~(see \cite{3sara99,3sara00}).
The $2^1$p--1$^1$s transitions of He\,{\sc i} are optically thick,
the same situation as for H.  This makes the electrons take longer to
reach the ground state and causes the recombination of He\,{\sc i} to
be slower than Saha equilibrium.  However, unlike for H, and despite
the optically thick $2^1$p--1$^1$s transition line, the $2^1$p--1$^1$s
rate dominates, as shown in Fig.~\ref{3ratesHeI}.  For He\,{\sc ii},
due to the fast two-photon transition rate (see Fig.~\ref{3HeIIrates}),
there is no `bottleneck' at the $n=2$ level in the recombination
process.  Hence He\,{\sc ii} recombination can be well approximated
by using the Saha equilibrium formula~\cite{3sara00}.

We can see the effect of the above differences in recombination
history on the lines: the width of the recombination peak of both H
and He\,{\sc i} is larger than that of He\,{\sc ii}.  Overall, the
spectral lines of He\,{\sc ii} are of much lower amplitude than those of H
(see Fig.~\ref{3distortHHeII}) with the distortion to the CMB about an
order of magnitude smaller.

The peaks of the line distortions from H and He\,{\sc ii} are located at
nearly the same wavelengths.  For hydrogenic ions the 1s--2p energy (and all
the others) scales as $Z^2$, where $Z$ is the atomic number.  Hence for
He\,{\sc ii} recombination takes place at $z\,{\simeq}\,6000$ rather than
the $z\,{\simeq}\,1500$ for hydrogen.  Hence the line distortion from
the 2p--1s transition of He\,{\sc ii} redshifts down to about $200\,\mu$m,
just like Ly$\,\alpha$.

\begin{figure} 
\begin{center}
\includegraphics[width=0.9\textwidth]{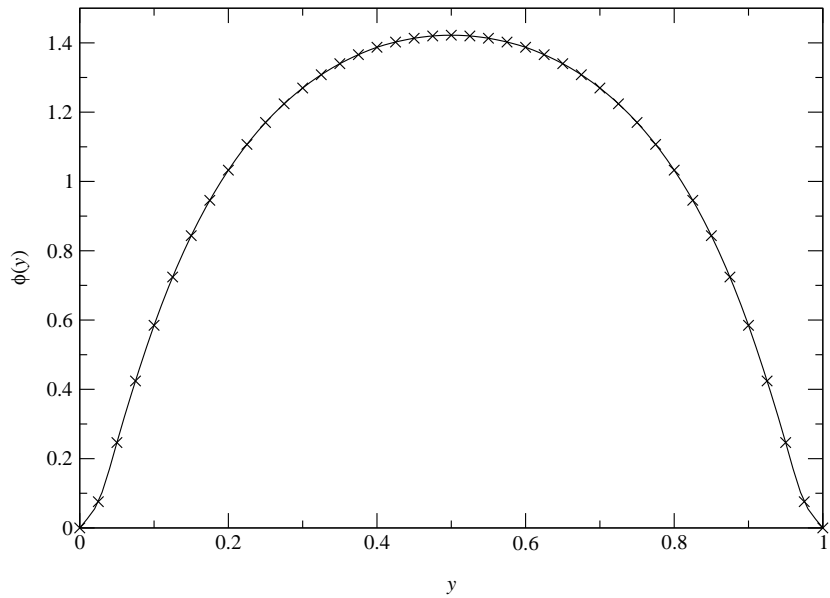}
\caption[The normalized emission spectrum for the two-photon
emission process (2$^1$s--1$^1$s) in He\,{\sc i}.]
{The normalized emission spectrum for the two-photon
emission process (2$^1$s--1$^1$s) in He\,{\sc i}.  Here $y= \nu /
\nu_{2\rm{s}-1{\rm s}}$, where $\nu_{2{\rm s} -1{{\rm s}}}=4.9849
\times 10^{15}\,$Hz.  The crosses are the calculated points from
Drake et al.~(1969)\,\cite{3drake69} and Drake~(1986)\,\cite{3drake86},
 while the line is a cubic spline fit.}
\label{3phi_HeI}
\end{center}
\end{figure}

The two-photon frequency spectrum of He\,{\sc ii} is the same as for H, since
they are both single-electron atoms~\cite{3tung84}.  However,
it is complicated to calculate the two-photon frequency spectrum of
He\,{\sc i} very accurately, since there is no exact wave-function for the
state of the atom.  Drake et al.~(1969)\,\cite{3drake69} used a variational 
method to calculate the two-photon frequency spectrum of He\,{\sc i} 
with values given up to 3 significant figures.
Drake~(1986)\,\cite{3drake86} presented another calculation, giving one more
digit of precision, and making the spectrum smoother, as shown in
Fig.~\ref{3phi_HeI}.  These two calculations differ by only about 1\%, which
makes negligible change to the two-photon He\,{\sc i} spectral line.

All of the H and He lines (for $n$\,=\,2 to $n$\,=\,1) are presented in
Fig.~\ref{3distortHHeII} and the sum is shown as a fractional
distortion to the CMB spectrum in Fig.~\ref{3ratioCMB}.
We find that in the standard cosmological model, for He\,{\sc i}
recombination, there are about 0.67 photons
created per helium atom in the `main' $2^1$p--$1^1$s peak, 0.70 per helium atom
in the `pre-recombination peak', and 0.66 in the two-photon process.  The
numbers for He\,{\sc ii} recombination are 0.62, 0.76 and 6.85 for these three
processes, respectively.

\section{Discussion}
\label{3discuss}
\subsection{Modifications in the recombination calculation}
\label{3sec:modifications}
There are several possible improvements that we could make to the line
distortion calculation.  However, as we will discuss below, we do not believe
that any of them will make a substantial difference to the amplitudes of the
lines.

\begin{figure} 
\begin{center}
\includegraphics[width=0.9\textwidth]{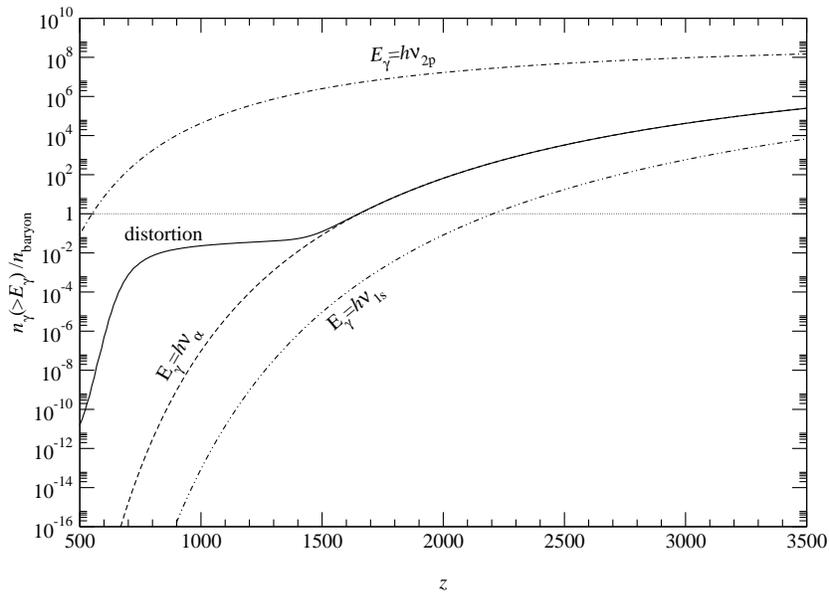}
\caption[The ratio of number of CMB photons with energy larger than
$E_{\gamma}$ ($n_{\gamma}(>E_{\gamma})$) to number of
baryons is plotted
against redshift $z$.]
{The ratio of number of CMB photons with energy larger than
$E_{\gamma}$ ($n_{\gamma}(>E_{\gamma})$) to number of
baryons ($n_{\rm B}$) is plotted
against redshift $z$.  The solid line includes the extra distortion photons
from the recombination of He{\small I}.  From the graph, we can see that
the recombination of H occurs approximately at the redshift when the ratio of
photons with energy ${>}\,h \nu _{\alpha}$ to baryons is about unity.  By
the time the helium recombination photons are a significant distortion to
the CMB tail above Ly$\,\alpha$ the density of the relevant photons has
already fallen by 2 orders of magnitude, and so the effects can make only a
small correction.}
\label{3noratio}
\end{center}
\end{figure}

In order to calculate the distortion lines to higher accuracy, we
should use the multi-level model without any thermal equilibrium
assumption among the bound states.  And we also need to take into
account the secondary spectral distortion in the radiation field,
i.e. we cannot approximate the background radiation field $\bar{J}$ as
a perfect blackbody spectrum.  This means, for example, that the extra
photons from the recombination of He\,{\sc i} may redshift into an energy
range that can photo-ionize H($n$\,=\,1)\,\cite{3dell93,3sara00}.
We can assess how significant this effect might be by considering the ratio
of the number of CMB background photons with energy larger than $E_{\gamma}$,
$n_{\gamma}(> E_{\gamma})$, to the number of baryons, $n_{\rm B}$, at
different redshifts (see Fig.~\ref{3noratio}).

Roughly speaking,
the recombination of H occurs at the redshift when
$n_{\gamma}(> h_{\rm p} \nu_{\alpha})/n_{B}$ is about equal to 1.
This is because at lower redshifts there are not enough high
energy background photons to photo-ionize or excite electrons
from the ground state to the upper states (even to $n$\,=\,2),
while at higher redshift, when such transitions are possible, there are huge
numbers of photons able to ionize the $n=2$ level.  The solid line in
Fig.~\ref{3noratio} shows the effect of the helium line distortions on the
number of high energy photons (above Ly$\,\alpha$) per baryon.
The amount of extra distortion photons with redshifted
energy larger than $h_{\rm P} \nu_{\alpha}$ coming from the recombination
of He{\sc i} is only about 1 per cent of the number of hydrogen atoms.
Their effect is therefore expected
to be negligibly small for $x_{\rm e}$.
We neglect the effect of the helium recombination photons on the hydrogen
line distortion, since it is clearly going to make a small correction
(at much less than the 10 per cent level).

As well as this particular approximation, there have been
some other recent studies which have suggested that it may be necessary to
make minor modifications to the recombination calculations presented in 
Seager et al.~(1999, 2000)\,\cite{3sara99,3sara00}.  Although
these proposed modifications would give only small changes to the recombination
calculation, it is possible that they could have much more significant
effects on the line amplitudes and shapes.  Recent papers have described 3
separate potential effects.

In the effective three-level model, Leung et al.~(2004)\,\cite{3leung04}
 argued that the adiabatic index of the matter should
change during the recombination process, as the ionized gas becomes
neutral, giving slight differences in the recombination history.
Dubrovich \& Grachev~(2005)\,\cite{3dub05} have claimed that the two-photon
rate between the lowest triplet state and the ground state and that between
the upper singlet states and the ground state should not be ignored in the
recombination of He{\sc i}.
And Chluba \& Sunyaev~(2005)\,\cite{3chluba05} suggested that one should also include
stimulated emission from the 2s state
of H, due to the low frequency photons in the CMB blackbody spectrum.
Even if all of these effects are entirely completely
correct, we find that the change to the amplitude of the main spectral
distortion is much less than 10\%.  We therefore leave the detailed
discussion of these and other possible modifications to a future work.

\subsection{Possibility of detection}
There is no avoiding the fact that detecting these CMB spectral
distortions will be difficult.  There are three main challenges to
overcome: (1) achieving the required raw sensitivity; (2) removing the
Galactic foreground emission; and (3) distinguishing the signal from
the CIB.

Let us start with the first point.  We can estimate the raw
sensitivity achievable in existing or planned experiments (even although
these instruments have {\it not\/} been designed for measuring the
line distortion).  Since the relevant wavelength range is essentially
impossible to observe from the ground, it will be necessary to go into space,
or at least to a balloon-based mission.  One existing experiment with
sensitivity at relevant wavelengths is BLAST\,\cite{3devlin04}
which has an array of bolometers operating at $250\,\mu$m
on a balloon payload.  The estimated sensitivity is $236\,$mJy in
$1\,$s, for a 30 arcsec FWHM beam, which corresponds to $\lambda
I_\lambda=1.2\times10^{-7}{\rm W}\,{\rm m}^{-2}{\rm sr}^{-1}$.
Comparing with equation~(\ref{peakLya}) for the peak intensity, it would take
${\sim}\,10^7$ such detectors running for a year to detect the line
distortion.  The SPIRE instrument on {\sl Herschel\/} will have a
similar bolometer array, but with better beamsize.  The estimated
sensitivity of $2.5\,$mJy at $5\sigma$ in 1 hour for a 17.4 arcsec
FWHM beamsize\,\cite{3griff00}
corresponds to
$\lambda I_\lambda=4.4\times10^{-8}{\rm W}\,{\rm m}^{-2}{\rm sr}^{-1}$
per detector for the $1\sigma$ sensitivity in $1\,$second.  So
detection of the line would still require ${\sim}\,10^6$ such detectors
operating for a year.

These experiments are limited by thermal emission from the instrument itself,
and so a significant advance would come from cooling the telescope.  This is
one of the main design goals of the proposed {\sl SAFIR\/}\,\cite{3leisawitz04}
and {\sl SPICA\/}\,\cite{3nakagawa04}
missions.  One can imagine improvements of a factor ${\sim}\,100$
for far-IR observations with a cooled mirror.  This would put us in the regime
where arrays of ${\sim}\,10^4$ detectors (of a size currently being
manufactured for sub-mm instruments) could achieve the desired sensitivity.

One could imagine an experiment designed to have enough spectroscopic
resolution to track the shape of the expected line distortion.  The minimum
requirement
here is rather modest, with only $\lambda/\delta\lambda\sim10$ in at least
3 bands.  An important issue will be calibration among the different
wavelengths, so that the non-thermal shape can be confidently measured.
To overcome this, one might consider the use of direct spectroscopic
techniques rather than filtered or frequency-sensitive bolometers.

Another way of quoting the required sensitivity is to say that any experiment
which measures the recombination line distortion would have to measure the
CIB spectrum with a precision of about 1 part in $10^5$, which is obviously
a significant improvement over what has been currently achieved.  A detection
of the line distortion might therefore naturally come out of an extremely
precise measurement of the CIB spectrum, which would also constrain other
high frequency distortions to the CMB spectrum.

Some of the design issues involved in such an experiment are discussed
by Fixsen \& Mather~(2002)\,\cite{3fixsen02}.  They describe a future experiment for measuring
deviations of the CMB spectrum from a perfect blackbody form, with an
accuracy and precision of 1 part in $10^6$.  This could provide upper
limits on Bose-Einstein distortion $\mu$ and Compton distortion $y$
parameters at the ${\sim}\,10^{-7}$ level~(the current upper
limits for $y$ and $\mu$ are $15 \times 10^{-6} $ and $9 \times
10^{-5}$, respectively; \cite{3fixsen96}).  The frequency coverage they
discuss is 2--120$\,{\rm cm}^{-1}$ (about 80--5{,}000$\,\mu$m), which
extends to much longer wavelengths than necessary for measuring the
line distortion.  The beam-size would be large, similar to FIRAS, but
the sensitivity achieved could easily be 100 times better.  An
experiment meant for detecting the line distortion would have to be
another couple of orders of magnitude more sensitive still.

Turning to the second of the major challenges, it will be necessary to detect
this line in the presence of the strong emission from our Galaxy.  At
$100\,\mu$m the Galactic Plane can be as bright as
${\sim}\,10^3{\rm MJy}\,{\rm sr}^{-1}$ which is about a billion times brighter
than the signal we are looking for!  Of course the brightness falls
dramatically as one moves
away from the Plane, but the only way to confidently avoid the
Galactic foreground is to measure it and remove it.  So any experiment designed
to detect the line distortion will need to cover some significant part of the
sky, so that it will be possible to extrapolate to the cosmological
background signal.
The spectrum of the foreground emission is likely to be smoother than
that of the line distortion, and it may be possible to use this fact to
effectively remove it.  However, it seems reasonable to imagine that the
most efficient separation of the signals will involve a mixture of spatial
and spectral information, as is done for CMB data~(see, for example, \cite{3patanchon04}).

In the language of spherical harmonics, the signal we are searching for is a
monopole, with a dipole at the ${\sim}\,10^{-3}$ level and smaller angular
scale fluctuations of even lower amplitude.  Hence we would expect to be
extrapolating the Galactic foreground signals so that we can measure the
overall DC level of the sky.  This is made much more difficult by the
presence of the CIB, which is also basically a monopole signal.  Hence spatial
information cannot be used to separate the line distortion from the CIB.
The measurement of the line distortion is therefore made much more difficult
by the unfortunate fact that the CIB is several orders of magnitude brighter
-- this is the third of the challenges in measuring the recombination lines.

The shape of the CIB spectrum is currently not very well characterised.
It was detected using data from the DIRBE and FIRAS experiments on
the {\sl COBE\/} satellite.  Estimates for the background
($\lambda I_{\lambda}$) are:
$9\,{\rm nW}\,{\rm m}^{-2}{\rm sr}^{-1}$ at $60\,\mu$m\,\cite{3miville02};
$23\,{\rm nW}\,{\rm m}^{-2}{\rm sr}^{-1}$ at $100\,\mu$m\,\cite{3lagach00};
$15\,{\rm nW}\,{\rm m}^{-2}{\rm sr}^{-1}$ at $140\,\mu$m\,\cite{3hauser98,3lagache99};
and $11\,{\rm nW}\,{\rm m}^{-2}{\rm sr}^{-1}$ at $240\,\mu$m\,\cite{3hauser98,3lagache99};
In each case the detections are only at the 3--$5\sigma$ level, and the
precise values vary between different prescriptions for data analysis
(see also \cite{3finkbeiner00,3hauserdwek01,3schlegel98}).
The short wavelength distortion of the CMB, interpretted as a measurement
of the CIB\,\cite{3puget96}
can be fit with a modified blackbody with
temperature $18.5\,$K and emissivity index 0.64 (although there is
degeneracy between these parameters), which we plotted in Fig.~\ref{3sumline}.

The CIB is thus believed to peak somewhere around $100\,\mu$m,
which is just about
where we are expecting the recombination line distortion.
The accuracy with which the CIB spectrum is known will have to
improve by about 5 orders of magnitude before the distortion will be
detectable.  Fortunately the spectral shape is expected to be significantly
narrower than that of the CIB -- the line widths are similar to the
$\delta z/z\sim0.1$ for the last scattering surface thickness, as opposed to
$\delta\lambda/\lambda\sim1$ for a modified blackbody shape (potentially
even wider than this, given that the sources of the CIB come from a
range of redshift $\Delta z\sim1$).

One issue, however, is how smooth the CIB will be at the level of detail with
which it will need to be probed.  It may be that emission lines, absorption
features, etc. could result in sufficiently narrow structure to obscure the
recombination features.  We are saved by 2 effects here: firstly the CIB
averaged over a large solid angle patch is the sum of countless galaxies,
and hence the individual spectral features will be smeared out; and
secondly, the far-IR spectral energy distributions of known galaxies do
{\it not\/} seem to contain strong features of the sort which might mimic
the recombination distortion (see, for example, \cite{3lagach05}).
As we learn more about the detailed far-IR spectra of individual galaxies
we will have a better idea of whether this places a fundamental limit on
our ability to detect the recombination lines.

Overall it would appear that the line distortion should be detectable
in principle, but will be quite challenging in practice.

\section{Conclusion}
We have studied the spectral distortion to the CMB due to the
Ly$\,\alpha$ and 2s--1s two-photon transition of H\,{\sc i} and the
corresponding lines of He\,{\sc i} and He\,{\sc ii}.  Together these lines
give a quite non-trivial shape to the overall distortion.
The strength and shape of the line distortions are very sensitive to the
details of the recombination processes in the atoms.  Although the
amplitude of the spectral line is much smaller than the Cosmic
Infrared Background, the raw precision required is within the grasp of
current technology, and one can imagine designing an experiment to
measure the non-trivial line shape which we have calculated.  The
basic detection of the existence of this spectral distortion would
provide incontrovertible proof that the Universe was once a hot plasma
and its amplitude would give direct constraints on physics at the
recombination epoch.

\newpage
\section{Remarks}
\label{ch3:remark}
Since this work was published, there have been other studies calculating
 the same spectral distortions with different approach in
 a different independent numerical
 code\,\cite{3RubinoMartin:2006ug,3RubinoMartin:2007}.
Rubi{\~n}o-Mart{\'{\i}}n~et~al.~(2006)\,\cite{3RubinoMartin:2006ug} 
 pointed out a correction in the treatment of the two-photon spectrum, 
 and found no pre-recombination peak in the H Ly\,$\alpha$ line distortion,
 in contradiction to the results of this chapter. 
As an addition to our published study, we now discuss these two issues.

\subsubsection{Normalization of the two-photon spectrum}
In our calculation of the two-photon line distortion, the emission
 spectrum $\phi (\nu)$ is normalized to 1~(see Figure~\ref{3phi_H}
 and Equation~(3.22)).  
However, the two-photon spectrum $\phi (\nu)$ should be normalized 
 to 2 (as pointed out by \cite{3RubinoMartin:2006ug}) because there 
 are two photons emitted in each electron transition from the 2s state 
 to the ground state.  
Due to this correction, the intensity of 
 the two-photon line distortion presented before should be doubled.
For H\,{\sc i}, since the amplitude of the distortion from Ly\,$\alpha$ 
 emission is about 10 times larger than the two-photon contribution,
 the overall shape and the peak location of the line spectrum remain almost 
 the same as before.  
The same correction should be made for the helium
 line distortion spectrum as well, and again the effect
 or the overall distortion from He is small.

\subsubsection{The pre-recombination peak}
Rubi{\~n}o-Mart{\'{\i}}n~et~al.~(2006)\,\cite{3RubinoMartin:2006ug}
 performed an independent calculation of the spectral line 
 distortions from H\,{\sc i} recombination with a multi-level
 atom model.  
The authors adopted the same procedure described in 
 Seager et al.~(2000)\,\cite{3sara00} but considered separate
 $l$-states within each $n$-shell of H\,{\sc i} with no thermal
 equilibrium assumption.
In this Chapter, we obtained a pre-recombination peak using 
 a 3-level atom model also based on the recombination model
 given by Seager et al.~(2000)\,\cite{3sara00}.
In contrast, Rubi{\~n}o-Mart{\'{\i}}n et al.~(2006)\,\cite{3RubinoMartin:2006ug}
 found {\it no} pre-recombination peak in their calculation.

As with earlier work\,\cite{3dell93}, 
 our pre-recombination peak was only found from
 numerical calculation and no explicit theoretical 
 argument for the formation of this peak was given.
We can consider the calculation a different way
 in order to understand the underlying physics.
Since the population of the hydrogen atom states
 is well described by the Boltzmann equations before 
 recombination~(say $z$\,$\lesssim$\,1700 for H\,{\sc i};
 see for example \cite{3sara00}),
 we now present an analytical estimate of the H\,{\sc i} 
 pre-recombination peak under the local 
 thermal equilibrium assumption in a 3-level atom model.

The pre-recombination peak was previously found in the
 calculation of the H\,{\sc i}  Ly\,$\alpha$ emission line,
 and also for the corresponding
 He\,{\sc i} and He\,{\sc ii} emission lines. 
Here we only discuss the case of H\,{\sc i}, since the 
 physics is basically the same for other species within 
 the standard recombination model\,\cite{3sara00}.
Since the spectrum of the photon emission in this 
 transition is narrowly peaked at the Ly\,$\alpha$ 
 frequency, the distortion shape is mainly controlled
 by the net Ly\,$\alpha$ emission rate
 $\Delta R_{\rm 2p-1s}$~(see Equation~(\ref{ILya})).
From Equation~(\ref{RLyH}), the net Ly\,$\alpha$ rate
can be rewritten as
\begin{equation}
\Delta R_{\rm 2p-1s} = \frac{p_{12} n_{\rm 1s} A_{21}}{1-e^{-h_{\rm P} \nu_{\alpha}/k_{\rm B} T_{\rm R}}}
\left( \frac{n_{\rm 2p}}{n_{\rm 1s}} - \frac{g_{\rm 2p}}{g_{\rm 1s}} 
e^{-h_{\rm P} \nu_{\alpha}/k_{\rm B} T_{\rm R}} \right) \, .
\end{equation}
\\*
Here we approximate $T_{\rm M}$\,$\simeq$\,$T_{\rm R}$.  
We can also write the net 2s--1s two-photon rate as
\begin{equation}
\Delta R_{\rm 2s-1s} = \Lambda^{\rm H}_{\rm 2s-1s} n_{\rm 1s} \frac{g_{\rm 2s}}{g_{\rm 2p}}
\left( \frac{n_{\rm 2p}}{n_{\rm 1s}} - \frac{g_{\rm 2p}}{g_{\rm 1s}} 
e^{-h_{\rm P} \nu_{\alpha}/k_{\rm B} T_{\rm R}} \right)  \, ,
\end{equation}
 by assuming that the 2p and 2s states are in thermal equilibrium.
From the above equations, we can see that these two rates 
 are controlled by the same difference, i.e.~the difference
 between the ratio $n_{\rm 2p}/n_{\rm 1s}$ and its local 
 thermal equilibrium value from the Boltzmann factor.  

\begin{figure} 
\begin{center}
\includegraphics[width=0.9\textwidth]{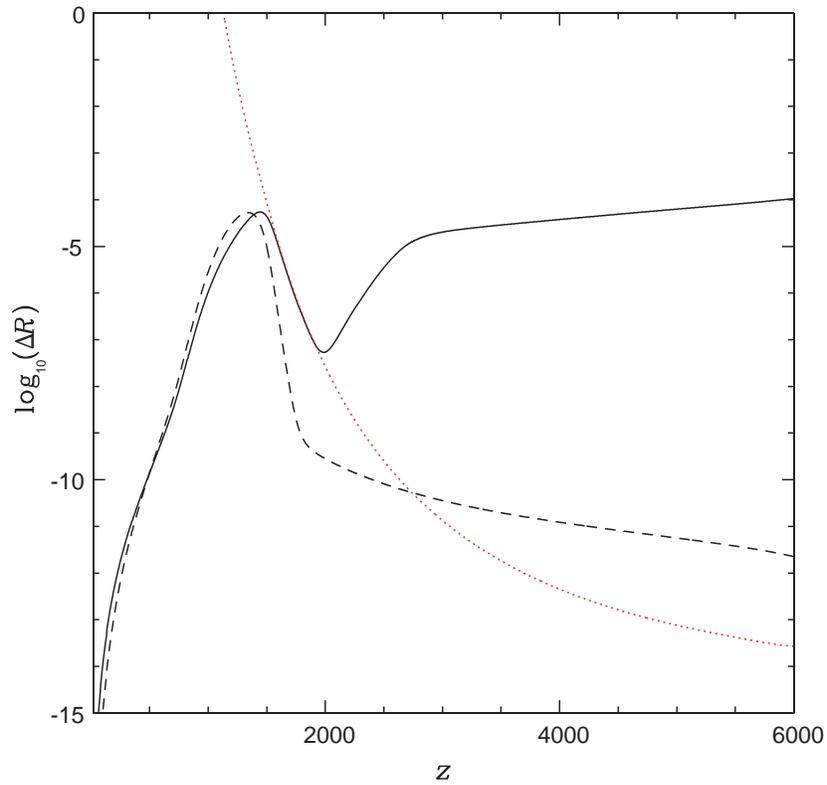}
\caption[A plot of the approximated Ly\,$\alpha$ rate calculated under
 thermal equilibrium assumption  at the time
 when the pre=recombination peak formed. ]
{The net Ly\,$\alpha$ transition rate $\Delta R_{\rm 2p-1s}$ (solid line)
and the net 2s--1s two-photon transition rate 
$\Delta R_{\rm 2s-1s}$ (dashed line) of H\,{\sc i} as 
a function of redshift $z$.  
These two curves are generated from the multi-level numerical recombination code. 
The dotted line (red) is the approximate analytical Ly\,$\alpha$ transition rate 
$\Delta R_{\rm 2p-1s}^{\rm LTE}$ from Equation~(\ref{3pre_Rlya}).}
\label{3prerecom01}
\end{center}
\end{figure}

Rubi{\~n}o-Mart{\'{\i}}n et al.~(2006)\,\cite{3RubinoMartin:2006ug}
 argued that $\Delta R_{\rm 2p-1s}$ is equal to zero at $z \geq 2000$
 because the states are in thermal equilibrium.  
This is not entirely true, since the expanding Universe 
 is a fundamentally out-of-equilibrium system; 
we will show that $\Delta R_{\rm 2p-1s}$
 is non-zero (although the rate is very low) even if 
 the population of the states in H\,{\sc i}
 is well approximated by the Boltzmann distribution 
 at each instant of time during the pre-recombination period.
From the Saha equation, we have
\begin{equation}
n_{i} = x_{\rm e} x_{\rm p} n_{\rm H}^2
\frac{g_i}{2}
\left( \frac{2\pi k_{\rm B} T_{\rm R}}{h_{\rm P}^2}\right)^{-3/2} 
\left( \frac{m_{\rm p} m_{\rm e}}{m_{\rm H}} \right)^{-3/2}
e^{h_{\rm P} \nu_{i,{\rm c}}/k_{\rm B} T_{\rm R}} \ ,
\label{3niLTE}
\end{equation}
\\*
where $\nu_{i,{\rm c}}$ is the frequency of the energy
difference between the $i$th state and the continuum.
We can take Equation~(\ref{3niLTE}) for $i$\,=\,1 (1s, the ground state),
 differentiate with respect to $z$ and substitute into Equation~(3.1),
 giving
\begin{equation}
\Delta R_{2\rm{p}-1\rm{s}}^{\rm LTE} + \Delta R_{2\rm{s}-1\rm{s}}^{\rm LTE}
 = n_{1 \rm s} H(z) \left( 
\frac{h_{\rm P} \nu_{1,{\rm c}}}{k_{\rm B} T_{\rm R}} - \frac{3}{2} 
- \frac{1+z}{x_{\rm e}} \frac{d x_{\rm e}}{d z}
\right) \ .
\label{3RlyLTE}
\end{equation}
The right-hand side of the above equation is dominated by
 the first two terms, since
\begin{equation}
\frac{h_{\rm P} \nu_{1,{\rm c}}}{k_{\rm B} T_{\rm R}} =
\frac{5.792\times 10^4}{1+z} \quad {\rm with}~T_{\rm R}=2.725(1+z)\,{\rm K},
\end{equation}
and
\begin{equation}
\frac{1+z}{x_{\rm e}} \frac{d x_{\rm e}}{d z} \lesssim 0.1 
\end{equation}
before the recombination of H\,{\sc i}~($z$\,$\simeq$\,1800).
This makes the sum of the two rates larger than zero
 and implies that there are net recombinations to
 the ground state even in the case that the number density
 of each state closely follows the thermal equilibrium 
 distribution.
Physically, the non-zero net recombination rate of H\,{\sc i} is due to the 
 decreasing number of high-energy photons in the expanding Universe. 
And as we know from Equation~(3.31) and (3.32), 
\begin{equation}
\frac{\Delta R_{\rm 2p-1s}}{\Delta R_{\rm 2s-1s}} \simeq 
\frac{3 p_{12} A_{21}}{\Lambda^{\rm H}_{\rm 2s-1s}} \sim 10^8 \quad {\rm at}~z> 1700\,.
\end{equation}
Before the H\,{\sc i} recombination ($z$\,$\gtrsim$\,1800 say), 
 the net Ly\,$\alpha$ rate dominates, because the neutral 
 hydrogen abundance is very low and the escape probability $p_{12}$ is 
 very close to~1.
Therefore, we can ignore $\Delta R^{\rm LTE}_{\rm 2s-1s}$ in 
Equation~(\ref{3RlyLTE}) and we have
\begin{equation}
\Delta R_{2\rm{p}-1\rm{s}}^{\rm LTE} \simeq 
n_{1 \rm{s}} H(z)  \left( 
\frac{h_{\rm P} \nu_{1\rm{s},{\rm c}}}{k_{\rm B} T_{\rm R}} - \frac{3}{2} \right)
\ .
\label{3pre_Rlya}
\end{equation}
In Figure~\ref{3prerecom01}, the approximate rate  
 $\Delta R^{\rm LTE}_{2\rm{p}-1\rm{s}}$ is plotted along 
 with the previous result from the numerical recombination code.  
We can see that $\Delta R_{2\rm{p}-1\rm{s}}^{\rm LTE}$ 
 matches the numerical rate $\Delta R_{\rm 2p-1s}$
 very well when the hydrogen is about to recombine at
 $z$\,=1800--2000.
These two rates are expected to depart at $z \simeq 1750$ 
 when the ground state goes out of thermal equilibrium 
 with the higher excited states due to the bottleneck at 
 the first excited state.  
On the other hand, we expect the thermal equilibrium assumption
 to be valid at even higher redshifts ($z$\,$>$\,2000). 
Under this assumption,
 we find no significant emission before recombination and 
 therefore, there is no pre-recombination peak.
Why does this result contradict with what we found in the numerical
 calculation?

In fact we found that the pre-recombination peak that we presented before
 arose due to a systematic error in the ODE (ordinary differential
 equation) solver.  
Any ODE solver allows us to find a numerical approximation to
 an exact (or real) solution of the equations to within some error.
We usually want the relative error to be small, and this is controlled
 by setting the required accuracy as an input parameter in the solver.
 In our case, the relative error of the number density is 
$\Delta n_i /n_i = (n_i^{\rm num} -n_i^{\rm real})/n_i^{\rm real}
\simeq 10^{-7} - 10^{-5}$ ($n_i^{\rm real}$ 
and $n_i^{\rm num}$ are the real and numerical values 
of $n_{\rm i}$, respectively).

\begin{figure} 
\begin{center}
\includegraphics[width=0.9\textwidth]{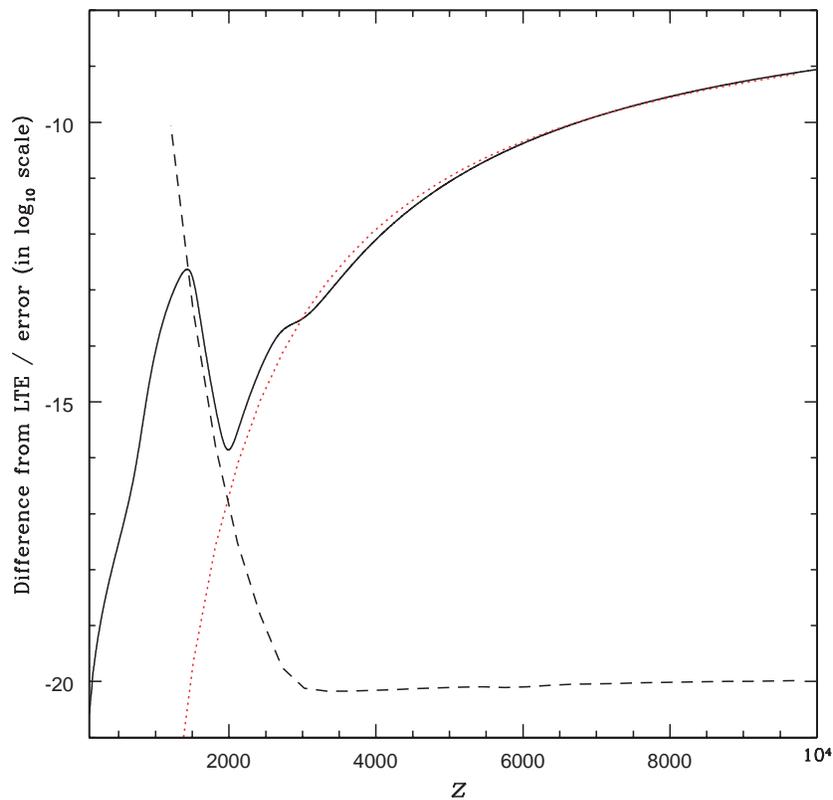}
\caption[The difference between the ratio 
 $n_{2 \rm p}/n_{1 \rm s}$ and its Boltzmann value
 as a function of redshift $z$.]
{The difference between the ratio $n_{2 \rm p}/n_{1 \rm s}$
 and its Boltzmann value as a function of redshift $z$.
  The solid line is $\Delta (n_{2 \rm p}/n_{1 \rm s})^{\rm num}$ 
 from the numerical recombination code, while the dashed line is the approximate
 value of the actual difference 
$\Delta (n_{2 \rm p}/n_{1 \rm s})^{\rm real}$ from Equation~(\ref{3app}).  
The dotted (red) line is the estimated numerical
 error $\epsilon_{12} (n_{2 \rm p}/n_{1 \rm s})^{\rm LTE}$ 
 with  $\epsilon_{12}=10^{-7.8}$.}
\label{3prerecom02}
\end{center}
\end{figure}

Now consider the effect of this error on the net rates.
The net rates are strongly controlled by the deviation
of the ratio $n_{\rm 2p}/n_{\rm 1s}$ from its Boltzmann 
value.  Note that
{\setlength\arraycolsep{1pt}
\begin{eqnarray}
\label{3eq_err}
\Delta \left( \frac{n_{2 {\rm p}}}{n_{1 \rm{s}} }\right)^{\rm num}
&=& \left( \frac{n_{2 {\rm p}}}{n_{1 \rm{s}}}\right)^{\rm num} - \
\left( \frac{n_{2 {\rm p}}}{n_{1\rm{s}}}\right)^{\rm LTE}   \\
&=& \left( \frac{n_{2 \rm p}^{\rm real} 
+ \Delta n_{2 \rm p}}{n_{1 \rm{s}}^{\rm real} + \Delta n_{1\rm{s}}}\right)
- \frac{g_{2 \rm p}}{g_{1 \rm s}} 
e^{-h_{\rm{p}}\nu_{\alpha}/k_{\rm B}T_{\rm{R}}} \nonumber \\
&\simeq &  \underbrace{
 \left[ \left( \frac{n_{2 {\rm p}}}{n_{1\rm{s}}}\right)^{\rm real}
-\frac{g_{2 \rm p}}{g_{1 \rm s}} e^{-h_{\rm{p}}\nu_{\alpha}/k_{\rm B}T_{\rm{R}}} \right]}
_{\Delta \left( \frac{n_{2 {\rm p}}}{n_{1\rm{s}}} \right)^{\rm real}} 
 + \underbrace{
\left( \frac{n_{2 {\rm p}}}{n_{1\rm{s}}}\right)^{\rm real} 
\left( \frac{\Delta n_{2\rm{p}}}{n_{2 {\rm p}} } 
- \frac{\Delta n_{1 \rm s}}{n_{1\rm s}} \right)}
_{\simeq  \left( \frac{n_{2 {\rm p}}}{n_{1\rm s}} \right)^{\rm LTE} 
\epsilon_{12}} \ , \nonumber
\end{eqnarray}}
\\*
where $\epsilon_{12} = \Delta n_{2 \rm p}/n_{2 \rm p}
- \Delta n_{1 \rm s} / n_{1 \rm s}$, should be 
the same order of magnitude as the uncertainty in $n_i$ 
(i.e.~$\Delta n_i / n_i$). 
In the above equation, the first bracket accounts for
 how much the first excited state and the ground
 state are out of equilibrium and this gives us the 
 actual net Ly\,$\alpha$ rate.
The second term is the error in the Ly\,$\alpha$
 rate due to the numerical errors in the number 
 densities.  
Somewhat surprisingly, it is directly proportional to
 the actual value of $n_{2 \rm p}/n_{1 \rm s}$, which
 increases with $z$.
In the pre-recombination epoch, we can approximate 
$(n_{2 \rm p}/n_{1 \rm s})^{\rm real}$ using Boltzmann
equations in order to calculate the error of the rate.  
For comparison, we use Equation~(3.31) to obtain the estimate 
\begin{equation}
\Delta \left( \frac{n_{2 {\rm p}}}{n_{1 \rm s}} \right)^{\rm real}
\simeq \frac{\Delta R_{2 \rm p - 1 \rm s}^{\rm LTE}}{p_{12} 
n_{1 \rm s} A_{21}}
\left( 1 - e^{-h_{\rm P} \nu_{\alpha}/k_{\rm B} T_{\rm R}} \right).
\label{3app}
\end{equation}
In Fig.~\ref{3prerecom02}, we separately plot the two terms in 
 Equation~(\ref{3eq_err}) as well as 
 $\Delta (n_{2 \rm p}/n_{1 \rm s})^{\rm num} $ from the 
 numerical code.
The estimated numerical error dominates at $z \gtrsim 2000$ 
 and it matches well with the 
 $\Delta (n_{2 \rm p}/n_{1 \rm s})^{\rm num}$ curve if we take 
 $\epsilon_{12} = 10^{-7.8}$, which is even smaller than
 the required accuracy in the ODE solver.  
This error term explains why there is an anomalous
 increasing trend of $\Delta (n_{2 \rm p}/n_{1 \rm s})^{\rm num}$
 at high $z$, while we expect the difference in the ratio to
 get smaller with increasing $z$ due to the
 tight thermal equilibrium relation between the states.
This estimated numerical error is directly proportional to 
 $n_{2 \rm p}/n_{1 \rm s}$ and decreases with decreasing $z$.
On the other hand, $ \Delta (n_{2 \rm p}/n_{1 \rm s})^{\rm real}$ is getting 
 larger and larger as the recombination of hydrogen begins.  
So at $z \simeq 2000$, 
 $\Delta (n_{2 \rm p}/n_{1 \rm s})^{\rm real}$ takes over. 
This explains why the $\Delta R_{2{\rm p}-1{\rm s}}^{{\rm num}}$
 and $\Delta R_{2\rm{p}-1\rm{s}}^{{\rm LTE}}$ values agree with
 each other only in the range of $z$\,$\simeq$\,1600--2000.
Overall, the pre-recombination peak that we found earlier
 seems to have beem caused by a systematic error.
This should serve as a warning for blindly accepting numerical result.

Physically, any possible pre-recombination peak or extra 
 emission through the H\,{\sc i} Ly\,$\alpha$ transition 
 would require channels for electrons in the ground state to get 
 back to higher excited states or the continuum, since there is 
 almost no net neutral hydrogen H\,{\sc i} formed due to these processes.  
In the three-level atom model, no such path
 exists since all the transitions to and from the ground state
 are connected with the first excited state~($n$\,=\,2).
We have also investigated this problem in a multi-level atom
 model based on the paper of Seager et al.~(2000)\,\cite{3sara00},
 and still find no such net  excitation to
 the higher excited states.
From this physical reasoning and the previous
 analysis of numerical errors, the pre-recombination 
 peak we found in the standard recombination calculation
 therefore appears to have been a false signal.

On the other hand, additional radiative processes not 
 considered in the standard recombination model of
 Seager et al.~(2000)\,\cite{3sara00} 
 provide the channels necessary for a pre-recombination peak.
In the recent studies which include addditional  
 continuum opacity of H\,{\sc i} in the He\,{\sc i}
 recombination\,\cite{3Kholupenko:2007qs,3Switzer:2007sn} evolution, 
 the extra high energy distortion photons from He\,{\sc i} 
 recombination can excite the electrons in the ground state
 for H\,{\sc i} atom before the recombination of H\,{\sc i} 
 at $z$\,=\,1600\,--\,2200.
This allows for direct ionization from the ground state,
 and a narrow  `pre-recombination peak' in the H\,{\sc i} 
 Ly\,$\alpha$ line is formed at 
 $z$\,$\simeq$\,1870~(see Figure~9 in \cite{3RubinoMartin:2007}).
This effect is, however, much smaller than the previous
 false signal. 
The amplitude of this pre-recombination peak is 
 about an order of magnitude smaller than the main peak
 of the H\,{\sc i} Ly\,$\alpha$ line formed during
 H\,{\sc i} recombination.

To conclude, the pre-recombination peaks found in our previous 
 studies\,\cite{3dell93,3Wong:2005yr}
 seem to have been false signals coming from a systematic 
 error in the numerical code~(for both H and He).  
By including the direct ionization from the ground state 
 of H\,{\sc i} due to the distortion photons from He\,{\sc i}
 recombination, we can find a pre-recombination peak 
 in the H\,{\sc i} Ly\,$\alpha$ line, but with a much reduced
 amplitude.

\chapter[Forbidden transitions]{Forbidden 
transitions\footnote[3]{A version of this chapter has been published: Wong W.~Y.
and Scott D.~(2007) `The effect of forbidden transitions
on cosmological hydrogen and helium recombination',
Monthly Notices of the Royal Astronomical Society, 375, 1441--1448.}}

\section{Introduction}
The release of the third year data from the Wilkinson Microwave Anisotropy 
Probe ({\sl WMAP}) has further improved the precision with which we can
constrain the cosmological parameters from the shape of 
the Cosmic Microwave Background (CMB) anisotropies $C_\ell$\,\cite{4spergel06}.
The {\sl Planck\/} satellite, scheduled for launch in 2008\,\cite{4planck06}, 
will provide even higher precision $C_\ell$ values and data down 
to smaller angular scales ($\ell \lesssim 2500$).  Higher precision in 
the observations requires increased accurarcy
from the theoretical calculations, in order for the correct cosmological parameters 
to be extracted.  It now seems crucial to obtain the $C_\ell$s down to 
at least the 1 percent level over a wide range of $\ell$.

{\sc cmbfast}\,\cite{4cmbfast} is the most commonly used Boltzmann code for calculating 
the $C_\ell$s, and it gives consistent results with other independent codes
(see \cite{4seljak03} and references therein).  The dominant 
uncertainty in obtaining accurate $C_\ell$s comes from details in the physics
of recombination, for example, the `fudge factor' in the {\sc recfast} 
routine\,\cite{4seager99, 4seager00}.
Calculations of cosmological recombination were first published by
Peebles~(1968)\,\cite{4peebles68} and Zeldovich et al.~(1968)\,\cite{4zks68}.
Seager et al.~(2000)\,\cite{4seager00} presented the most detailed multi-level 
calculation and introduced a fudge factor to reproduce the results within an effective
three-level atom model.  Although the multi-level calculation 
already gives reasonable accuracy, the required level of accuracy continues to
increase, so that today any effect which is $\sim$\,1 per cent over a range of 
multipoles is potentially significant.  Several modifications have been recently 
suggested to give per cent level changes in the ionization fraction 
and/or the $C_\ell$s (see Section~4.4 for details).
Most of these modifications have been calculated only with an effective 
three-level code, and so the results may be different in the 
multi-level calculation, since there is no thermal equilibrium assumed 
between the upper states.  Here we want to focus on one of these modifications, 
namely the extra forbidden transitions proposed by 
Dubrovich \& Grachev~(2005)\,\cite{4dub05}, which
we study using a multi-level code.  

In the standard calculations of recombination, one
considers all the resonant transitions, but only one forbidden transition, 
which is the 2S--1S two-photon transition, and this can be included 
for both H and He.  Dubrovich \& Grachev~(2005)\,\cite{4dub05} 
suggested that one should also include 
the two-photon transitions from higher excited S and D states to the 
ground state for H and He\,{\sc i}, and 
also the spin-forbidden transition between the triplet $2^3$P$_1$ and 
singlet ground state $1^1$S$_0$ for He\,{\sc i}.  They showed 
that the recombination of both H\,{\sc i} and He\,{\sc i} sped up in the 
three-level atom model.  The suggested level of change is large enough 
to bias the determination of the cosmological parameters\,\cite{4lewis06}. 

In this chapter we try to investigate the effect of the extra forbidden 
transitions suggested by Dubrovich \& Grachev~(2005)\,\cite{4dub05}
 in the multi-level atom model without assuming thermal equilibrium 
 among the higher excited states.
The outline of this chatper is as follows.
In Section\,\ref{sec:c4model} 
we will describe details of the rate equations in our numerical model.
In Section\,\ref{sec:c4result} we will present results on the ionization 
fraction $x_{\rm e}$ and the anisotropies $C_\ell$, and 
assess the importance of the addition of the forbidden transitions. Other 
possible improvements of the recombination calculation will 
be discussed in Section\,\ref{sec:c4discuss}.  
And finally in Section\,\ref{sec:c4conclu} we will present our conclusions.

\section{Model}
\label{sec:c4model}
Here we follow the formalism of the multi-level calculation
performed by Seager et al.~(2000)\,\cite{4seager00}.  
We consider 100 levels for H\,{\sc i}, 103 levels for He\,{\sc i}, 
10 levels for He\,{\sc ii}, 1 level for He\,{\sc iii}, 
1 level for the electrons and 1 level for the protons.
For H\,{\sc i}, we only consider discrete $n$ levels and assume that the angular sub-levels
($l$-states) are in statistical equilibrium within a given shell.  
For both He\,{\sc i} and He\,{\sc ii}, we consider all the 
$l$-states separately. The multi-level He\,{\sc i} atom includes all
states with $n \leq 10$ and $l \leq 7$.  Here we give a summary of 
the rate equations for the number density of each energy level $i$, 
and the equation for the change of matter temperature $T_{\rm M}$.  
The rate equation for each state with respect to redshift $z$ is
{\setlength\arraycolsep{2pt}
\begin{equation}
 (1+z) \frac{dn_i}{dz} = -\frac{1}{H(z)} 
   \left[ \left( n_{\rm e} n_{\rm c} R_{{\rm c}i} - n_i R_{i {\rm c}} \right) 
	+ \sum^N_{j=1} \Delta R_{j-i} \right] + 3 n_i , 
\end{equation}} 
\\*
where $n_i$ is the number density of the $i$th excited atomic state, $n_{\rm e}$
is the number density of free electrons, and $n_{\rm c}$ is the number density
of continuum particles such as a proton, He$^+$, or He$^{2+}$. 
Additionally $R_{{\rm c}i}$ is the photo-recombination rate, $R_{i {\rm c}}$ is the 
photo-ionization rate, $\Delta R_{j-i}$ is the net bound-bound rate for 
each line transition, and $H(z)$ is the Hubble parameter.
We do not include the collisional rates, as they have been shown to be
negligible~\cite{4seager00}. 

For He\,{\sc i}, we update the atomic data for the energy 
levels\,\cite{4MWD06}, the oscillator strength for resonant 
transitions\,\cite{4Drake:2007}and the photo-ionization 
cross-section spectrum.   We use the photo-ionization cross-section given by 
Hummer \& Storey (1998)\,\cite{4hs98} for $n$\,$\leq$\,10 and $l$\,$\leq$\,4, 
and adopt the hydrogenic approximation for states with $l$\,$\geq$\,5\,\cite{4sh91}.
It is hard to find published accurate and complete data for the photo-ionization
cross-section of He\,{\sc i} with large $n$ and $l$.  For example, 
a recent paper by Bauman et al.~(2005)\,\cite{4bauman05} claimed that 
they had calculated the photo-ionization cross-section up to 
$n$\,$=$\,27 and $l$\,$=$\,26, although, no numerical values were provided.

For the matter temperture $T_{\rm M}$, we only include the 
adiabatic and Compton cooling terms in the rate
and it is given by Equation~(\ref{eqTM}).
Seager et al.~(2000)\,\cite{4seager00} considered all 
the resonant transitions and only one forbidden transition,
 namely the 2S--1S two-photon transition, 
in the calculation of each atom, (for He\,{\sc i}, 
2S\,$\equiv 2^{1}$S$_{0}$ and 1S\,$\equiv 1^{1}$S$_{0}$).  
The 2S--1S two-photon transition rate is given by 
\begin{equation}
\Delta R_{\rm 2S \rightarrow 1S} = \Lambda_{\rm 2S-1S} \left( 
n_{2\rm S} - n_{1 \rm S} \frac{g_{2 \rm S}}{g_{1 \rm S}}
e^{-h_{\rm P} \nu_{2\rm S-1\rm S}/k_{\rm B} T_{\rm M}}\right), 
\end{equation}
where $\Lambda_{\rm 2S-1S}$ is the spontaneous rate of the corresponding
 two-photon transition, $\nu_{2\rm S-1\rm S}$ is the frequency 
between levels 2S and 1S and $g_i$ is the degeneracy of the energy level
$i$.
Here we include the following extra forbidden transitions, which
were first suggested by Dubrovich \& Grachev~(2005)\,\cite{4dub05}.
The first ones are the two-photon transitions from $n$S and $n$D to 1S for H, plus 
$n^{1}$S$_{0}$ and $n^{1}$D$_{2}$ to $1^{1}$S$_{0}$ for He\,{\sc i}.
For example, for H\,{\sc i}, we can group together the $n$S and $n$D states coming from
 the same level, so that we can write the two-photon transition rate as  
{\setlength\arraycolsep{1pt}
\begin{equation}
\Delta R^{\rm H}_{n {\rm S} + n {\rm D}  \rightarrow 1{\rm S}} =
\Lambda^{\rm H}_{n {\rm S} + n {\rm D}}
\left( n_{n {\rm S} + n {\rm D}} - n_{1 \rm S} 
\frac{g_{n {\rm S} + n {\rm D}}}{g_{1 \rm S}}
e^{-h_{\rm P} \nu_{n1}/k_{\rm B} T_{\rm M}}\right).
\end{equation}}
\\
Here $n$ (without a subscript) is the principle quantum number of the state,
$n_{n {\rm S} + n {\rm D}}$ is the total number density of the excited
atoms in either the $n$S or $n$D states,
and $\Lambda^{\rm H}_{n {\rm S} + n {\rm D}}$ is the effective  
spontaneous rate of the two-photon transition from $n {\rm S} + n {\rm D}$ to 1S, 
which is approximated by the following formula\,\cite{4dub05}:
\begin{equation}
\Lambda^{\rm H}_{n {\rm S} + n {\rm D}} =
\frac{54 \Lambda^{\rm H}_{2 \rm S-1S} }{g_{n{\rm S} + n {\rm D}}}
\left( \frac{n-1}{n+1} \right)^{2n} \frac{11 n^2 - 41}{n} \, ,
\end{equation}
where $ \Lambda^{\rm H}_{\rm 2S-1S}$ is equal to 
8.2290\,s$^{-1}$\,\cite{4goldman89,4santos98}. 
The latest value of $ \Lambda^{\rm H}_{\rm 2S-1S}$ is equal to 
 8.2206\,s$^{-1}$\,\cite{4labzowsky05} and does not bring any 
noticeable change to the result.  Here $g_{n{\rm S} + n {\rm D}}$
is equal to 1 for $n$\,$=$\,2, and 6 for $n$\,$\ge$\,3. 
This spontaneous rate is estimated by considering only the non-resonant
two-photon transitions through one intermediate state $n$P.  
Dubrovich \& Grachev~(2005)\,\cite{4dub05} ignored the 
resonant two-photon transition contributions,
 since the escape probability of these emitted photons is very low.
The above formula for $\Lambda^{\rm H}_{n {\rm S} + n {\rm D}}$ is 
valid up to $n$\,$\simeq$\,$40$, due to the dipole approximation used,  
although it is not trivial to check how good this approximate rate is.  Besides the 
2S--1S two-photon rate, only the non-resonant two-photon rates from 3S to 1S and
3D to 1S are calculated accurately and available in the literature.
Cresser et al.~(1986)\,\cite{4ctsc86} evaluated $\Lambda^{\rm H}_{\rm 3S}$ and 
$\Lambda^{\rm H}_{\rm 3D}$ by including the non-resonant
transitions through the higher-lying intermediate $n$P states ($n$\,$\ge$\,$4$),
 which are equal to 8.2197\,s$^{-1}$ 
and 0.13171\,s$^{-1}$, respectively.  These values were confirmed 
by Florescu~(1988)\,\cite{4fsm88} and agreed to three significant figures.
Using these values, we find that $\Lambda^{\rm H}_{n {\rm S} + n {\rm D}}$ is equal 
to 1.484\,s$^{-1}$, which is an order of magnitude smaller than 
the value from the approximated rate coming from equation~(4.4).  The 
approximation given by Dubrovich \& Grachev~(2005)\,\cite{4dub05} 
therefore seems to be an overestimate.  This leads us instead 
to consider a scaled rate $\tilde{\Lambda}^{\rm H}_{n {\rm S} + n {\rm D}}$,
which is equal to $\Lambda^{\rm H}_{n {\rm S} + n {\rm D}}$ multiplied
by a factor to bring the approximated two-photon rates of H\,{\sc i}~(equation~(4.4)) 
with $n$\,$=$\,3 into agreement with the numerical value given above, i.e.
\begin{equation}
\tilde{\Lambda}^{\rm H}_{n {\rm S} + n {\rm D}} =
0.0664 \ \Lambda^{\rm H}_{n {\rm S} + n {\rm D}}.
\end{equation}

Note that the use of the non-resonant rates is an approximation.
The resonant contributions are suppressed in practice because of
optical depth effects, and in a sense some of these contributions
are already included in our multi-level calculation.  Nevertheless,
the correct way to treat these effects would be in a full 
radiative transfer calculation, which we leave for a future study.
For He\,{\sc i}, we treat $n^{1}$S$_{0}$ and  $n^{1}$D$_{2}$ 
separately and use equation~(4.3) for calculating the transition rates.  
The spontaneous rate $\Lambda^{\rm HeI}_{n {\rm S}/n {\rm D}} $ 
is estimated by Dubrovich \& Grachev~(2005)\,\cite{4dub05}
 by assuming a similar form to that used for 
 $\Lambda^{\rm H}_{n {\rm S} + n {\rm D}}$:
\begin{equation}
\Lambda^{\rm HeI}_{n {\rm S}/n {\rm D}} =
\frac{1045 A^{\rm HeI} }{g_{n{\rm S} + n {\rm D}}}
\left( \frac{n-1}{n+1} \right)^{2n} \frac{11 n^2 - 41}{n} \, ,
\end{equation}
where $A^{\rm HeI}$ is a fitting parameter~(which is still uncertain 
 both theoretically and experimentally). According to 
Dubrovich \& Grachev~(2005)\,\cite{4dub05},
 resonable values of $A$ range from 10 to 12\,s$^{-1}$, 
and we take $A$\,$=$\,11\,s$^{-1}$ here.  In our calculation, we include 
these extra two-photon rates up to $n$\,$=$\,40
for H and up to $n=10$ for He\,{\sc i}.

The other additional channel included is the spin-forbidden 
transition between the triplet $2^3$P$_1$ and singlet 
$1^1$S$_0$ states in He\,{\sc i}. This is an intercombination/
semi-forbidden electric-dipole transition which emits a single 
photon and therefore we can calculate the corresponding 
net rate by using the bound-bound resonant rate expression, i.e. 
{\setlength\arraycolsep{2pt}
\begin{equation}
\Delta R_{2^3{\rm P}_1-1^1{\rm S}_0} = p_{2^3{\rm P}_1, 1^1{\rm S}_0}
\left( n_{2^3{\rm P}_1} R_{2^3{\rm P}_1, 1^1{\rm S}_0}
- n_{2^1{\rm S}_0} R_{1^1{\rm S}_0, 2^3{\rm P}_1}
\right),
\end{equation}}
\\ where 
{\setlength\arraycolsep{2pt}
\begin{eqnarray}
&&  R_{2^3{\rm P}_1,1^1{\rm S}_0} = 
A_{2^3{\rm P}_1-1^1{\rm S}_0} + B_{2^3{\rm P}_1- 1^1{\rm S}_0} \bar{J}, \\
&&  R_{1^1{\rm S}_0,2^3{\rm P}_1} = B_{1^1{\rm S}_0-2^3{\rm P}_1} \bar{J}, \\
&& p_{2^3{\rm P}_1, 1^1{\rm S}_0} = \frac{1-e^{-\tau_{\rm s}}}{\tau_{\rm s}}, 
\quad \rm{with} \\
&& \tau_{\rm s} = \frac{A_{2^3{\rm P}_1- 1^1{\rm S}_0}\lambda^3_{2^3{\rm P}_1, 
1^1{\rm S}_0}
}{8 \pi H(z)} \left[ \frac{g_{2^3{\rm P}_1} }
{g_{ 1^1{\rm S}_0}} n_{1^1{\rm S}_0} - n_{2^3{\rm P}_1}  \right] .
\end{eqnarray}} 
\\
Here $A_{2^3{\rm P}_1-1^1{\rm S}_0}$, $B_{2^3{\rm P}_1- 1^1{\rm S}_0}$ 
and $B_{1^1{\rm S}_0-2^3{\rm P}_1}$ are the
Einstein coefficients, $p_{2^3{\rm P}_1, 1^1{\rm S}_0}$ is the
Sobolev escape probability, $\tau_{\rm s}$ is the Sobolev optical 
depth~(see \cite{4seager00} and references therein), 
$\lambda_{2^3{\rm P}_1, 1^1{\rm S}_0}$ is the wavelength of the energy 
difference between states $2^3{\rm P}_1$ and $1^1{\rm S}_0$,
and $\bar{J}$ is the blackbody intensity with temperature $T_{\rm R}$. 

This $2^3$P$_1$--$1^1$S$_0$ transition is not the lowest transition 
between the singlet and the triplet states.  
The lowest one is the magnetic-dipole transition 
between $2^3$S$_1$ and $1^1$S$_0$, 
with Einstein coefficient \mbox{$A_{2^3{\rm S}_1-1^1{\rm S}_0}$ =  
$1.73 \times 10^{-4}$\,s$^{-1}$}\,\cite{4lin77}.  However, 
this is much smaller than \mbox{$A_{2^3{\rm P}_1-1^1{\rm S}_0} 
= 177.58$\,s$^{-1}$}\,\cite{4Drake:2007,4lach01},
so this transition can be neglected.  Note that 
Dubrovich \& Grachev~(2005)\,\cite{4dub05} used an older value of 
\mbox{$A_{2^3{\rm P}_1-1^1{\rm S}_0} = 233$\,s$^{-1}$} \cite{4lin77} 
in their calculation. 

We use the Bader-Deuflhard semi-implicit numerical integration 
scheme (see Section 16.6 in \cite{4nr}) to solve the above rate equations.
All the numerical results are carried out using the $\Lambda$CDM model with
cosmological parameters: $\Omega_{\rm b}$\,$=$\,0.04; $\Omega_{\rm C}$\,$=$\,0.2;
$\Omega_{\Lambda}$\,$=$\,0.76; $\Omega_{\rm K}$\,$=$\,0; 
$Y_{\rm p}$\,$=$\,0.24; $T_0$\,$=$\,2.725\,K
and $h$\,=\,0.73 (consistent with those in \cite{4spergel06}). 

\begin{figure}
\begin{center}
\includegraphics[width=0.8\textwidth]{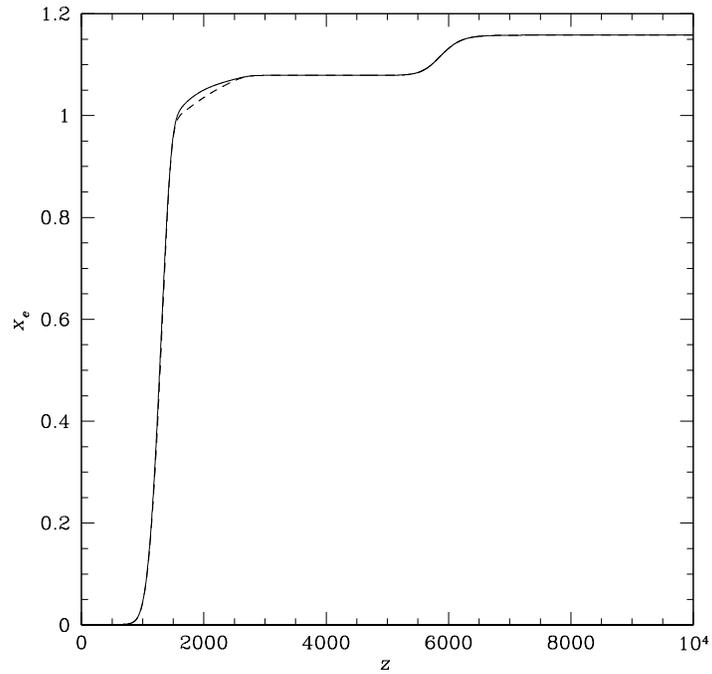}
\caption[The ionization fraction $x_{\rm e}$ as a function of redshift $z$
with extra forbidden transitions.]
{The ionization fraction $x_{\rm e}$ as a function of redshift $z$.
The solid line is calculated using the original multi-level code of
Seager et al.~(2000)\,\cite{4seager00}, while the dashed line 
includes all the extra forbidden transitions discussed here.}
\label{4graphxe}
\end{center}
\end{figure}

\begin{figure} 
\begin{center}
\includegraphics[width=0.8\textwidth]{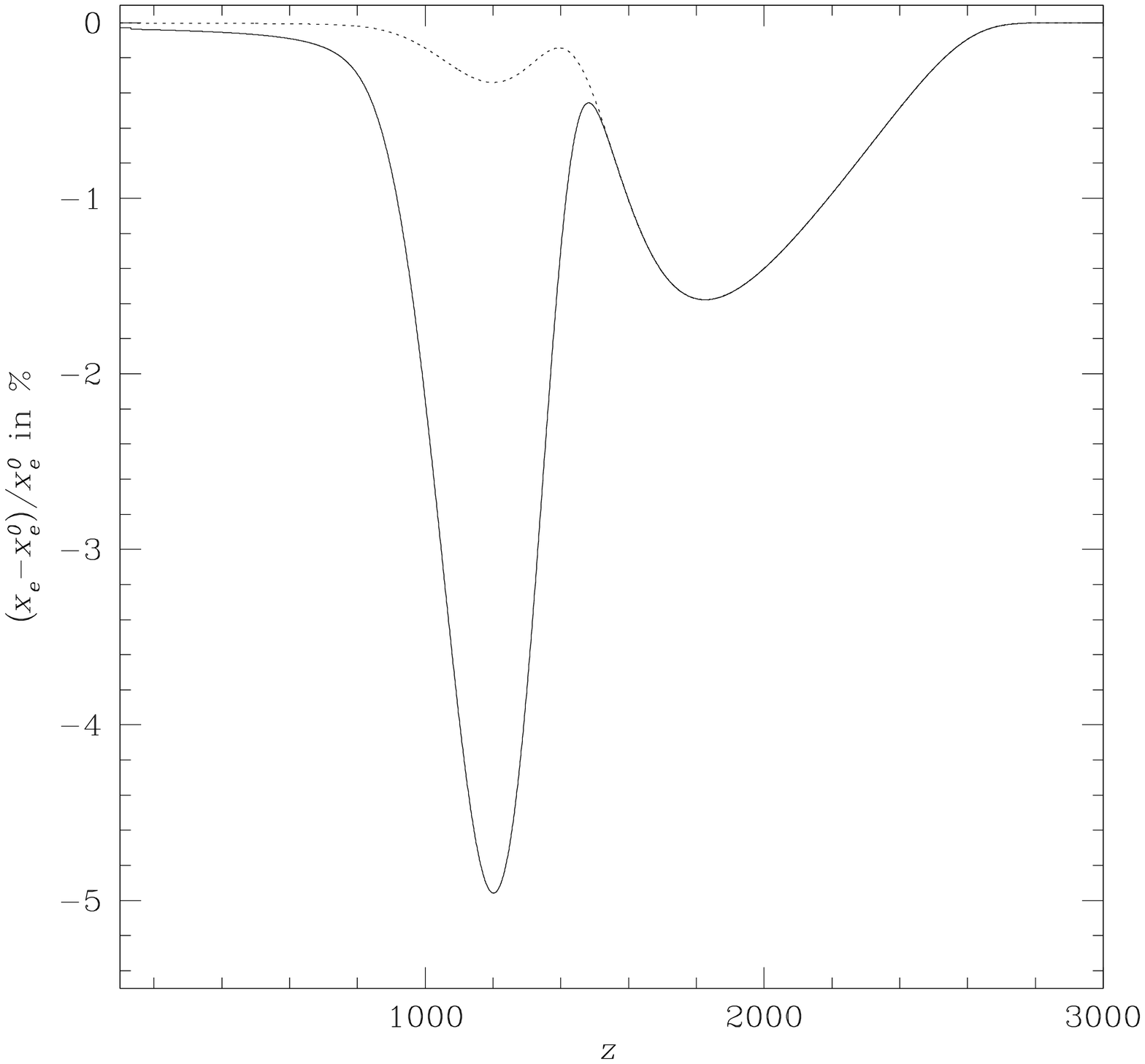}
\caption[The fractional difference (`new' minus `old') in $x_{\rm e}$ 
between the two models plotted in Fig.~\ref{4graphxe}
as a function of redshift $z$.]
{The fractional difference (`new' minus `old') in $x_{\rm e}$ 
between the two models plotted in Fig.~\ref{4graphxe}
as a function of redshift $z$.  The solid and dotted lines are the models
with the two-photon rates for H\,{\sc i} given by 
Dubrovich \& Grachev~(2005)\,\cite{4dub05} and the scaled 
one given by equation~(4.6), respectively.  
Both curves are calculated using all the He\,{\sc i} forbidden 
transitions as discussed in the text.}
\label{4gdxe}
\end{center}
\end{figure}

\section{Result}
\label{sec:c4result}
The recombination histories calculated using the previous multi-level
code\,\cite{4seager00} and the code in this paper are shown
in Fig.~\ref{4graphxe}, where $x_{\rm e} \equiv n_{\rm e}/ n_{\rm H}$
is the ionization fraction relative to hydrogen.  As we have included 
more transitions in our model, and these give electrons more 
channels to cascade down to the ground state, we expect the 
overall recombination rate to speed up, and that this will be noticeable 
if the rates of the extra forbidden transitions are significant.  
From Fig.~\ref{4graphxe}, we can see that the recombination to 
He\,{\sc i} is discernibly faster in the new calculation.  
Fig.~\ref{4gdxe} shows the difference in $x_{\rm e}$ with 
and without the extra forbidden transitions.
The dip at around $z=1800$ corresponds to the recombination
of He\,{\sc i} and the one around $z=1200$ is for H\,{\sc i}.  
Overall, the addition of the forbidden transitions claimed by
Dubrovich \& Grachev~(2005)\,\cite{4dub05} leads to greater 
than 1\% change in $x_{\rm e}$ over the redshift range where
the CMB photons are last scattering. 

\begin{figure} 
\begin{center}
\includegraphics[width=0.8\textwidth]{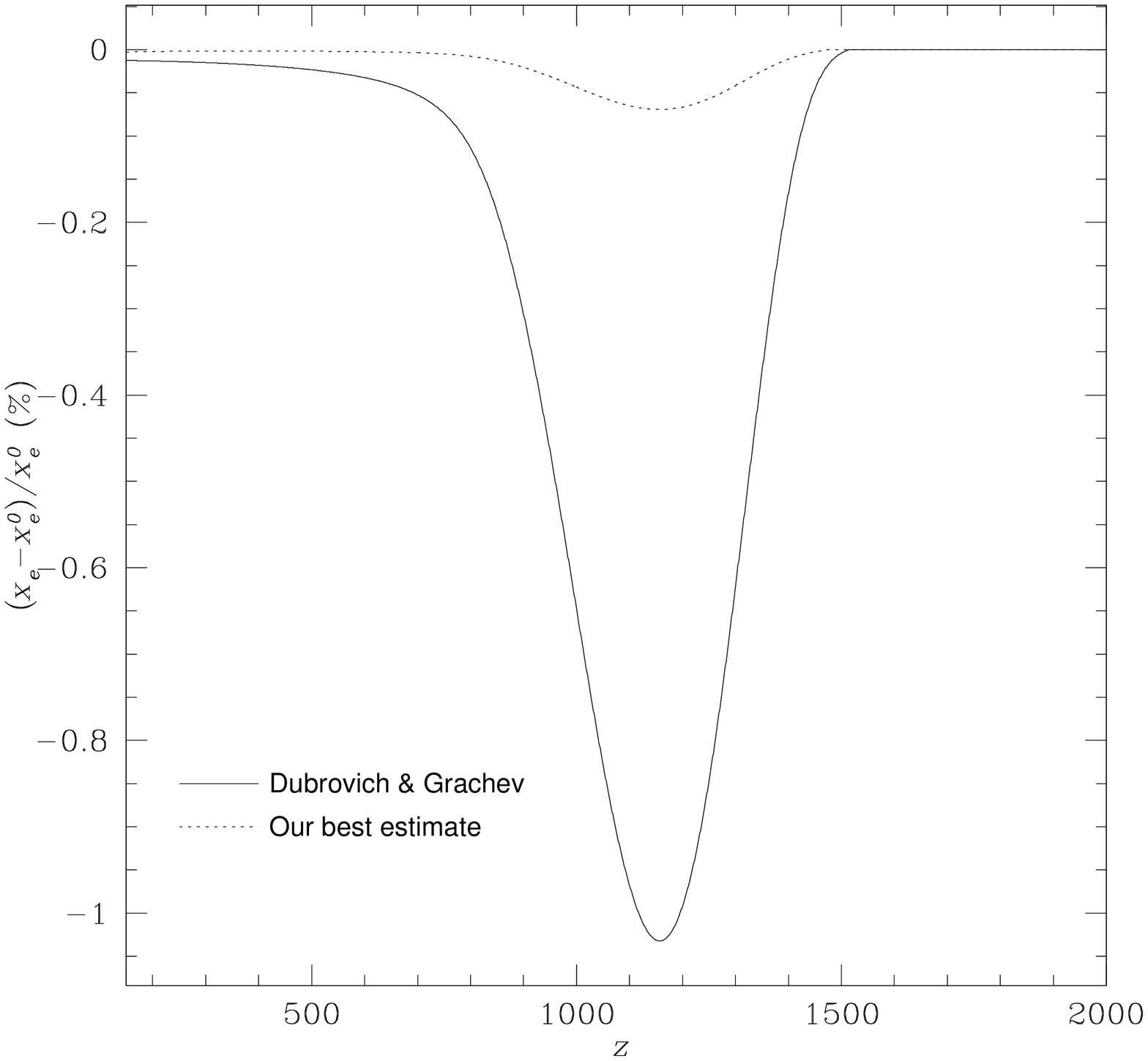}
\caption[Fractional change in $x_{\rm e}$ with the addition of the two-photon
transition from 3S and 3D to 1S for H\,{\sc i}.]
{Fractional change in $x_{\rm e}$ with the addition of the two-photon
transition from 3S and 3D to 1S for H\,{\sc i}.  The solid line is calculated with
the approximate rate given by Dubrovich \& Grachev~(2005)\,\cite{4dub05}
while the dashed line is calculated with the numerical rates given
by Cresser et al.~(1986)\,\cite{4ctsc86}.}
\label{4Hn3}
\end{center}
\end{figure}

\begin{figure} 
\begin{center}
\includegraphics[width=0.8\textwidth]{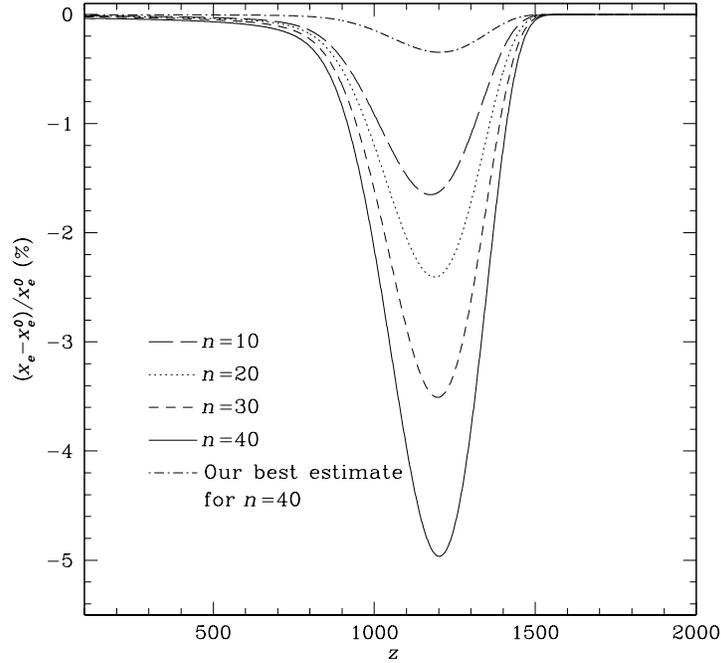}
\caption[Fractional change in $x_{\rm e}$ with the 
addition of different forbidden transitions for H\,{\sc i}.]
{Fractional change in $x_{\rm e}$ with the 
addition of different forbidden transitions for H\,{\sc i}.  The long-dashed, dotted,
dashed and solid lines include the two-photon transitions up to $n=10$, 20,
30 and 40, respectively, using the approximation for the rates given by equation~(4.5).
The dot-dashed line is calculated with the scaled rate from equation~(4.6).}
\label{4gdxeH}
\end{center}
\end{figure}

In the last Section, we found that the approximated two-photon rate 
given by Dubrovich \& Grachev~(2005)\,\cite{4dub05} for H\,{\sc i} 
with $n$\,$=$\,3 was overestimated by more 
than a factor of 10.  By considering only this extra two-photon transition,
 the approximate rate gives more than a per cent difference in $x_{\rm e}$, 
while with the more accurate numerical rates, the
change in $x_{\rm e}$ is less than 0.1 per cent~(as shown in Fig.~\ref{4Hn3}). 
 Based on this result, we do not need to include this two-photon transition, 
 as it brings much less than a per cent effect on $x_{\rm e}$.
For estimating the effect of the extra two-photon transitions for 
higher $n$, we use the scaled two-photon rate given by equation~(4.6).
The result is plotted in Fig.~\ref{4gdxeH}.  The change in $x_{\rm e}$ with
the scaled two-photon rates is no more than 0.4 per cent, while the one
with the Dubrovich \& Grachev~(2005)\,\cite{4dub05} approximated rates 
brings about a 5 per cent change.
%
%
For He\,{\sc i}, Dubrovich \& Grachev~(2005)\,\cite{4dub05}
 included the two-photon transitions from $n$\,$=$\,6 to 40,
  since they claimed that the approximate 
formula (equation~4.7) is good for $n$\,$>$\,6.  In our calculation, we use  
$\Lambda^{\rm HeI}_{n {\rm S}/n {\rm D}}$ from the approximate formula
 for the two-photon transitions of $n$\,$=$\,3 to 10, since this is the best one 
can do for now (and the formula at least gives the right order of magnitude).
The addition of the singlet-triplet $2^3{\rm P}_1$--$1^1{\rm S}_0$
transition and the $n^1{\rm S}_0$--$1^1{\rm S}_0$ and $n^1{\rm D}_2$--$1^1{\rm S}_0$ 
two-photon transitions with \mbox{$n$\,$=$\,3$-$10} cause more than
1 per cent changes in $x_{\rm e}$ (as shown in Fig.~2).  The 
$2^3{\rm P}_1$--$1^1{\rm S}_0$ transition has the biggest 
effect on $x_{\rm e}$.  

\begin{figure} 
\begin{center}
\includegraphics[width=0.8\textwidth]{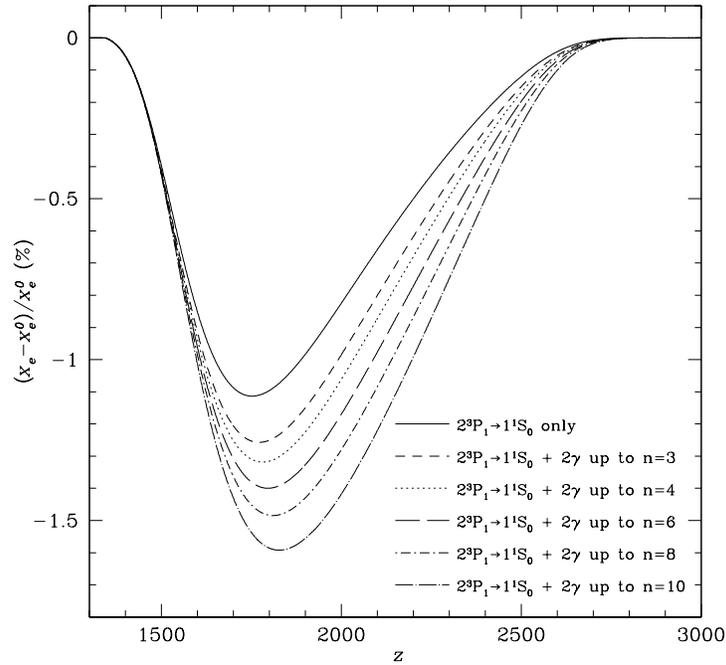}
\caption[Fractional change in $x_{\rm e}$ with the 
addition of different forbidden transitions for He\,{\sc i} as a function 
of redshift.]
{Fractional change in $x_{\rm e}$ with the 
addition of different forbidden transitions for He\,{\sc i} as a function 
of redshift.  The solid line corresponds to the calculation with only the
$2^3$P$_1$--$1^1$S$_0$ spin-forbidden transition.
The short-dashed, dotted, long-dashed, dot-dashed and long dot-dashed lines 
include both the spin-forbidden transition and the two-photon ($2 \gamma$) 
transition(s) up to $n=3, 4$, 6, 8 and 10, respectively.}
\label{4gdxeHeI}
\end{center}
\end{figure}

\begin{figure}
\begin{center}
\includegraphics[width=0.8\textwidth]{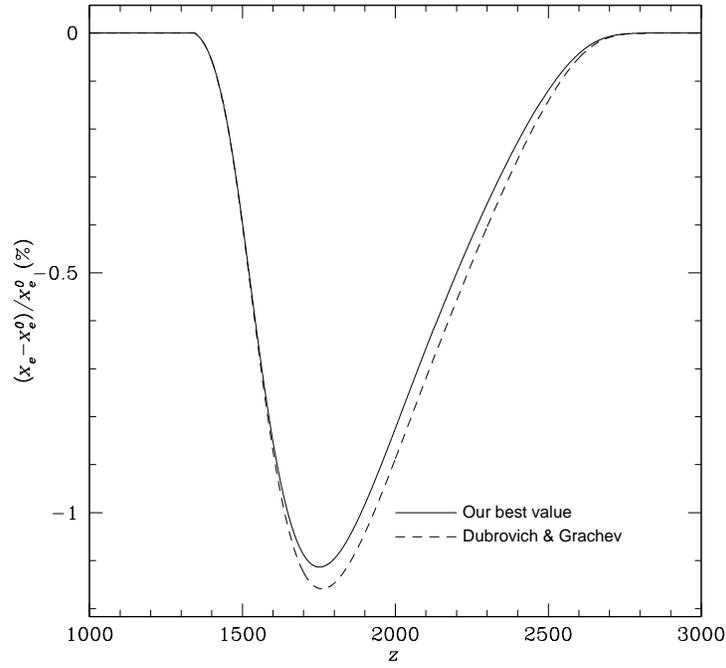}
\caption[Fractional change in $x_{\rm e}$ with only the He\,{\sc i}
$2^3$P$_1$--$1^1$S$_0$ forbidden transition.]
{Fractional change in $x_{\rm e}$ with only the He\,{\sc i}
$2^3$P$_1$--$1^1$S$_0$ forbidden transition.  The solid 
line is 
computed with our best value $A_{2^3{\rm P}_1-1^1{\rm S}_0}$\,=\,177.58\,s$^{-1}$
 from Lach \& Panchucki~(2001)\,\cite{4lach01} 
and the dashed line is 
calculated with the rate $A_{2^3{\rm P}_1-1^1{\rm S}_0}$\,=\,233\,s$^{-1}$ from
Dubrovich \& Grachev~(2005)\,\cite{4dub05}.
}
\label{4gdxe23P}
\end{center}
\end{figure}

Fig.~\ref{4gdxeHeI}  shows the fractional difference
in $x_{\rm e}$ using different combinations of additional forbidden transitions.
We can see that the $2^3{\rm P}_1$--$1^1{\rm S}_0$ transition
alone causes more than a 1 per cent change in $x_{\rm e}$, and the addition 
of each two-photon transition only gives about another 0.1 per cent change.  The extra
two-photon transitions from higher excited states (larger $n$) have a lower
effect on $x_{\rm e}$ compared with that from small $n$, and we 
checked that this trend continues to higher $n$.  However, 
the convergence is slow with increasing $n$.  Therefore,
one should also consider these two-photon transitions with 
\mbox{$n$\,$>$\,10} for He\,{\sc i}, and the precise result will 
require the use of accurate rates, rather than an approximate 
formula such as equation~(4.7).  For the $2^3{\rm P}_1$--$1^1{\rm S}_0$
transition, Dubrovich \& Grachev~(2005)\,\cite{4dub05} adopted 
an older and slightly larger rate, 
and this causes a larger change of the ionization fraction 
(about 0.5 per cent more compared with that calculated with our 
best rate), as shown in Fig.~\ref{4gdxe23P}.

\subsection{The importance of the forbidden transitions}
\begin{figure}
\begin{center}
\includegraphics[width=0.8\textwidth]{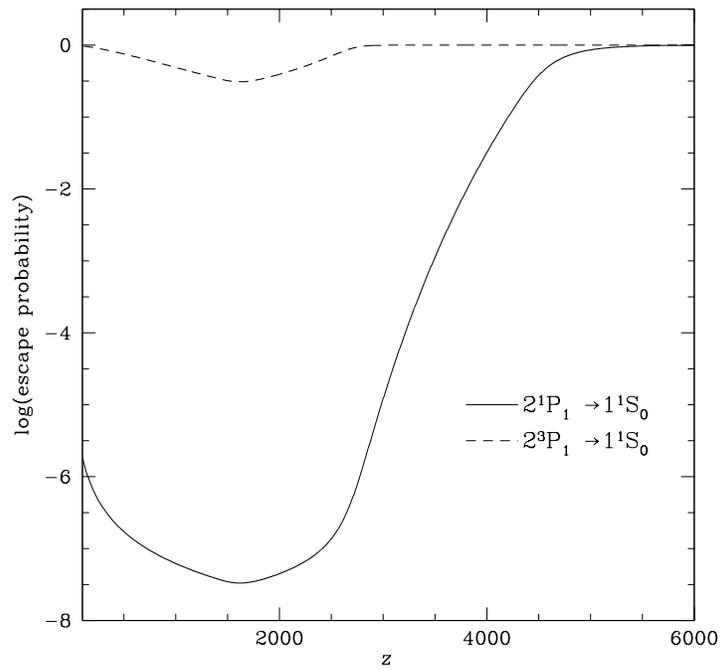}
\caption[Escape probability $p_{ij}$
of the resonant transition between 
He\,{\sc i} $2^1$P$_1$ and $1^1$S$_0$ and  the spin-forbidden transition between
He\,{\sc i} $2^3$P$_1$ and $1^1$S$_0$  as a function of redshift.]
{Escape probability $p_{ij}$ as a function of redshift.
The solid line corresponds to the resonant transition between 
He\,{\sc i} $2^1$P$_1$ and $1^1$S$_0$,
while the dashed line refers to the spin-forbidden transition between
He\,{\sc i} $2^3$P$_1$ and $1^1$S$_0$.}
\label{4gescp}
\end{center}
\end{figure}

\begin{figure} 
\begin{center}
\includegraphics[width=0.8\textwidth]{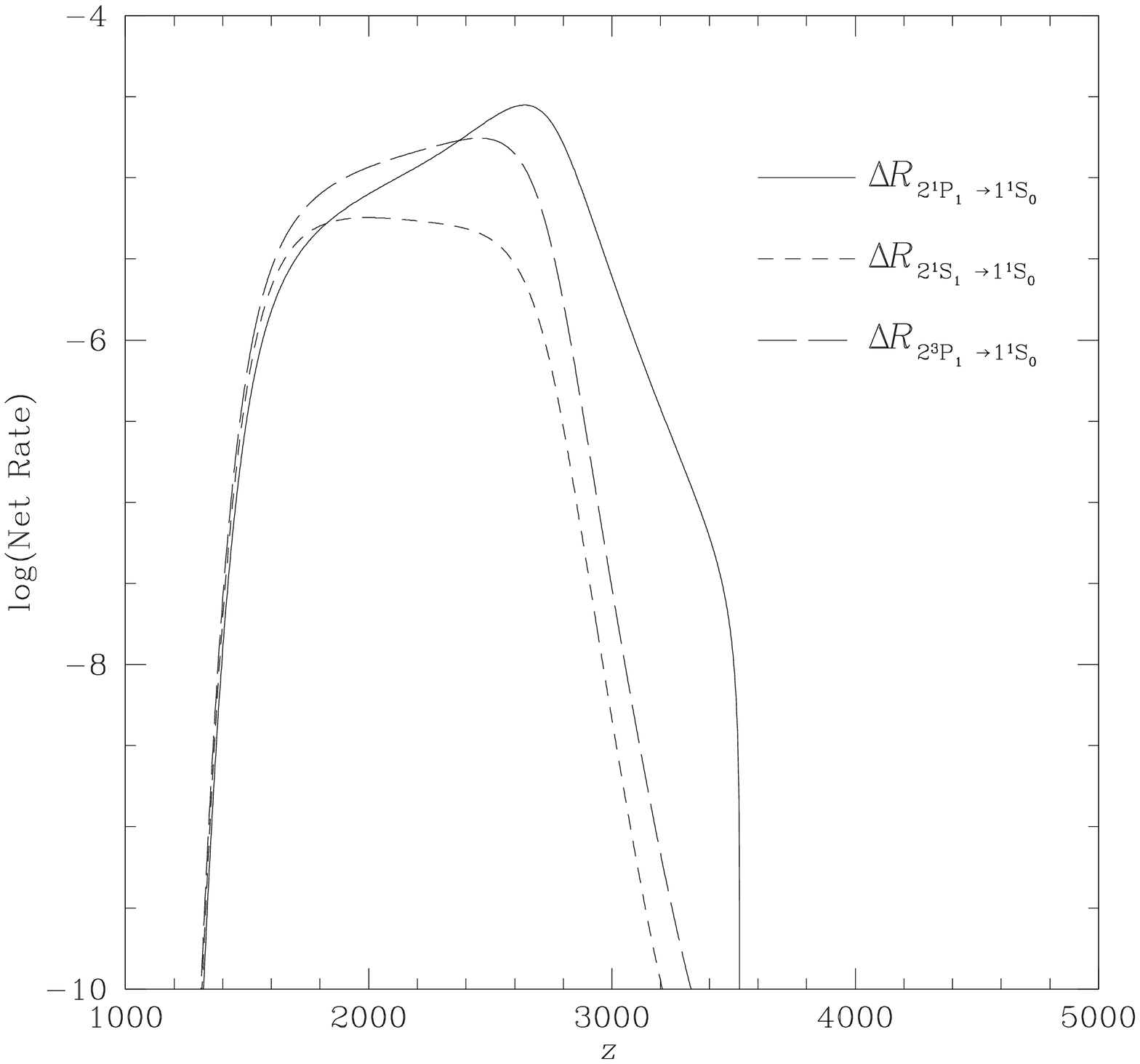}
\caption[Net bound-bound rates 
of the resonant transition between 
He\,{\sc i} $2^1$P$_1$ and $1^1$S$_0$ and  the spin-forbidden transition between
He\,{\sc i} $2^3$P$_1$ and $1^1$S$_0$  as a function of redshift.]
{Net bound-bound rates for He\,{\sc i} as a function of redshift.
The solid line is the resonant transition between $2^1$P$_1$ and $1^1$S$_0$,
the short-dashed line is the two-photon transition between $2^1$S$_1$ and $1^1$S$_0$.
And the long-dashed line is the spin-forbidden transition between
$2^3$P$_1$ and $1^1$S$_0$.}
\label{4grateHeI}
\end{center}
\end{figure}

One might wonder why the semi-forbidden transitions are significant
in recombination {\it at all}, since the spontaneous rate (or the Einstein $A$
coefficient) of the semi-forbidden transitions are about 6 orders of 
magnitude (a factor of $\alpha^2$, where $\alpha$ is the fine-structure 
constant) smaller than those of the 
resonant transitions. Let us take He\,{\sc i} as an example for explaining 
the importance of the spin-forbidden $2^3{\rm P}_1$--$1^1{\rm S}_0$
transition in recombination.  The spontaneous rate 
is equal to 177.58\,s$^{-1}$ for this semi-forbidden transition, which is much
smaller than $1.7989 \times 10^9$\,s$^{-1}$ for the 
$2^1{\rm P}_1$--$1^1{\rm S}_0$ resonant transition.  But when we 
calculate the net rate [see equation~(4.7)], we also need to include 
the effect of absorption of the emitted photons by the surrounding 
neutral atoms, and we take this into account by multiplying the net 
bound-bound rate by the Sobolev escape probability $p_{ij}$\,\cite{4seager00}.  
If $p_{ij}$\,=\,1, the emitted line photons can 
escape to infinity, while if $p_{ij}$\,=\,0 the photons
will all be reabsorbed and the line is optically thick.  
Fig.~\ref{4gescp} shows that the escape probability of the 
$2^1{\rm P}_1$--$1^1{\rm S}_0$ resonant transition is about 7 orders
of magnitude smaller than the spin-forbidden transition.  This makes the 
two net rates roughly comparable, as shown in Fig.~\ref{4grateHeI}. 
From equation~(4.11), we can see that the easier it is to emit a photon, the 
easier that photon can be re-absorbed, because the optical depth $\tau_{\rm s}$ 
is directly proportional to the Einstein $A$ coefficient.  So when radiative 
effects dominate, it is actually natural to expect that some forbidden 
transitions might be important (although this is not true in a regime where 
collisonal rates dominate which is often the case in astrophysics).  
In fact for today's standard cosmological
model, slightly more than half of all the hydrogen atoms in the Universe
recombined via a forbidden transition\,\cite{4wong06}.  Table~4.1 shows 
that this is also true for helium.
\begin{table*} 
\centering
\begin{minipage}{105mm}
\caption{The percentage of electrons cascading down in each channel
  from $n=2$ states to the $1^1$S$_0$ ground state for He\,{\small I}.}
\begin{tabular}{@{}cccc@{}}
\hline
& $2^1$S$_0 \rightarrow 1^1$S$_0 $ & $2^1$P$_1 \rightarrow 1^1$S$_0 $ &
$2^3$P$_1 \rightarrow1^1$S$_0$ \\
& (two-photon) & (resonant) & (spin-forbidden) \\
Seager et al.~(2000) & 30.9\% & 69.1\% & -- \\
this work & 17.3\% & 39.9\% & 42.8\% \\
\hline
\label{tab:HeI}
\end{tabular}
\end{minipage}
\end{table*}

In the previous multi-level calculation\,\cite{4seager00}, there was no 
direct transition between the singlet and triplet states.  The only 
communication between them was via the continuum, through the 
photo-ionization and photo-recombination transitions.  
Table~4.1 shows how many electrons cascade down through each channel from $n=2$ 
states to the ground state.  In the previous calculation, about 70\% of the 
electrons went down through the $2^1{\rm P}_1$--$1^1{\rm S}_0$ resonant transition.
In the new calculation, including the spin-forbidden transition between the 
triplets and singlets, there are approximately the same fraction 
of electrons going from the $2^1{\rm P}_1$ and $2^3{\rm P}_1$ states
to the ground state~(actually slightly more going from $2^3$P$_1$ 
in the current cosmological model).  This shows that we should certainly 
include this forbidden transition in future calculations. Our estimate
is that only about 40\% of helium atoms reach the ground state without
going through a forbidden transition. 

How about the effect of other forbidden transitions in He\,{\sc i}
recombination?  We have included all the semi-forbidden electric-dipole
transitions with $n$\,$\leq$\,$10$ and $l$\,$\leq$\,$7$, and with
 oscillator strengths larger than $10^{-6}$ given by 
Drake \& Morton~(2007)\cite{4Drake:2007}.  
There is no significant change found in the ionization fraction. Besides 
the $2^3$P$_1$--$1^1{\rm S}_0$ transition, all the other extra 
semi-forbidden transitions are among the higher excited 
states where the resonant transitions dominate.  This is because these 
transition lines are optically thin and the escape probabilities are close 
to 1. 

\subsection{Effects on the anisotropy power spectrum}
The CMB anisotropy power spectrum $C_{\ell}$ depends on the 
detailed profile of the evolution of the ionization fraction 
$x_{\rm e}$. This determines the thickness of the photon
last scattering surface, through the visibility function 
\mbox{$g(z) \equiv e^{- \tau} d \tau/dz$}, where $\tau$ is the 
Thomson scattering optical depth 
\mbox{($\tau = c\, \sigma_{\rm T} \int n_{\rm e} (dt/dz)\,dz$)}. 
The function $x_{\rm e}(z)$ sets the epoch when the tight coupling 
between baryons and photons breaks down, i.e. when the photon 
diffusion length becomes long, and the visibility 
function fixes the time when the fluctuations are effectively frozen 
in (see \cite{4hu95,4seager00} and references therein).
The addition of the extra forbidden transitions
speeds up both the recombination of H\,{\sc i} and He\,{\sc i}, and hence
we expect that there will be changes in $C_\ell$.

\begin{figure}
\begin{center}
\includegraphics[width=0.8\textwidth]{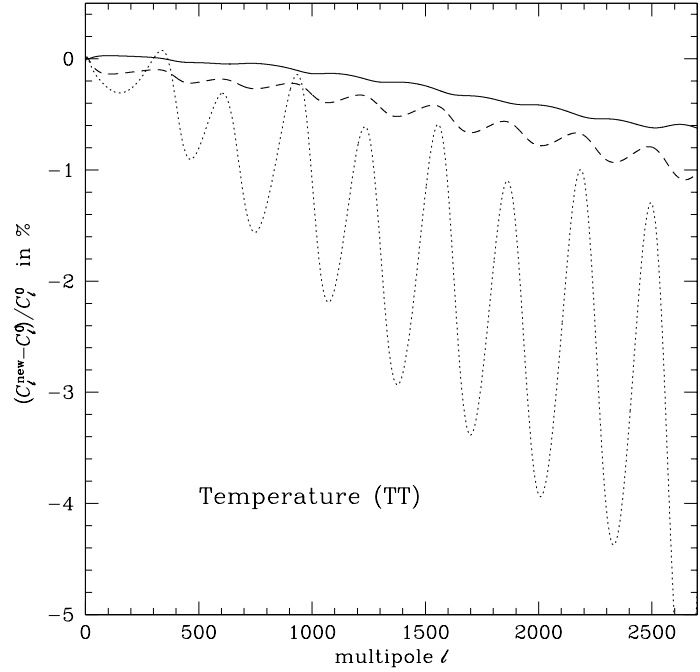}
\caption[Relative change in the temperature ({\sl TT}\/) angular power spectrum
due to the addition of the forbidden transitions.]
{Relative change in the temperature ({\sl TT}\/) angular power spectrum
due to the addition of the forbidden transitions.  The solid line 
includes the spin-forbidden transition and also the two-photons transitions
up to $n=10$ for He\,{\sc i}, the dotted line includes all the 
above transitions and also the two-photon transitions up to $n=40$ for H\,{\sc i}
calculated with the approximate rates given by Dubrovich \& Grachev~(2005)\,\cite{4dub05}. 
The dashed line is computed with the same forbidden transitions as the dotted 
line, but with our scaled rates~(and represents our best current estimate).}
\label{4gclTT}
\end{center}
\end{figure}

\begin{figure}
\begin{center}
\includegraphics[width=0.8\textwidth]{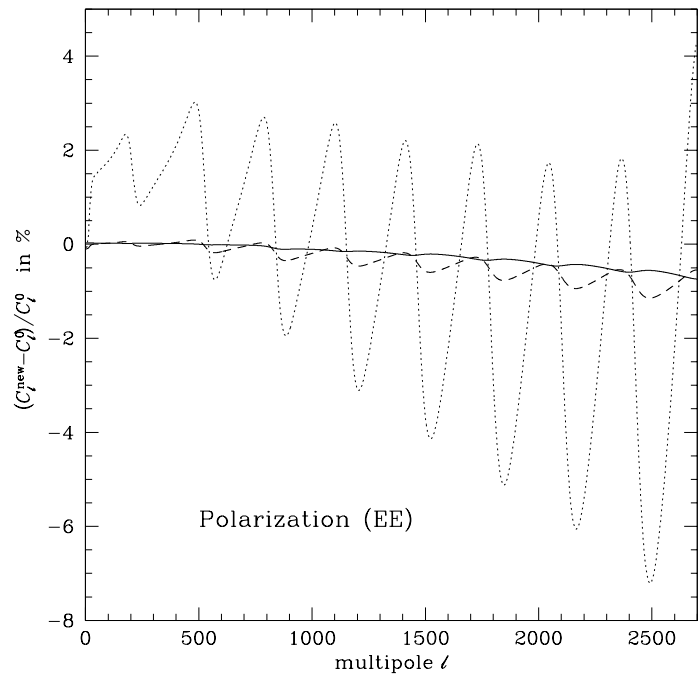}
\caption[Relative change in the polarization ({\sl EE}\/) angular power spectrum
due to the addition of the forbidden transitions, with the curves
the same as in Fig.~\ref{4gclTT}.]
{Relative change in the polarization ({\sl EE}\/) angular power spectrum
due to the addition of the forbidden transitions, with the curves
the same as in Fig.~\ref{4gclTT}.}
\label{4gclEE}
\end{center}
\end{figure}

In order to perform the required calculation,
we have used the code {\sc cmbfast}\,\cite{4cmbfast} and modified it  
to allow the input of an arbitrary recombination history. 
Figs.~\ref{4gclTT} and \ref{4gclEE} show the relative changes in 
the CMB temperature ({\sl TT}\/) and polarization ({\sl EE}\/) anisotropy 
spectra, respectively, with different combinations of extra 
forbidden transitions.  The overall decrease of free electrons 
brings a suppression of $C_\ell$ over a wide range of $\ell$. 

For He\,{\sc i}, there is 
less $x_{\rm e}$ at $z$\,$\simeq$\,$1400-2500$, which leads to an 
earlier relaxation of tight coupling.  Therefore, both the
photon mean free path and the diffusion length are longer.  Moreover,
the decrease of $x_{\rm e}$ in the high-$z$ tail results in increased
damping, since the effective damping scale is an average over the 
visibility function.  This larger damping scale leads to  
suppression of the high-$\ell$ part of the power spectrum.  From 
Figs.~\ref{4gclTT} and \ref{4gclEE}, we can see a decrease of 
$C_\ell$ (for both {\sl TT} and {\sl EE}) toward high $\ell$ 
for He\,{\sc i}, with the maximum change being about 0.6 percent.

For H\,{\sc i}, the change of $C_\ell$ is due to the decrease in $x_{\rm e}$
at $z$\,$\simeq$\,$600-1400$ (see Fig.~\ref{4gdxe}).  
There are two basic features in the curve of change in $C_{\ell}$ 
(the dotted and dashed lines in Fig.~\ref{4gclTT}). Firstly, the power
spectrum is suppressed with increasing $\ell$, due to the lower 
$x_{\rm e}$ in the high-$z$ tail ($z$\,$>$\,1000). 
Secondly, there are a series of wiggles, showing that the 
locations of the acoustic peaks are slightly shifted.  This is due to 
the change in the time of generation of the $C_\ell$s in 
the low-$z$ tail.  $C^{EE}_\ell$ actually shows an increase 
for $\ell \leq 1000$ (see Fig.~\ref{4gclEE});
this is caused by the shift of the center of the visibility 
function to higher $z$, leading to a longer diffusion length.
Polarization occurs when the anisotropic hot and cold photons are 
scattered by the electrons.  The hot and cold photons can 
interact with each other through multiple scatterings within
 the diffusion length, and therefore,
a longer diffusion length allows more scatterings and leads to a 
higher intensity of polarization at large scales.  

With the approximate rates used by Dubrovich \& 
Grachev~(2005)\,\cite{4dub05}, the maximum relative 
change of $C^{TT}_\ell$ is about 4 percent and for 
$C^{EE}_\ell$ it is about 6\%.  
The overall change is thus more than 1\% over a wide 
range of $\ell$.  However, if we adopt the scaled two-photon rate given
by equation~(4.6), the relative changes of $C^{TT}_\ell$ and 
$C^{EE}_\ell$ are no more than 1 per cent.  Note that
we do not plot the temperature-polarization 
correlation power spectrum here, since there is no dramatically
different change found (and relative differences are less meaningful 
since $C^{TE}_\ell$ oscillates around zero).

\section{Discussion}
\label{sec:c4discuss}
In our model we only consider the semi-forbidden transitions with 
$n$\,$\leq$\,10 and $l$\,$\leq$\,7 for He\,{\sc i} and 
the two-photon transitions from the higher S and D states to the ground
state for H and He\,{\sc i}.  It would be desirable to perform 
a more detailed investigation of all the other forbidden transitions, which may 
provide more paths for the electrons to cascade down to the ground state
and speed up the recombination process.  
In this paper we have tried to focus on the forbidden transitions 
which are likely to be the most significant.  However we caution that, if the 
approximations used are inadequate, or other transitions prove to be 
important, then our results will not be accurate.

There are several other approximations that we have adopted in order
to perform our calculations.  For example,
we consider the {\it non-resonant} two-photon rates for
higher excited rates.  The two-photon transitions from higher excited 
states ($n$\,$\ge$\,3) to the ground state are more complicated than 
the 2S--1S transition, because of the resonant intermediate states.
For example, for the 3S--1S two-photon transition, the spectral 
distribution of the emitted photons shows infinities (resonance
peaks) at the frequencies corresponding to the 3S--2P and 2P--1S
transitions\,\cite{4tung84}.  Here, we use only the non-resonant
rates, by assuming a smooth spectral distribution of the 
emitted photons; this probably gives a lower limit on the change of
$x_{\rm e}$ and $C_\ell$ coming from these extra forbidden transitions.
The correct way to treat this would be to consider the rates and 
feedback from medium using the full spectral distribution of the 
photons and radiative transfer;  this will have to wait for a
future study.

Besides the consideration of more forbidden transitions, there are 
many other improvements that could be made to the recombination 
calculation by the time when this work was published.
In particular, Rubi{\~n}o-Mart{\'{\i}}n et al.~(2006)\,\cite{4rub06}
showed that a multi-level calculation of
the recombination of H\,{\sc i} with the inclusion of separate $l$-states
can give more than 20 per cent difference in the population 
of some levels compared with the thermal equilibrium assumption
for each $n$-shell.  The latest calculation, considering up to 100 
shells, is presented by Chluba et al.~(2006)\,\cite{4CRMS06}, 
but does not include all the forbidden transitions studied here. 
A more complete calculation should be done by combining the 
forbidden transitions in a code with full
angular momentum states, and we leave this to a future study.  
There are also other elaborations which could be included in future
calculations, which we now describe briefly.

The rate equation we use for all the two-photon transitions only
includes the spontaneous term, assuming there is no interaction
with the radiation background (see equation~(4.3)).  
Chluba \& Sunyaev~(2005)\,\cite{4chluba05}
suggested that one should also consider the stimulated effect
of the 2S--1S two-photon transition for H, due to photons 
in the low frequency tail of the CMB blackbody spectrum.  
Leung et al.~(2004)\,\cite{4leung04} additionally argued that the change of the adiabatic 
index of the matter should also be included, arising due to the neutralization
of the ionized gas.  These two modifications have been studied only 
in an effective three-level atom model, and more than a percent change
in $x_{\rm e}$ was claimed in each case~(but see Chapter~5\,\cite{4wong06b} 
for arguments against the effect claimed by Leung et al.~2004\,\cite{4leung04} ).

For the background radiation field $\bar{J}$, we approximated it
with a perfect blackbody Planck spectrum.  This approximation
is not completely correct for the recombination of H\,{\sc i}, since the He line 
distortion photons redshift into a frequency range that can in principle
photo-ionize the neutral H~\cite{4dell93,4seager00, 4wong06}.  
Althought we expect this secondary distortion effect to bring the 
smallest change on $x_{\rm e}$ among all the modifications suggested here, 
it is nevertheless important to carry out the calculation 
self-consistently, particularly for the spectral line distortions.
In order to obtain an accurate recombination history, 
we therefore need to perform a full multi-level calculation with 
seperate $l$-states and include at least all the improvements 
suggested above, which we plan to do in a future study.

For completeness we also point out that the accuracy of the physical 
constants is important for recombination
as well.  The most uncertain physical quantity in the recombination
calculation is the gravitational constant $G$.  The value of $G$ used
previously in the
{\sc recfast} code is $6.67259$\,$\times$\,$10^{-11}$m$^{3}$kg$^{-1}$s$^{-2}$ and 
the latest value~(e.g. from the Pariticle Data Group\,\cite{4pdg06})
is $6.6742$\,$\times$\,$10^{-11}$m$^{3}$kg$^{-1}$s$^{-2}$.  Another quantity 
we need to modify is the atomic mass ratio of $^4$He and $^1$H, 
$m_{\rm ^4He}/m_{\rm ^1H}$, which was previously taken to be 
equal to 4~(as pointed out by Steigman 2006\,\cite{4steigman06}).
By using the atomic masses given by Yao et al.~(2006)\,\cite{4pdg06}, 
the mass ratio is equal to 3.9715.  The overall change in $x_{\rm e}$
is no more than 0.1 per cent after updating these two constants in both
{\sc recfast} and multi-level code.  

\section{Conclusion}
\label{sec:c4conclu}
In this paper, we have computed the cosmological recombination 
history by using a multi-level code with the addition of the 
$2^3$P$_1$ to $1^1$S$_0$ spin-forbidden
transition for He\,{\sc i} and the two-photon transitions 
from $n$S and $n$D states to the ground state for both
H\,{\sc i} and He\,{\sc i}.  With the approximate rates from
Dubrovich \& Grachec~(2005)\,\cite{4dub05},
we find that there is more than a per cent decrease in the ionization 
fraction, which agrees broadly with the result they claimed.  
However, the only available accurate numerical value of two-photon rate 
with $n\geq 3$ is for the 3S to 1S and 3D to 1S transitons for H.  
We found that the approximate rates from
Dubrovich \& Grachec~(2005)\,\cite{4dub05} 
were overestimated, and instead we considered a scaled rate in order 
to agree with the numerical $n=3$ two-photon rate.  With this scaled 
rate, the change in $x_{\rm e}$ is no more than 0.5 per cent.

Including these extra forbidden transitions,
the change in the CMB anisotropy power spectrum is more 
than 1 per cent, which will potentially 
affect the determination of cosmological parameters in future
CMB experiments.  Since one would like the level of theoretical
uncertainty to be negligible, it is essential to include these
forbidden transitions in the recombination calculation.  In addition,
we still require accurate spontaneous rates to be calculated
for the two-photon transitions and also a code which includes
at least all the modifications suggested in Section~4.4, 
in order to obtain the $C_\ell$s down to the 1 per cent level.  
Achieving sub-percent accuracy in the calculations is challenging!

However, the stakes are high -- the determination of the parameters
which describe the entire Universe -- and so further work will be 
necessary.  Systematic deviations of the sort we have shown would 
potentially lead to incorrect values for the spectral tilt derived
from Planck and even more ambitions future CMB experiments, and 
hence incorrect inferences about the physics which produced the 
density perturbations in the very early Universe.  It is 
amusing that in order to understand physics at the $10^{15}$\,GeV 
energy scale we need to understand eV scale physics in exquisite 
detail!

\chapter[Reheating of matter]{Matter 
temperature\footnote[4]{A version of this chapter has 
been posted on the e-prints ArXiv:  Wong W.~Y. and Scott D.~(2006)
`Comment on ``Recombination induced softening and reheating of the cosmic plasma"',
ArXiv e-prints, arXiv:astro-ph/0612322.}}
\label{ch5:TM}

\section{Introduction}
Detailed calculations of the process through which the early
Universe ceased to be a plasma are increasingly important
because of the growing precision of microwave anisotropy 
experiments.  The standard way to calculate the evolution 
of the  matter temperature during the process of cosmological 
recombination is to consider the expansion of radiation and 
matter separately, and include the relevant interactions,
specifically Compton scattering~(see Equation~(\ref{eqTM})) and photoionization
cooling, as corrections~\cite{5peebles71,5peebles93,5seager00}.  
The matter is treated as a perfect gas 
which is assumed to envolve adiabatically.  
Recently, Leung et al.~(2004)\,\cite{5leung04} suggested that we need to 
use a generalized adiabatic index, since the gas was initially 
ionized and so the number of species (ion + electron vs atom) 
changes in the recombination process.  In their derivation, they 
considered the effect of photoionization, recombination and 
excitation on the matter, but assumed that the matter was 
undergoing an adiabatic process. However,
an adiabatic approximation for only the ionized matter is not valid 
in this case, because the change of entropy of the matter is not zero.    
Moreover, the photons released from the recombination of atoms mostly 
escape as free radiation\,\cite{5Wong:2005yr}, instead of reheating the matter,
since the heat capacity of the radiation is much larger
than that of the matter~(see, for example, \cite{5peebles93,5seager00}).

Here, we try to study this problem in a consistent way by 
considering both the radiation and ionizing hydrogen as components 
in thermal equilibium and under adiabatic expansion.  This is not
{\it exactly} the way things happened during recombination, 
but this will give us the maximum effect of the heat if it 
is all shared by the radiation and matter.

\section{Discussion}
For simplicity, we consider that the matter consists only of 
hydrogen (including helium does not change the physical picture).
By assuming that the radiation field and the matter are in 
thermal equilibrium, the total internal energy per unit mass 
of the system is
{\setlength\arraycolsep{2pt}
\begin{eqnarray}
E^{\rm int} &=& \frac{1}{n_{\rm H} m_{\rm H}} \left[ a T^4
+ \frac{3}{2}(n_{\rm H} + n_{\rm e})k_{\rm B}T 
+ n_{\rm p} \epsilon^{\rm H}_{\rm ion}
+ \sum_i n^{\rm H}_i \epsilon^{\rm H}_i \right] \nonumber \\ 
&=& \frac{1}{m_{\rm H}} \left[ \frac{a T^4}{n_{\rm H}}
+ \frac{3}{2}(1 + x_{\rm e})k_{\rm B} T 
+ x_{\rm p} \epsilon^{\rm H}_{\rm ion}
+ \sum_i x^{\rm H}_i \epsilon^{\rm H}_i \right],
\label{e1}
\end{eqnarray}} 
\\
where $n^{\rm H}_i$ is the number density of neutral atoms in 
the $i$th state, $n_{\rm p}$ is the number density of free protons,
$n_{\rm H} \equiv n_{\rm p}+\sum_i n^{\rm H}_i$ is the total number 
density of neutral and ionized hydrogens, and $n_{\rm e}$ is the 
number density of free electrons.  Additionally $\epsilon^{\rm H}_{\rm ion}$ 
and $\epsilon^{\rm H}_i$ are the ionization energy for the 
ground state and the $i$th state of hydrogen, respectively,  
the $x$s are the fractional number densities 
normalized by $n_{\rm H}$, $T$ is the temperature of the whole system,
$a$ is the radiation constant and $k_{\rm B}$ is Boltzmann's constant.

In equation~(\ref{e1}), the first term is the radiation energy, the
second term is the kinetic energy of the matter, and the last two terms
are the excitation energy of the atoms.  Here the energy of the ground
state is set to be equal to zero~\cite{5mihalas84}.  No matter 
what energy reference is chosen, the change of energy should 
be the same, i.e.
{\setlength\arraycolsep{2pt}
\begin{eqnarray}
dE^{\rm int} &=& \frac{1}{m_{\rm H}} \bigg[ \frac{4 a T^3}{n_{\rm H}}dT 
 - \frac{ a T^4}{n_{\rm H}^2}dn_{\rm H} 
+ \frac{3}{2}(1 + x_{\rm e})k_{\rm B} dT  \nonumber \\
&& \qquad \qquad  + \frac{3}{2}k_{\rm B} T dx_{\rm e} 
+\epsilon^{\rm H}_{\rm ion}  dx_{\rm p} 
+ \sum_i  \epsilon^{\rm H}_i dx^{\rm H}_i \bigg].
\label{de1}
\end{eqnarray}}
\\
We know that the radiation and matter are not {\it exactly} in thermal 
equilibrium (the two temperatures are not precisely the same) 
during the cosmological recombination of hydrogen, because
the recombination rate is faster than the rate of expansion
and cooling of the Universe. Nevertheless, the radiation
background and matter are tightly coupled and it is a good 
approximation to treat the two as if they were in thermal 
equilibrium~(for example, Peebles~1971\,\cite{5peebles71} P.232).
This simple approach allows us to estimate how much of the heat 
released is shared with the radiation field and the matter during the 
recombination of hydrogen.  In the expansion of the Universe, 
the whole system (radiation plus matter) is under an adiabatic process.
However, this is not the case for the ionizing matter on its own, because
the change of entropy of the matter is not zero.  
For an adiabatic process we have 
{\setlength\arraycolsep{2pt}
\begin{eqnarray}
dE^{\rm int} &=& P \frac{d\rho}{\rho^2} \\
&=& \frac{1}{m_{\rm H}}\left[ (1+ x_{\rm e})k_{\rm B} T + 
\frac{1}{3} \frac{a T^4}{n_{\rm H}} \right] \times \frac{3 dz}{1+z} ,
\label{de2}
\end{eqnarray}}
\\
where $P$ and $\rho$ are the pressure and mass density of the system
and $z$ is redshift.  
By equating equations~(\ref{de1}) and (\ref{de2}), we have
{\setlength\arraycolsep{2pt}
\begin{eqnarray}
\frac{1+z}{T} \frac{dT}{dz} &=&
\frac{3(1+x_{\rm e})k_{\rm B} T + \frac{4 a T^4}{n_{\rm H}}}  
{\frac{3}{2}(1+x_{\rm e})k_{\rm B}T + \frac{4 a T^4}{n_{\rm H}} }  
 - \frac{1+z}{\frac{3}{2}(1+x_{\rm e})k_{\rm B}T + \frac{4 a T^4}{n_{\rm H}}} 
\nonumber \\
&& \qquad \quad \times \left[ \frac{3}{2}k_{\rm B}T \frac{dx_{\rm e}}{dz}  
 + \epsilon^{\rm H}_{\rm ion}  \frac{dx_{\rm p}}{dz} 
+ \sum_i  \epsilon^{\rm H}_i \frac{dx^{\rm H}_i}{dz} 
\right].
\label{f1}
\end{eqnarray}}
\\*
In order to see whether we can ignore the radiation field, we need 
to compare the two terms in the denominator, i.e. the radiation 
energy and the kinetic energy of the matter.   If the matter energy
were much greater than the radiation energy, then we would  have  
{\setlength\arraycolsep{3pt}
\begin{eqnarray}
\frac{1+z}{T} \frac{dT}{dz} \simeq
2 - \frac{1+z}{\frac{3}{2}(1+x_{\rm e})k_{\rm B}T } 
\left[ \frac{3}{2}k_{\rm B}T \frac{dx_{\rm e}}{dz} 
+ \epsilon^{\rm H}_{\rm ion}  \frac{dx_{\rm p}}{dz} 
+ \sum_i  \epsilon^{\rm H}_i \frac{x^{\rm H}_i}{dz} 
\right],
\label{matter}
\end{eqnarray}}
\\*
which is the result given by Leung et al.~(2004)\,\cite{5leung04}.
However, in the current cosmological model
with $T_0 = 2.725$, $Y_{\rm p} = 0.24$, $h=0.73$ and 
$\Omega_{\rm b}=0.04$~(for example, \cite{5spergel06}), we have
{\setlength\arraycolsep{2pt}
\begin{eqnarray}
\frac{E_{\rm matter}^{\rm int}}{E_{\rm radiation}^{\rm int}} \simeq
\frac{n_{\rm H}k_{\rm B}T}{a T^4}
= \frac{n_{\rm H,0}k_{\rm B}}{a T_0^3} \simeq 1.6 \times 10^{-10}.
\end{eqnarray}}
\\*
So, the radiation energy is {\it much} larger than both the 
kinetic energy of matter and also the total heat released 
during recombination. Hence, we definitely cannot ignore the 
radiation field.  In such a case, the second term in 
equation~(\ref{f1}) is much smaller than the first term, because
{\setlength\arraycolsep{2pt}
\begin{eqnarray}
\frac{(1+z) \epsilon^{\rm H}_{\rm ion}}
{\frac{3}{2}(1+x_{\rm e})kT + \frac{4 a T^4}{n_{\rm H}}} \simeq
\frac{(1+z)n_{\rm H} \epsilon^{\rm H}_{\rm ion}}{a T^4} 
\sim 10^{-6}.
\end{eqnarray}}
\\*
Hence the energy change due to the recombination process is taken up mostly 
by the radiation field, since there are many more photons than baryons.
In other words, most of the extra photons (or heat) escape to the 
photon field, with just a very small portion ($\sim$\,$10^{-10}$) 
reheating the matter.  Therefore, the change of the temperature of the 
system can be approximated as
{\setlength\arraycolsep{2pt}
\begin{eqnarray}
\frac{1+z}{T} \frac{dT}{dz} \simeq\ 1 \pm \delta,
\end{eqnarray}}
\\*
where $\delta < 10^{-6}$.
This gives us back the usual formula for the radiation temperature,
which is consistent with the result that the matter temperature
closely follows the radiation temperatre~(for example, \cite{5peebles68,5seager00}).
Leung, Chan \& Chu~(2004)\,\cite{5leung04} assumed that the extra heat
is shared by the matter only, and hence that the second term of 
equation~(\ref{matter}) is significant, because
{\setlength\arraycolsep{2pt}
\begin{eqnarray}
\frac{(1+z)\epsilon^{\rm H}_{\rm ion}}{(1+x_{\rm e})kT} 
\simeq 6 \times 10^4. 
\end{eqnarray}}
\\
By comparing this and the ratio given in equation~(5.8), we can see that
the factor is about 10 orders of magnitude larger if we ignore the radiation 
field.  Another way to understand this overestimate is that the 
adiabatic approximation for matter {\it only} is not valid in their 
derivation, because the entropy of the matter is changing 
(i.e. $dS_{\rm matter}/dz > 0$).
The Leung et al.~(2004)\,\cite{5leung04} paper ignored the last term (the sum 
of the excitation energy terms) in equation~(5.6), which physically 
 means that there is a photon with energy equal to 
$\epsilon^{\rm H}_{\rm ion}$ ($\sim$13.6\,eV) emitted when a proton and
electron recombine, and the energy of this distortion photon 
is used up to heat the matter.  This is actually not true for the 
recombination of hydrogen, since there is no direct recombination
to the ground state~\cite{5peebles68,5zks68,5seager00}
and there are about 5 photons per neutral hydrogen atom produced 
for each recombination~\cite{5chluba06}.  

Note that what we calculate above is in the thermal equilibrium 
limit and it assumes that all the distortion photons are
{\it thermalized} with the radiation background and the matter.
However, in the standard recombination calculation, 
most of these distortion photons escape to infinity 
with tiny energy loss to the matter through Compton 
scattering.
The maximum fraction of energy loss by the distortion 
photons after multiple scatterings ($\Delta E_{\gamma}/E_{\gamma}$) 
is very low~\cite{5switzer05}.  An approximate estimate is
{\setlength\arraycolsep{2pt}
\begin{eqnarray}
\frac{\Delta E_{\gamma}}{E_{\gamma}} 
&\simeq& \frac{\epsilon^{\rm H}_{\rm ion}}{m_{\rm e}c^2} \, \tau \\
&\simeq & \frac{13.6 \, {\rm eV}}{511 \, {\rm keV}} \times 30 
\quad {\rm at} \  z \simeq 1500, \ {\rm when} \ x_{\rm e} \simeq 0.9 
\nonumber \\
& \simeq& 8 \times 10^{-4}, \nonumber
\end{eqnarray}}
where $m_{\rm e}$ is the mass of electron, $c$ is the speed of light and 
$\tau$ is the optical depth.
Therefore, the $\epsilon^{\rm H}_{\rm ion} dx_{\rm p}/dz$ term
is in practise suppressed by 
at least $10^{-4}$ (since $\tau$ decreases when more neutal
hydrogen atoms form at lower redshift).
Hence, although there is {\it some \/}heating of the matter, 
the ratio of the heat shared by the matter and the radiation 
is very small, and the effect claimed by 
Leung et al.~(2004)\,\cite{5leung04} is negligible
for the recombination history and also for the microwave
anisotropy power spectra.

\section{Conclusion}
By considering a simple model consisting of the radiation background 
and the ionizing gas under equilibrim adiabatic expansion, 
we show that the effect claimed by Leung et al.~(2004)\,\cite{5leung04}
 is hugely overestimated.
The appropriate method for calculating the matter temperature
is to deal with Compton and Thomson scattering between the 
background photons, distortion photons and matter in detail.
In general the Compton cooling time of the baryons off the CMB
is very much shorter than the Hubble time until $z \sim 200$, hence
it is extremely hard for any heating process to make the matter and 
radiation temperatures differ significantly at much earlier times.

\chapter[How well do we understand cosmological recombination?]
{How well do we understand recombination?\footnote[5]{A version 
of this chapter has been published: Wong W.~Y., 
Moss A. and Scott D.~(2008) `How well do we understand cosmological
recombination?',  Monthly Notices of the Royal Astronomical Society, 386, 1023-1028.}}

\section{Introduction}
{\sl Planck}~\cite{6Planck:2006}, the third generation 
Cosmic Microwave Background~(CMB) satellite 
will be launched in 2008; it will measure the CMB temperature 
and polarization anisotropies $C_\ell$ at multipoles $\ell=1$
to $\simeq 2500$ at much higher precision than has been possible
before.  In order to interpret these high fidelity 
experimental data, we need to have a correspondingly high 
precision theory. Understanding precise details of the recombination 
history is the major limiting factor in calculating the $C_\ell$ 
to better than 1 per cent accuracy.  An assessment of the level 
of this uncertainty, in the context of the expected 
{\sl Planck} capabilities, will be the subject of this chapter. 

The general physical picture of cosmological recombination 
was first given by Peebles~(1968)\,\cite{6Peebles:1968} 
and Zeldovich et al.~(1968)\,\cite{6Zeldovich:1968}.
They adopted a simple three-level atom model for hydrogen~(H), 
with a consideration of the Ly\,$\alpha$ and lowest order 2s--1s 
two-photon rates.  Thirty years later, Seager et al.\,\cite{6Seager:1999km} 
performed a detailed calculation by following all the 
resonant transitions and the lowest two-photon transition
in multi-level atoms for both hydrogen and helium in a
 blackbody radiation background.  
Lewis et al.~(2006)\,\cite{6Lewis:2006ym} first 
discussed how the uncertainties in recombination might 
bias the constraints on  
cosmological parameters coming from {\sl Planck}; 
this study was mainly motivated by the effect of 
including the semi-forbidden and high-order two-photon 
transitions~\cite{6Dubrovich:2005fc}, which had been
ignored in earlier calculations.

\begin{figure} 
  \begin{center}
    \includegraphics[width=0.8\textwidth]{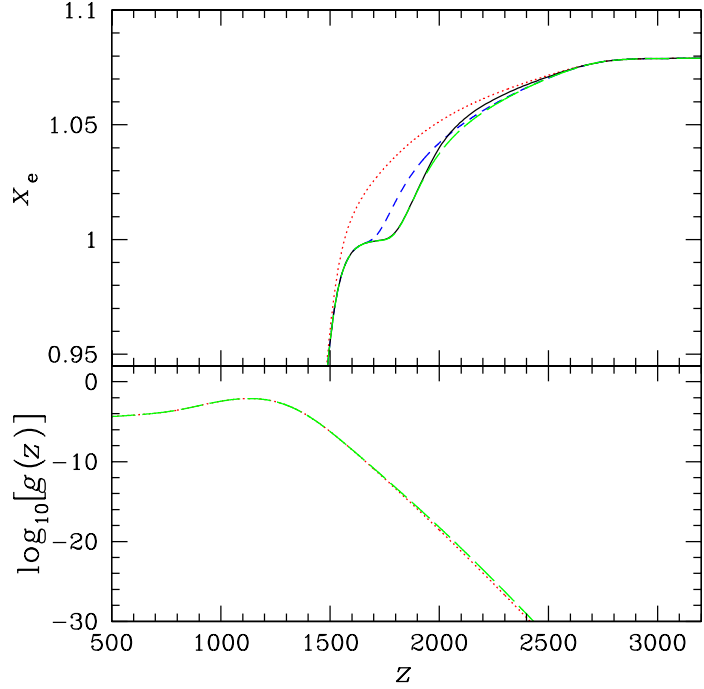}
    \caption[Ionization fraction $x_{\rm e}$ and the visibility function
as a function of redshift $z$ with different He\,{\sc i} scenarios.]
{Top panel: Ionization fraction $x_{\rm e}$ 
as a function of redshift $z$.
The dotted\,(red) line is calculated using the original RECFAST code.  
The solid\,(black) line is the numerical result from \cite{6Switzer:2007sq},
while the dashed\,(blue) and long-dashed\,(green) lines are both 
evaluated based on the modification given by 
\cite{6Kholupenko:2007qs} -- the dashed one 
has $b_{\rm He}=0.97$ (the value used in the original paper) and 
the long-dashed one has $b_{\rm He}=0.86$.
Bottom panel: The visibility function $g(z)$ versus redshift $z$.  
The two curves calculated~(dotted and long-dashed) correspond to the 
same recombination models in the upper panel.
The cosmological parameters used for these two graphs are, 
$\Omega_{\rm b} = 0.04592$, $\Omega_{\rm m} = 0.27$, 
$\Omega_{\rm C}= 0.22408$, $T_0=2.728$\,K, 
$H_0=71$ \,kms$^{-1}$ Mpc$^{-1}$ and $f_{\rm He}=0.079$.}
    \label{ch6fig1}
  \end{center}
\end{figure}

There have been many updates and improvements in the 
modelling of recombination since
then.  Switzer \& Hirata~(2008)\,\cite{6Switzer:2007sn} presented a multi-level
calculation for neutral helium~(He\,{\sc i}) recombination 
including evolution of the radiation field, which had been 
assumed to be a perfect blackbody in previous studies.
Other issues discussed recently include the continuum opacity due to 
neutral hydrogen\,(H\,{\sc i})~(see also \cite{Kholupenko:2007qs}), 
the semi-forbidden transition $2^3$p--$1^1$s
~(the possible importance of which was first proposed by \cite{6Dubrovich:2005fc}),
the feedback from spectral distortions between $2^1$p--$1^1$s 
and $2^3$p--$1^1$s lines, and the radiative line transfer.  
In particular, continuum absorption of the $2^1$p--$1^1$s line 
photons by neutral hydrogen causes helium recombination 
to end earlier than previously estimated~(see Fig.~\ref{ch6fig1}).
Hirata \& Switzer~(2008)\,\cite{6Hirata:2007sp} also found that the high order two-photon 
rates have a  negligible effect on He\,{\sc i},
and the same conclusion was made by other groups for hydrogen 
as well~\cite{6Chluba:2007qk,6Wong:2006iv}, largely because the 
approximate rates adopted by Dubrovich \& Grachev~(2005)\,\cite{6Dubrovich:2005fc} had been 
overestimated.  The biggest remaining uncertainty in He\,{\sc i}
recombination is the rate of the $2^3$p--$1^1$s transition, which causes
a variation equal to about 0.1 per cent in the ionization fraction
$x_{\rm e}$~\cite{6Switzer:2007sq}.

For hydrogen, Chluba et al.(2006)\,\cite{6Chluba:2006bc} improved the multi-level 
calculation by considering seperate angular momentum $\ell$ 
states.  This brings about a 0.6 per cent change in $x_{\rm e}$ 
at the peak of the visibility function, and about 
1 per cent at redshifts $z < 900$.  The effect of the induced 
2s--1s two-photon rate due to the radiation background\,\cite{6Chluba:2005uz}
is partially compensated by the feedback of the Ly\,$\alpha$ photons
\,\cite{6Kholupenko:2006jm}, and the net maximum effect on $x_{\rm e}$
is only 0.55 per cent at $z \simeq 900$.  
The high-order two photon transitions bring about a 0.4 per cent
change in $x_{\rm e}$ at $z \simeq 1160$\,\cite{6Chluba:2007qk,6Wong:2006iv}.
There are also 0.22 per cent changes in $x_{\rm e}$ at $z \simeq 1050$ when
one considers the Lyman series feedback up to $n=30$, and there is additionally
the possibility of direct recombination, although this 
has only a roughly $10^{-4}$ per cent effect\,\cite{6Chluba:2007yp}.

The list of suggested updates on $x_{\rm e}$ is certainly not complete
yet, since some additional effects, such as the convergence of 
including higher excited states and the feedback-induced corrections 
due to the He\,{\sc i} spectral distortions,  may enhance or cancel 
other effects.  In general we still need to develop a complete 
multi-level code for hydrogen, with detailed interactions between the
atoms and the radiation field.  However, what is really important here is
establishing how these effects propagate into possible
systematic uncertainties in the estimation of cosmological 
parameters.

Since the uncertainties in cosmological recombination discussed 
in the Lewis et al.~(2006)\,\cite{6Lewis:2006ym} paper have 
been reduced or updated, it is
time to revisit the topic on how the new effects or remaining 
uncertainties might affect the constraints on cosmological parameters in 
future experiments.
The recent full version of the He\,{\sc i} recombination 
calculation\,\cite{6Hirata:2007sp,6Switzer:2007sn,6Switzer:2007sq} 
takes too long to run to be included within the current
Boltzmann codes for $C_\ell$.  So instead, in this paper, we try to 
reproduce the updated ionization history by modifying 
{\sc recfast}\,\cite{6Seager:1999bc} using a simple parametrization
based on the fitting formulae provided by 
Kholupenko et al.~(2007)\,\cite{6Kholupenko:2007qs}.
We then use the C{\sc osmo}MC\,\cite{6Lewis:2002ah} code to 
investigate how much this impacts the constraints on cosmological 
parameters for an experiment like {\sl Planck}.

\section{Recombination model}
In this paper, we modify {\sc recfast} based on the fitting 
formulae given by Kholupenko et al.~(2007)\,\cite{6Kholupenko:2007qs} 
for including the effect of the continuum opacity of neutral hydrogen 
for He\,{\sc i} recombination.  The basis set of rate equations
 of the ionization fraction of H\,{\sc i} and He\,{\sc i} used in 
{\sc recfast} are: {\setlength\arraycolsep{1pt}
\begin{eqnarray}
&& H(z)(1+z) {dx_{\rm p}\over dz}  = 
 \Big(x_{\rm e}x_{\rm p} n_{\rm H} 
 \alpha_{\rm H} - \beta_{\rm H} (1-x_{\rm p}) 
   {\rm e}^{-h_{\rm P}\nu_{\rm H2s}/kT_{\rm M}}\Big) C_{\rm H}, \\
\label{eqHeI_xe}
&& H(z)(1+z) {dx_{\rm He II}\over dz} =  
\left(x_{\rm He II}x_{\rm e} n_{\rm H} \alpha_{\rm HeI}
   - \beta_{\rm HeI} (f_{\rm He}-x_{\rm He II})
   {\rm e}^{-h_{\rm P} \nu_{{\rm HeI}, 2^1{\rm s}}/kT_{\rm M}}\right) C_{\rm HeI} \nonumber \\
  && \ + \left(x_{\rm He II}x_{\rm e} n_{\rm H} \alpha^{\rm t}_{\rm HeI}
   - \frac{g_{{{\rm HeI}, 2^3{\rm s}}}}{g_{{\rm HeI}, 1^1{\rm s}}} 
   \beta^{\rm t}_{\rm HeI} (f_{\rm He}-x_{\rm He II})
   {\rm e}^{-h_{\rm P} \nu_{{\rm HeI},2^3{\rm s}}/kT_{\rm M}}\right)C_{\rm HeI}^{\rm t} ,
\end{eqnarray}}
\\
where
{\setlength\arraycolsep{1pt}
\begin{eqnarray}
&& C_{\rm H} = {1 + K_{\rm H} \Lambda_{\rm H} n_{\rm H}(1-x_{\rm p})
    \over 1+K_{\rm H} (\Lambda_{\rm H} + \beta_{\rm H})
     n_{\rm H} (1-x_{\rm p}) }, \\
\label{6Cfactor}
&& C_{\rm HeI} = {1 + K_{\rm HeI} \Lambda_{\rm He} n_{\rm H}
  (f_{\rm He}-x_{\rm He II}){\rm e}^{h_{\rm P} \nu_{\rm ps}/kT_{\rm M}}
  \over 1+K_{\rm HeI}
  (\Lambda_{\rm He} + \beta_{\rm HeI}) n_{\rm H} (f_{\rm He}-x_{\rm He II})
  {\rm e}^{h_{\rm P} \nu_{\rm ps}/kT_{\rm M}} },  \\
&& C_{\rm HeI}^{\rm t} = {1  \over 1+K^{\rm t}_{\rm HeI}
  \beta^{\rm t}_{\rm HeI} n_{\rm H} (f_{\rm He}-x_{\rm He II})
  {\rm e}^{h_{\rm P} \nu^{\rm t}_{\rm ps}/kT_{\rm M}}}.
\end{eqnarray}}
\\
Note that $x_{\rm e}$ is defined as the ratio of free electons
per H atom and so \mbox{$x_{\rm e} > 1$} during He recombination. 
We follow the exact notation used in 
Seager et al.~(1999)\,\cite{6Seager:1999bc} and 
we do not repeat the definitions of all symbols, except those that
did not appear in that paper.  
The last term in 
equation~(\ref{eqHeI_xe}) is added to the original  $dx_{\rm He II}/dz$
rate for the recombination of He\,{\sc i} through the triplets by 
including the semi-forbidden transition from the $2^3$p state to
the $1^1$s ground state.  This additional term can be easily 
derived by considering an extra path for 
electrons to cascade down in He\,{\sc i} by going from the continuum
through $2^3$p to ground state, and assuming that the rate of change of 
the population of the $2^3$p state is negligibly small. 
 The superscript `t' stands for triplets, so that, for example,
$\alpha^{\rm t}_{\rm HeI}$ is the Case B He\,{\sc i} recombination
coefficient for triplets.  Based on the data given by 
Hummer \& Storey~(1998)\,\cite{6Hummer:1998}, 
$\alpha^{\rm t}_{\rm HeI}$ is fitted with the same functional 
form used for the $\alpha_{\rm HeI}$ 
singlets~(see equation~(4), in \cite{6Seager:1999bc}),
with different values for the  parameters: 
 $p=0.761$; $q=10^{-16.306}$; $T_1=10^{5.114}$\,K; and $T_2=3$\,K. 
This fit is accurate to better than 1 per cent for temperatures between 
$10^{2.8}$ and $10^{4}$\,K. Here $\beta^{\rm t}_{\rm HeI}$ is the 
photoionization coefficient for the triplets and is calculated
from $\alpha^{\rm t}_{\rm HeI}$ by
{\setlength\arraycolsep{1pt}
\begin{eqnarray}
\beta^{\rm t}_{\rm HeI} = \alpha^{\rm t}_{\rm HeI}
\left(\frac{2 \pi m_{\rm e} k_{\rm B} T_{\rm M}}{h_{\rm P} ^2}\right)^{3/2}
\frac{2 g_{\rm He^+}}{g_{\rm HeI, 2^3s}} e^{-h_{\rm P} \nu_{\rm 2^3s,c}/kT_{\rm M}},
\end{eqnarray}}
\\
where $g_{\rm He^+}$ and $g_{\rm HeI, 2^3s}$ are the degeneracies of
He\,{\sc ii} and of the He\,{\sc i} atom with electron in the $2^3$s state,
and $h_{\rm P} \nu_{\rm 2^3s,c}$ is the ionization energy of the $2^3$s state.  

The correction factor $C_{\rm HeI}$ accounts for the 
slow recombination due to the bottleneck of the He\,{\sc i}
$2^1$p--$1^1$s transition among singlets.  We can also derive
the corresponding correction factor $C_{\rm HeI}^{\rm t}$
for the triplets.
The $K_{\rm H}$, $K_{\rm HeI}$ and $K^{\rm t}_{\rm HeI}$ quantities
are the cosmological redshifting of the H Ly\,$\alpha$, 
He\,{\sc i} $2^1$p--$1^1$s  and He\,{\sc i} $2^3$p--$1^1$s 
transition line photons, respectively.  The factor $K$  
used in {\sc recfast} is a good approximation
 when the line is optically thick ($\tau \gg 1$) and
the Sobolev escape probability $p_{\rm S}$ is roughly equal to $1/\tau$.
In general, we can relate $K$ and $p_{\rm S}$ through the following 
equations~(taking He\,{\sc i} as an example):
{\setlength\arraycolsep{1pt}
\begin{eqnarray}
&& K_{\rm HeI} = \frac{g_{{\rm HeI}, 1^1{\rm s}}}{g_{{\rm HeI}, 2^1{\rm p}}}
\frac{1}{ n_{{\rm HeI}, 1^1{\rm s}}  
A^{\rm HeI}_{ 2^1{\rm p}-1^1{\rm s}} p_{\rm S}} \quad {\rm and}\\
&& K^{\rm t}_{\rm HeI} = \frac{g_{{\rm HeI}, 1^1{\rm s}}}{g_{{\rm HeI}, 2^3{\rm p}}}
\frac{1}{ n_{{\rm HeI}, 1^1{\rm s}}  
A^{\rm HeI}_{ 2^3{\rm p}-1^1{\rm s}} p_{\rm S}} \ , 
\label{eqKHeI}
\end{eqnarray}}
where 
$A_{{\rm HeI}, 2^1{\rm p}-1^1{\rm s}}$ and $A_{{\rm HeI}, 2^3{\rm p}-1^1{\rm s}}$ 
are the Einstein $A$ coefficients of the He\,{\sc I} $2^1$p--$1^1$s 
and  He\,{\sc I} $2^3$p--$1^1$s transitions, respectively.  Note 
that $A_{{\rm HeI}, 2^3{\rm p} -1^1{\rm s}}= 
g_{{\rm HeI}, 2^3{\rm P}_1}/g_{{\rm HeI}, 2^3{\rm p}}
 \times A_{{\rm HeI}, 2^3{\rm P}_1-1^1{\rm s}} 
= 1/3 \times 177.58\,$s$^{-1}$~\,\cite{6Lach:2001}.   
For He\,{\sc i} $ 2^1$p--$1^1$s, 
we replace $p_{\rm S}$ by the new escape probability $p_{\rm esc}$, 
to include the effect of the continuum opacity due to H, 
based on the approximate formula suggested by 
Kholupenko et al.~(2007)\,\cite{6Kholupenko:2007qs}. 
Explicitly this is
{\setlength\arraycolsep{1pt}
\begin{eqnarray}
&& p_{\rm esc} = p_{\rm S} + p_{\rm con, H} \, , \\
{\rm where} &&  \nonumber \\
&& p_{\rm S} = \frac{1 - e^{-\tau}}{\tau} \quad {\rm and} \\
\label{pconHe}
&& p_{\rm con, H} = \frac{1}{1 + a_{\rm He} \gamma^{b_{\rm He}}}, \\
{\rm with \ } && \nonumber \\
&& \gamma = \frac{\frac{g_{{\rm HeI}, 1^1{\rm s}}}{g_{{\rm HeI}, 2^1{\rm p}}}
A^{\rm HeI}_{2^1{\rm p}-1^1{\rm s}} (f_{\rm He} - x_{\rm HeII})c^2}
{8 \pi^{3/2} \sigma_{{\rm H},1{\rm s}}(\nu_{\rm HeI,2^1 p}) 
\nu_{\rm HeI,2^1{\rm p}}^2 \Delta \nu_{\rm D,2^1p} 
(1 - x_{\rm p})}\, , \nonumber
\end{eqnarray}}
\\
where $\sigma_{{\rm H},1s}(\nu_{\rm HeI,2^1p})$ is the H\,{\sc i} ionization 
cross-section at frequency $\nu_{\rm HeI,2^1p}$ and 
$\Delta \nu_{\rm D,2^1p} = \nu_{\rm HeI,2^1p} \sqrt{2 k_{\rm B} 
T_{\rm M}/m_{\rm He} c^2}$ 
is the thermal width of the He\,{\sc i} $2^1$p--$1^1$s line.
The $\gamma$ factor in $p_{\rm con, H}$ is approximately the 
ratio of the He\,{\sc i} $2^1$p--$1^1$s transition rate to the 
H\,{\sc i} photoionization rate.  When $\gamma \gg 1$, the effect 
of the continuum opacity due to neutral hydrogen on 
the He\,{\sc i} recombination is negligible.
Here $a_{\rm He}$ and $b_{\rm He}$ are fitting parameters, which are equal to
0.36 and 0.97, based on the results from 
Kholupenko et al.~(2007)\,\cite{6Kholupenko:2007qs}.

\begin{figure}
  \begin{center}
    \includegraphics[width=0.8\textwidth]{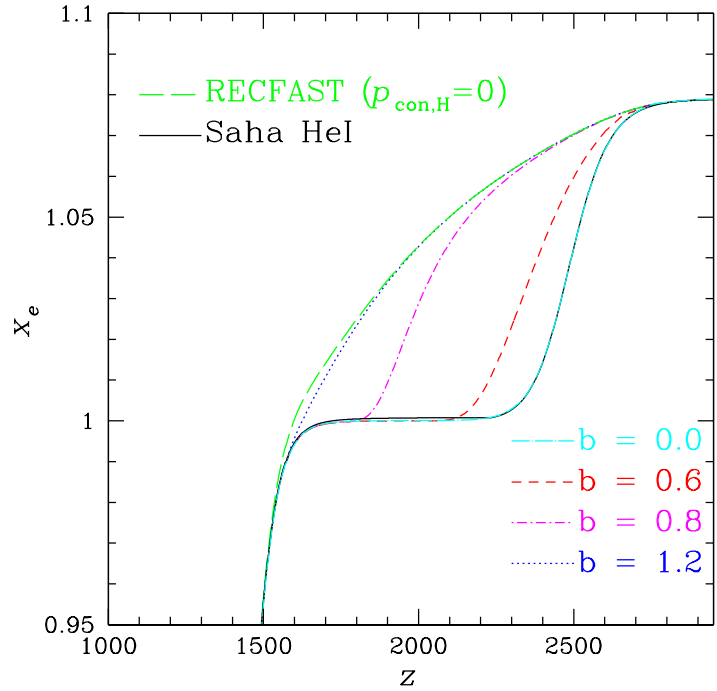}
    \caption[Ionization fraction $x_{\rm e}$ as a function of 
redshift $z$ calculated with the modified He\,{\sc i} recombination 
of different values of the helium fitting 
parameter $b_{\rm He}$]
{Ionization fraction $x_{\rm e}$ as a function of 
redshift $z$ calculated based on the modified He\,{\sc i} recombination 
discussed here with different values of the helium fitting 
parameter $b_{\rm He}$. 
The curve with $b_{\rm He}=0$~(long-dashed, cyan) overlaps  
the line using Saha equilibrium recombination~(solid, black).
 The cosmological parameters used in this graph are
the same as for Fig.~\ref{ch6fig1}.}
   \label{ch6fig2}
  \end{center}
\end{figure}

We now try to reproduce these results with our modified {\sc recfast}.
Fig.~\ref{ch6fig1} (upper panel) shows the numerical result 
of the ionization fraction $x_{\rm e}$ from different 
He\,{\sc i} recombination calculations.  The results from 
Kholupenko et al.~(2007)\,\cite{6Kholupenko:2007qs} and 
Switzer \& Hirata~(2008)\,\cite{6Switzer:2007sq}
both demonstrate a significant speed up of He\,{\sc i} recombination
compared with the original {\sc recfast}.  We do not expect these 
two curves to match each other, since 
Kholupenko et al.~(2007)\,\cite{6Kholupenko:2007qs}
just included the effect of the continuum opacity 
due to hydrogen, which is only one of the main improvements stated 
in Switzer \& Hirata~(2008)\,\cite{6Switzer:2007sq}.
Nevertheless, we can regard the 
Kholupenko et al.~(2007)\,\cite{6Kholupenko:2007qs} study 
as giving a simple fitting model in a three-level atom to 
account for the speed-up of the He\,{\sc i} recombination.
Fig.~\ref{ch6fig2} shows how the ionization history changes 
with different values of the fitting parameter $b_{\rm He}$
(with $a_{\rm He}$ fixed to be 0.36).
  When $b_{\rm He}$ is larger than 1.2, the effect of the 
neutral H is tiny and the fit returns to the situation
with no continuum opacity.  However, if $b_{\rm He}$ 
is smaller than 1, the effect of the continuum opacity 
becomes more significant.  Note that when $b_{\rm He}=0$, 
both the escape probability $p_{\rm esc}$ and the
correction factor $C_{\rm HeI}$ are close to unity.
This means that almost all the emitted photons can escape
 to infinity and so the ionization history returns to Saha 
equilibrium for He\,{\sc i} recombination. 

%

This simple fitting formula can reproduce quite well 
the detailed numerical result for the 
ionization history at the later stages of He\,{\sc i} 
recombination.  From Fig.~\ref{ch6fig1}, we can see 
that our model with $b_{\rm He}=0.86$ matches with the
 numerical result at $z \lesssim 2000$~\cite{6Switzer:2007sq}. 
Although our fitting model does not agree so well with 
the numerical results for the earlier stages of He\,{\sc i} 
recombination, the effect on the $C_\ell$ is 
neligible.  This is because the visibility function 
$g(z) \equiv e^{- \tau} d \tau/dz$,
is very low at $z > 2000$ (at least 16 orders of magnitude 
smaller than the maximum value of $g(z)$), as shown in 
the lower panel of
Fig.~\ref{ch6fig1}.  Our fitting approach also appears 
to work well for other cosmological 
models~(Switzer \& Hirata, private communication).  

\begin{figure}
  \begin{center}
    \includegraphics[width=0.8\textwidth]{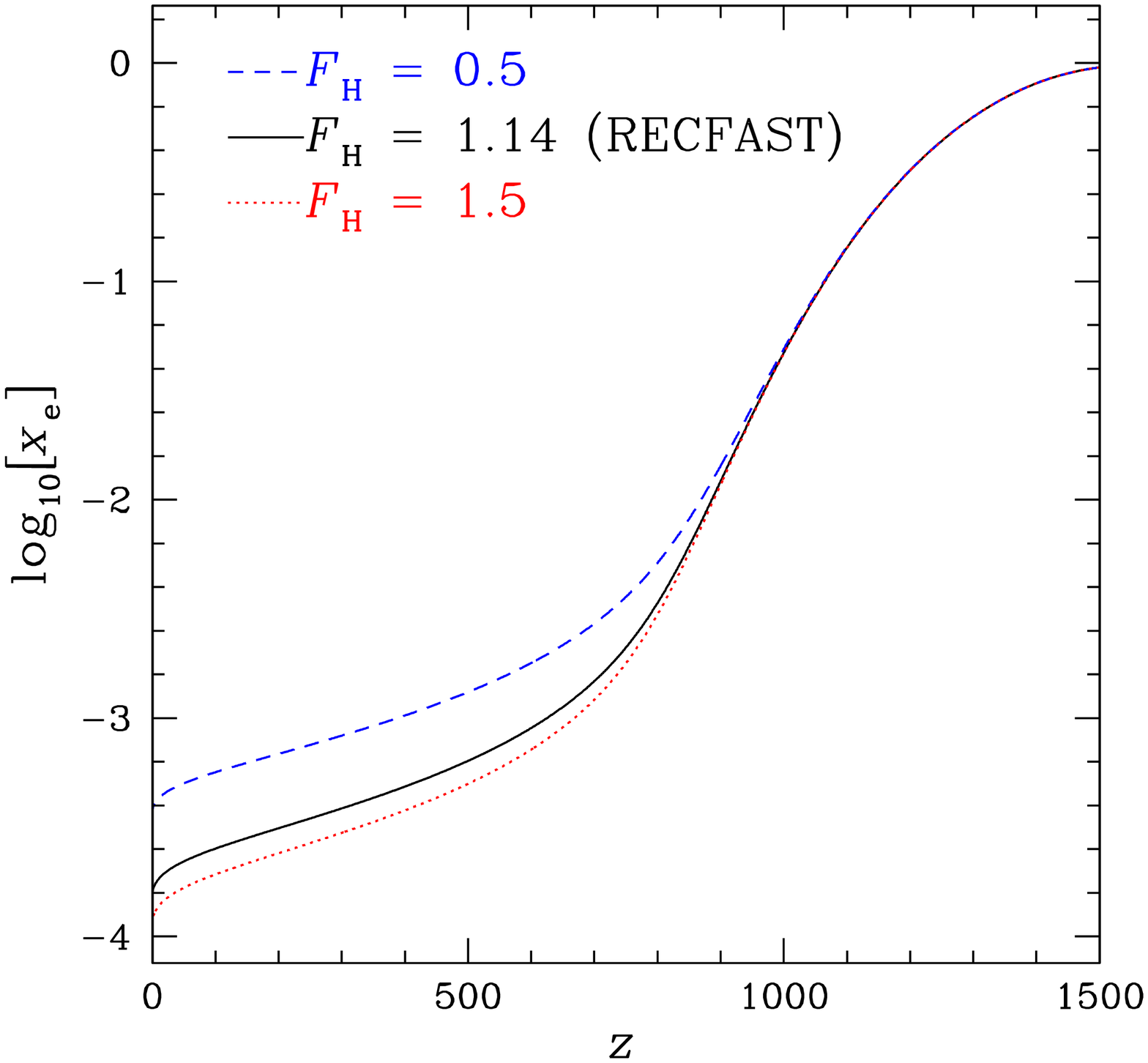}
\caption[The ionization fraction $x_{\rm e}$ as a function of 
redshift $z$ calculated with different values of the hydrogen fudge factor
$F_{\rm H}$.]
{The ionization fraction $x_{\rm e}$ as a function of 
redshift $z$ calculated with different values of the hydrogen fudge factor
$F_{\rm H}$.  The cosmological parameters used in this graph are
the same as in Fig.~\ref{ch6fig1}.}
   \label{ch6fig3}
  \end{center}
\end{figure}

In this paper, we employ the fudge factor $F_{\rm H}$ 
for H~(which is the extra factor multiplying
$\alpha_{\rm H}$) and the He\,{\sc i} parameter 
$b_{\rm He}$ in our model to represent the remaining 
uncertainties in recombination.  For He\,{\sc i}, the 
factors $a_{\rm He}$ and $b_{\rm He}$ in 
equation~(\ref{pconHe}) are highly correlated.
We choose to fix $a_{\rm He}$ and use $b_{\rm He}$ as 
the free parameter in this paper; this is because
it measures the power dependence of the ratio of 
the relevant rates $\gamma$ in the escape probability 
due to the continuum opacity $p_{\rm con, H}$.
For hydrogen recombination, all the individual updates
suggested recently give an overall change less 
than 0.5 per cent in $x_{\rm e}$ around 
the peak of the visibility function.  Only the effect of 
considering the separate $\ell$-states causes more than
a 1 per cent change, and only for the final stages 
of hydrogen recombination ($z \lesssim 900$).
Therefore, we think it is sufficient
to represent this uncertainty with the usual fudge factor
$F_{\rm H}$, which basically controls the speed of the 
end of hydrogen recombination (see Fig.~\ref{ch6fig3}).
The best-fit to the current recombination calculation 
has $F_{\rm H}\simeq 1.14$.

\section{Forecast data}
We use the C{\sc osmo}MC\,\cite{6Lewis:2002ah} code to 
perform a Markov Chain Monte Carlo (MCMC) 
calculation for sampling the posterior distribution with 
given forecast data.  The simulated {\sl Planck} data and 
 likelihood function are generated based on the 
settings suggested in Lewis et al.\,\cite{6Lewis:2006ym}. 
We use full polarization information for {\sl Planck} by considering 
the temperature $T$ and $E$-type polarization
anisotropies for $\ell \leq 2400$, and assume that they are 
statistically isotropic and Gaussian.  The noise is 
also isotropic and is based on a simplified model with
$N^{TT}_{\ell} = N^{EE}_{\ell}/4 = 2 \times 10^{-4} \mu$K$^2$, 
having a Gaussian beam of 7 arcminutes
(Full Width Half Maximum, \cite{6Planck:2006}).
For our fiducial model, we adopt the best values of the six 
cosmological parameters in a $\Lambda$CDM model from the {\sl WMAP} 
three-year result\,\cite{6Spergel:2006hy}.
The six parameters are the baryon density 
$\Omega_{\rm b}h^2$\,=\,0.0223, the cold dark matter density 
$\Omega_{\rm C}h^2$\,=\,0.104, the present Hubble parameter 
 $H_0$\,=\,73\,km\,s$^{-1}$Mpc$^{-1}$, the constant scalar 
adiabatic spectral index $n_{\rm s}$\,=\,0.951, the scalar amplitude 
(at $k$\,=\,0.05\,Mpc$^{-1}$) 
$10^{10}A_{\rm s}$\,=\,3.02 and the optical depth due to 
reionization (based on a sharp transition) $\tau$\,=\,0.09. 
For recombination, we calculate the ionization history 
using the original R{\sc ecfast} with the fudge 
factor for hydrogen recombination $F_{\rm H}$ set to 1.14
and the helium abundance equal to 0.24.

\section{Results}
\begin{figure}
\begin{center}
\includegraphics[width=1.0\textwidth]{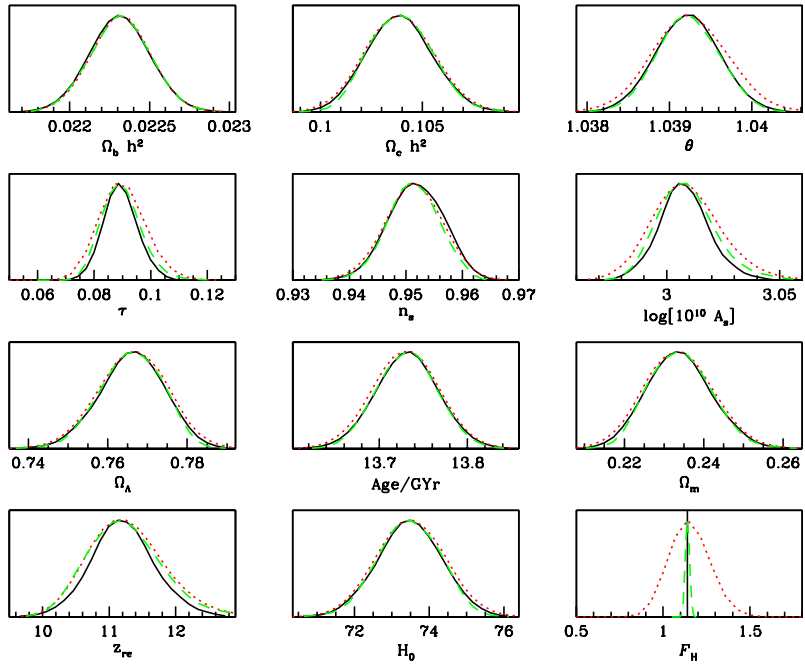}
\caption[Marginalized posterior distributions for forecast 
{\sl Planck} data varying the hydrogen recombination only.]
{Marginalized posterior distributions for forecast 
{\sl Planck} data varying the hydrogen recombination only.
All the curves are generated using the original R{\sc ecfast} code.
The solid\,(black) curve uses fixed $F_{\rm H}$, while the
dotted\,(red) and dashed\,(green) allow for varying $F_{\rm H}$ 
with Gaussian distributions centred at 1.14, with 
$\sigma = 0.1$ and 0.01, respectively.  Note that using a flat
prior (between 0 and 1.5) for $F_{\rm H}$ gives the same spectra 
as the case with $\sigma = 0.1$ (the red dotted line).}
\label{ch6fig4}
\end{center}
\end{figure}

Fig.~\ref{ch6fig4} shows the parameter constraints from our forecast
{\sl Planck} likelihood function using the original {\sc recfast} 
code with varying $F_{\rm H}$ and adopting different priors.
For the {\sl Planck} forecast data, $F_{\rm H}$ can be well
constrained away from zero (the same result as in\,\cite{6Lewis:2006ym}) 
and is bounded by a nearly Gaussian distribution
with $\sigma$ approximately equal to 0.1.
When we only vary $F_{\rm H}$ with different priors 
(compared with fixing it to 1.14), it basically does not change 
the size of the error bars on the cosmological parameters, 
except for the scalar adiabatic amplitude $10^{10}A_{\rm s}$.  
From Fig.~\ref{ch6fig2}, 
we can see that the factor $F_{\rm H}$
controls the speed of the final stages of H\,{\sc i} recombination,
when most of the atoms and electrons 
have already recombined.  Changing $F_{\rm H}$ affects the 
optical depth $\tau$ from Thomson scattering, which determines 
the overall normalization amplitude of $C_\ell$ ($\propto e^{-2 \tau}$) 
at angular scales below that subtended by the size of the horizon at last
scattering~($\ell \gtrsim 100$).  This is the reason 
why varying $F_{\rm H}$ affects the uncertainty in $A_{\rm s}$, 
since $A_{\rm s}$ also controls the overall amplitude of $C_\ell$
(see the upper right panel in Fig.~\ref{ch6fig6} for the marginalized 
distribution for $F_{\rm H}$ and $A_{\rm s}$).
The modified recombination model also changes the peak value
(but not really the width) of the adiabatic spectral index 
$n_{\rm s}$ distribution, as one can see by comparing the dotted and 
dashed curves in Fig.~\ref{ch6fig4}.

Based on all the suggested effects on H\,{\sc i} recombination, the 
uncertainty in $x_{\rm e}$ is at the level of a few per cent
 at $z \lesssim 900$, which corresponds to roughly a 1
per cent change in $F_{\rm H}$.  In Fig.~\ref{ch6fig4}, 
we have also tried to take this uncertainty into 
account by considering a prior on $F_{\rm H}$ with $\sigma=0.01$ 
(the long-dashed curves).  We find that the result is almost 
the same as for the case using $\sigma=0.1$ for the $F_{\rm H}$ prior.
On the other hand, the error bar (measured using the 68 per cent 
confidence level, say) of $A_{\rm s}$ is increased by 40 
and 16 per cent with $\sigma=0.1$ and 0.01, respectively.

\begin{figure}
\begin{center}
\includegraphics[width=1.0\textwidth]{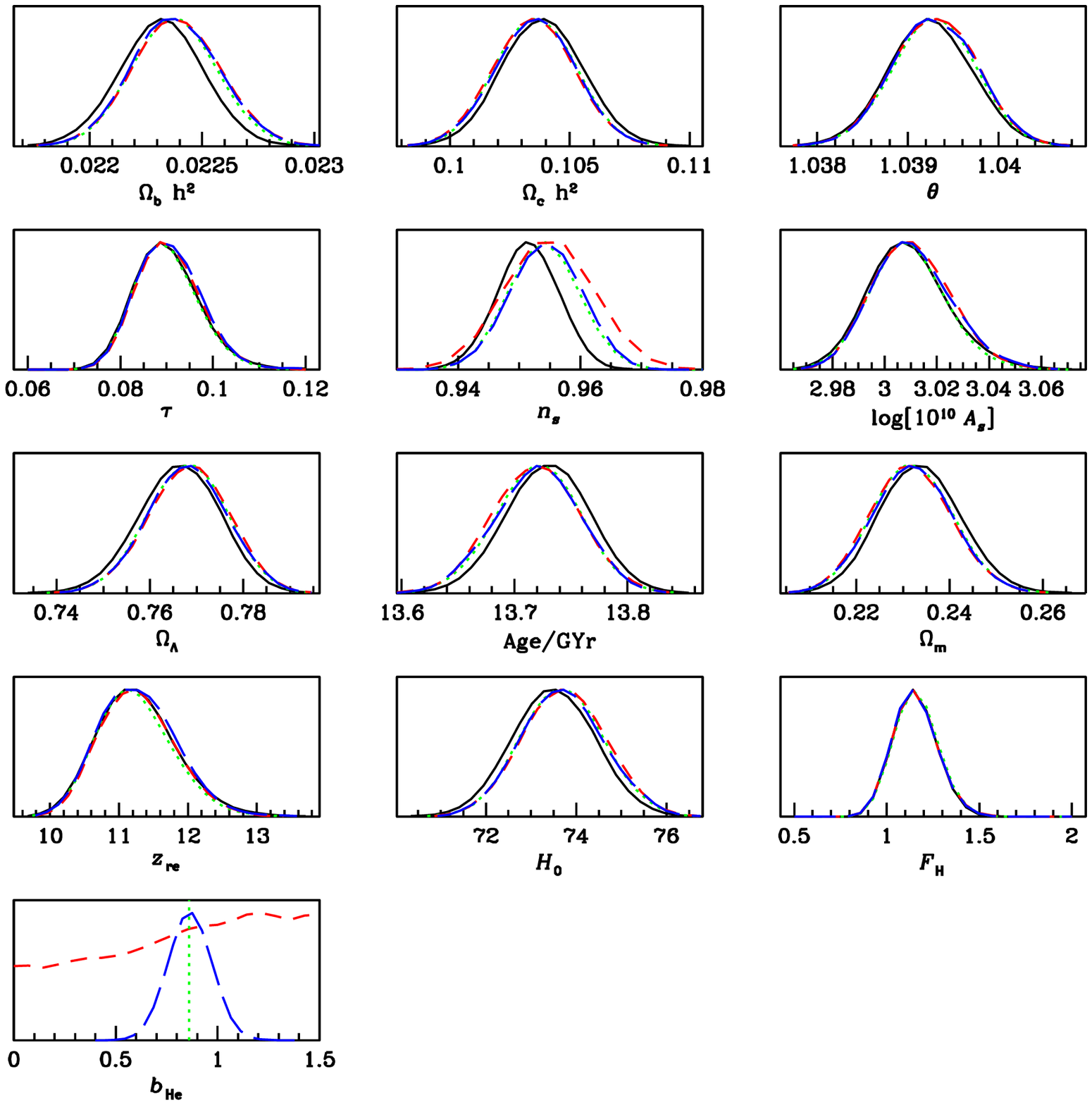}
\caption[Marginalized posterior distributions for forecast 
{\sl Planck} data with hydrogen and helium phonomenological parameters
both allowed to vary.]
{Marginalized posterior distributions for forecast 
{\sl Planck} data with hydrogen and helium phonomenological parameters
both allowed to vary.
The solid\,(black) curve shows the constraints using the 
original {\sc recfast} code and allowing $F_{\rm H}$ to be  
a free parameter.  The other curves also allow for the 
variation of $F_{\rm H}$ and use the fitting function 
for He\,{\sc i} recombination described in Section 2:
the dotted\,(green) line sets $b_{\rm He}$ equal to 0.86; the dashed\,(red)
 one is with a flat prior for $b_{\rm He}$ from 0 to 1.5; and the 
long-dashed\,(blue) one is with a narrow prior for $b_{\rm He}$, consisting of 
a Gaussian centred at 0.86 and with $\sigma=0.1$.}
\label{ch6fig5}
\end{center}
\end{figure}

Fig.~\ref{ch6fig5} shows the comparison of the constraints in
the original and modified versions of {\sc recfast}, with
both H\,{\sc i} and He\,{\sc i} parameterized.  
By comparing the solid and dotted curves in Fig.~\ref{ch6fig5}, 
we can see that only the peaks of the spectra of the cosmological 
parameters are changed, but not the width of the distributions,
when switching between the original and modified {\sc recfast} codes.
Allowing $b_{\rm He}$ to float in the 
modified recombination model only leads to an increase in the error bar 
for spectral index $n_{\rm s}$ among all the parameters, 
including $F_{\rm H}$.  For the dashed curves, we used a very 
conservative prior for $b_{\rm He}$, namely a flat spectrum 
from 0 to 1.5~(i.e. from Saha recombination to the old 
R{\sc ecfast} behaviour).  We can see that the value of $b_{\rm He}$ 
is poorly constrained, because the CMB is only weakly sensitive 
to the details of He\,{\sc i} recombination.  
Nevertheless, this variation allows for faster He\,{\sc i}
recombination than in the original R{\sc ecfast} code and this 
skews the distribution of $n_{\rm s}$ towards higher 
values~(see also the upper left panel in Fig.~\ref{ch6fig6}).  
This is because a faster He\,{\sc i} recombination leads to 
fewer free electrons before H\,{\sc i} recombination and this increases 
the diffusion length of the photons and baryons.  This 
in turn decreases the damping scale of the acoustic oscillations
at high $\ell$, which therefore gives a higher value of $n_{\rm s}$.
In addition, this variation in $b_{\rm He}$ 
increases the uncertainty (at the 68 per cent confidence level) 
of $n_{\rm s}$ by 11 per cent.

Based on the comprehensive study of 
Switzer \& Hirata~(2008)\,\cite{6Switzer:2007sq},
the dominant remaining uncertainty in He\,{\sc i} recombination
is the $2^3$p--$1^1$s transition rate, which causes about 
a 0.1 per cent variation in $x_{\rm e}$ at $z \simeq 1900$.
For our fitting procedure this corresponds to about a 1 
per cent change in $b_{\rm He}$.  We try to take this 
uncertainty into account in our calculation by adopting 
a prior on $b_{\rm He}$ which is peaked at 0.86 with width~(sigma) 
liberally set to 0.1.  From Fig.~\ref{ch6fig5}, one can see that the 
error bar on $n_{\rm s}$ is then reduced to almost the 
same size as found when fixing $b_{\rm He}$ equal to 0.86 
(the dotted and long-dashed curves).  This means that, for
the sensitivity expected from {\sl Planck},
it is sufficient if we can determine $b_{\rm He}$ 
to better than 10 per cent accuracy.

As well as the individual marginalized uncertainties,
we can also look at whether there are degeneracies among the 
parameters.  From Fig.~\ref{ch6fig6}, we see that $F_{\rm H}$ and 
$b_{\rm He}$ are quite independent.
This is because the two parameters govern recombination 
at very different times.  As discussed before, $b_{\rm He}$
 controls the speed of He\,{\sc i} recombination, which affects 
the high-$z$ tail of the visiblity function, while $F_{\rm H}$ 
controls the low-$z$ part.

\begin{figure}
\begin{center}
\includegraphics[width=0.8\textwidth]{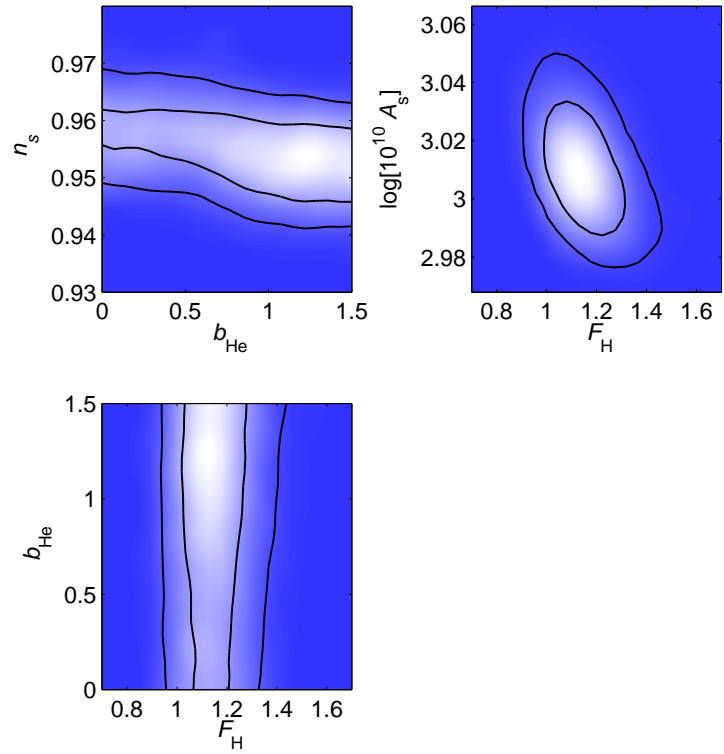}
\caption[Projected 2D likelihood for the four parameters
$n_{\rm s}$, $A_{\rm s}$, $F_{\rm H}$ and $b_{\rm He}$.]
{Projected 2D likelihood for the four parameters
$n_{\rm s}$, $A_{\rm s}$, $F_{\rm H}$ and $b_{\rm He}$.
Shading corresponds to the marginalized probabilities 
with contours at 68 per cent and 95 per cent confidence.}
\label{ch6fig6}
\end{center}
\end{figure}

\section{Discussion and conclusions}
In this paper, we modify {\sc recfast} by introducing
one more parameter $b_{\rm He}$ (besides the hydrogen fudge 
factor $F_{\rm H}$) to mimic the recent numerical results for 
the speed-up of He\,{\sc i} recombination.  
By using the C{\sc osmo}MC code with forecast {\sl Planck}
data, we examine the variation of these two factors to 
account for the remaining dominant uncertainties 
in the cosmological recombination calculation.  For He\,{\sc i}, 
the main uncertainty comes from the $2^3$p--$1^1$s 
rate\,\cite{6Switzer:2007sq}, which corresponds to 
about a 1 per cent change in $b_{\rm He}$.  We find that
this level of variation has a negligible effect on the 
determination of the cosmological parameters.  Therefore, 
based on this simple model, if the existing studies have 
properly considered all the relevant physical radiative processes 
in order to provide $x_{\rm e}$ to 0.1 per cent accuracy during 
He\,{\sc i} recombination, then we already have numerical
 calculations which are accurate enough for {\sl Planck}.

For H, since there is still no comprehensive 
model which considers all the interactions between
the atomic transitions and the radiation background,
we consider the size of the updates as 
an indication of the existing level of uncertainty.
We represent this uncertainty by varying the 
fudge factor $F_{\rm H}$, because the largest update on 
$x_{\rm e}$ occurs at $z \lesssim 900$, and comes from  
a consideration of the separate angular momentum
states~\cite{6Chluba:2006bc}.  We find that
$F_{\rm H}$ needs to be determined to better than 
1 per cent accuracy in order to have negligible 
effect on the determination of cosmological parameters
with {\sl Planck}. 

 Hydrogen recombination is of course
important for the formation of the CMB anisotropies 
$C_{\ell}$, since it determines the detailed profile 
around the peak of the visibility function $g(z)$.
A comprehensive numerical calculation of the recombination of H\,{\sc i}
(similar to He\,{\sc i}) to include at least all the recent
 suggestions for updates on the evolution of 
 $x_{\rm e}$ is an urgent task.  We need to determine that
 the phenomenological parameters $F_{\rm H}$ and $b_{\rm He}$
 are fully understood at the $\lesssim 1$ per cent level before we 
 can be confident that the uncertainties in the details of 
 recombination will have no significant effect on the
 determination of cosmological parameters from {\sl Planck}.

\newpage

\chapter{Summary and Future work}
\section{Effects of distortion photons}
In this thesis, we have presented the detailed profile
 of the spectral distortion to the Cosmic Microwave 
 Background~(CMB) due to the H\,{\sc i} Ly\,$\alpha$ 
 and 2s--1s two-photon transitions,
 and the corresponding lines of He\,{\sc i} and He\,{\sc ii}.
The main peak of the distortion is from the Ly\,$\alpha$ line 
 and is located at $\lambda$\,=\,170\,$\mu$m in the standard
 cosmological $\Lambda$CDM model.
Although the detection of these spectral distortions 
 will be quite challenging due to the presence of the
 Cosmic Infrared Background~(CIB), they would provide a direct
 probe for the detailed physical processes during the 
 recombination epoch.
These high energy distortion photons also have significant 
 effects on the recombination of lithium\,\cite{7Switzer:2005} 
 and formation of the primordial molecules\,\cite{7Hirata:2006}
 in the cosmological `dark ages' at redshift $z$\,$<$\,500.  
Recently, Switzer \& Hirata~(2005)\,\cite{7Switzer:2005} showed 
 that the distortion photons from H\,{\sc i} strongly suppress 
 and delay the formation of neutral lithium (Li\,{\sc i}).  
They found that neutral lithium is three orders of magnitude
 smaller than found in previous studies, which assumed a perfect 
 blackbody radiation background~(see \cite{7Galli:1998,7Lepp:2002,7Puy:2002} for reviews).  
This dramatically reduces the optical depth of Li\,{\sc i} and makes
 the effects of Li\,{\sc i} scattering on the CMB anisotropies
 unobservable\,\cite{7Switzer:2005}.

Despite the effect of these spectral distortions reducing
 the strength of some potentially observable anisotropy
 effects, there may be other, related effects which 
 {\it are} detectable.
Basu et al.~(2004)\,\cite{7Basu:2004} and
Hern{\'a}ndez-Monteagudo \& Sunyaev~(2005)\,\cite{7Hernadez:2005}  
 have shown that other sources of line scattering
 might lead to interesting signatures from the $z$\,$\sim$3--25 
 universe.
In a seperate study\,\cite{7Rubino:2005} it was suggested 
 that the spectral lines themselves, each with a different 
 effective visibility function, could lead to anisotropy 
 signatures which probe different epoches. 
Although all of these effects are relatively weak, 
 as the sensitivity of experiments increases,
 it seems likely that these subtle effects, which
 are essentially mixed anisotropy and spectral 
 signatures, will become of increasing importance. 
 
The primordial molecules~(for example, H$_2$, HD and LiH) are important
 in the formation of the first stars, since molecular cooling plays 
 a significant role in the first collapse of baryonic matter, 
 when the amplitudes of structures grow non-linear and 
 virialize\,\cite{7Galli:1998,7Lepp:2002}.
With the addition of the distortion photons, the 
 abundance of primordial H$_2$ was found to be about
 75\% less compared with previous studies\,\cite{7Hirata:2006}.  
Note that the cooling of gas is more effective through the
 HD dipole radiation than through the quadrupole radiation 
 from H$_2$, and therefore understanding the formation of HD 
 may be very important.
Since the main route for the formation of HD is 
 H$_2$\,+\,D$^+$\,$\rightarrow$\,HD\,+\,H$^+$, 
 it will be worth performing a follow-up calculation
 for HD with the updated populations of H$_2$.  

\section{A single numerical code for recombination}
From the above discussion, it is clear that the detailed 
 spectrum of the distortion photons can have strong 
 influence on the formation of primordial molecules.  
The distortion spectrum in turn depends strongly on 
 details of the radiative  processes in cosmological
 recombination.  
But the main motivation of improving the recombination calculation
 is to obtain an accurate visibility function for
 CMB anisotropies.
In anticipation of upcoming CMB experiments which
 push to smaller angular scales with higher 
 sensitivity~(for example, {\sl Planck}\,\cite{7Planck:2006},
 ACT\,\cite{7Kosowsky:2003sw} and SPT\,\cite{7Ruhl:2004}),
 it is crucial to understand all the relevant physical
 processes during recombination which may contribute 
 more than (say) 0.1\% to the ionization fraction $x_{\rm e}$,
 in order not to bias the cosmological parameter extraction.
In this thesis, we studied the effect on recombination of the He\,{\sc i} 
 2$^3$P$_1$--1$^1$S$_0$ spin-forbidden transition and also the higher
 order non-resonant two-photon transitions~($n$S--1S and $n$D--1S) 
 of H\,{\sc i} and He\,{\sc i} in a multi-level atom model.
We found that more than 40\% of electrons cascade down to the ground
 state through the 2$^3$P$_1$--1$^1$S$_0$ spin-forbidden transition 
 from the $n$\,=\,2 state, and the inclusion of this transition brings
 more than a 1\% change in $x_{\rm e}$ compared with previous studies.
We also adopted improved two-photon rates for the transitions 
 from 3S to 1S and 3D to 1S by including all the non-resonant 
 poles through the high-lying intermediate
 $n$P states~($n$\,$>$\,4)\,\cite{7ctsc86,7fsm88}.  
Our best estimated H\,{\sc i} non-resonance two-photon rates are 
 lower than the ones from Dubrovich \& Grachev~(2005)\,\cite{7Dubrovich:2005fc}
 due to destructive interference in the matrix element; 
 and so from this effect we found no more than a 0.5\% 
 change in $x_{\rm e}$.

Although in Chapter~4, we only considered the effect of 
 some of these specific additional transitions, 
 there have been many other recent updates on recombination
 calculation, as discussed in Chapter~2 and the discussion 
 sections in Chapter~3, 4 and 6.
Most of the suggested improvements are concerned
 with consistently treating the radiative interactions
 between matter and the surrounding photons.
We revisited one of the previous studies\,\cite{7Leung:2003je},
 which claimed that the matter was reheated by the distortion photons
 from recombination and that this delayed the H\,{\sc i} recombination.  
We found that the energy transfer between the distortion 
 photons and the matter (through Compton scattering) 
 is very inefficient, and the resulting effect on $x_{\rm e}$ 
 is no more than $10^{-6}$. 
This is much lower than the previous estimate and 
 hence this effect can be safely ignored.

Many suggestions for improvements to recombination have
 been carried out in different independent numerical codes,
 and therefore the overall effect of all the modifications
 is still uncertain.  
Recently, there has been a comprehensive study of helium 
 recombination\,\cite{7Hirata:2006,7Switzer:2007sn,7Switzer:2007sq},
 which includes most of the physical processes relevant of 
 calculating $x_{\rm e}$ at the 0.1\% level. 
Since 92\% of the atoms in the Universe are hydrogen, 
 it follows that H\,{\sc i} recombination is considerably
 more important in determing the detailed profile of the 
 last scattering surface for CMB photons. 
So a remaining task is to perform a similar systematic
 study for H\,{\sc i} recombination, or even a full 
 calculation combining the H and He cases.

Once all the relevant corrections for the detailed numerical
 recombination calculation have been solidified, we need to
 incorporate a modified approximate version of these effects
 into a fast code similar to {\sc recfast}\,\cite{7Seager:1999bc} for
 incorporating into the Boltzmann codes~(for example, 
{\sc cmbfast}\,\cite{7Seljak:1996} and {\sc camb}\,\cite{7Lewis:2000})
 which are used for calculating the CMB anisotropies, $C_{\ell}$s.
This is because the current detailed numerical recombination 
 calculations take far too long (typically more than a day) to 
 yield results for a single cosmological model.
In the previous chapter, we introduced an extra parameter 
 $b_{\rm He}$ in the current {\sc recfast} to approximately 
 model the speed-up of He\,{\sc i} recombination due to 
 the continuum opacity of H\,{\sc i}.  
This modified {\sc recfast} can be considered as the first 
 step in parametrizing the other recent result from the detailed 
 numerical codes into a simple three-level atom calculation.

We also studied how varying $b_{\rm He}$ along with the
 existing hydrogen fudge factor $F_{\rm H}$ might account for
 some of the remaining uncertainties in recombination.
Using the C{\sc osmo}MC code with {\sl Planck} forecast 
 data~($\ell$\,$\leq$\,2500), we found that we need to determine 
 the effective value of $b_{\rm He}$ to better than 10\% and 
 $F_{\rm H}$ to better than 1\%.
The current He\,{\sc i} recombination studies seem to 
 already calculate $x_{\rm e}$ accurately enough for {\sl Planck}, 
 but we still require a comprehensive study for H\,{\sc i} to 
 reach the same level of accuracy.
Note that these two phenomenological parameters mainly affect
 the determination of  the scalar amplitude $A_{\rm s}$ and 
 the spectral index $n_{\rm s}$ of the primordial perturbation spectrum.  
There are other CMB experiments, such as the Atacama Cosmology Telescope 
 (ACT)\,\cite{7Kosowsky:2003sw} which will be able to 
 measure $C_{\ell}$s over a wide range
 of angular scales (1000\,$<$\,$\ell$\,$<$10000); such 
 measurements can put tight constraints on the tilt of the
 temperature power spectrum, which is characterized by the 
 primodial spectral index $n_{\rm s}$.
For these and even better future experiments, we may need to 
 determine these two phenomemological 
 parameters ($F_{\rm H}$ and $b_{\rm He}$) to better than
 the \,1\% level in order to obtain the correct 
 inferences about inflationary models.
Alternatively, we should systematically account for all the 
 relevant updates on recombination, in additional to the
 one recent correction which we included in the modified
 \mbox{{\sc recfast}} code.  
There is still much work to be done!



\end{document}